\documentclass[iop]{emulateapj-rtx4}
\usepackage{graphicx}

\shorttitle{All Weather Calibration of Wide Field Optical and NIR Surveys}  
\shortauthors{Burke et al.}

\begin{document}
\title{All Weather Calibration of Wide Field Optical and NIR Surveys}

\author{David L. Burke$^1$, Abhijit Saha$^2$, Jenna Claver$^2$, T. Axelrod$^3$, Chuck Claver$^2$,  \\
        Darren DePoy$^4$, \v{Z}eljko Ivezi\'{c}$^5$,   Lynne Jones$^5$,   \\
                R. Chris Smith$^6$, Christopher W. Stubbs$^7$}
\affil{$^1$SLAC National Accelerator Laboratory, Menlo Park, CA, 94025 USA}
\affil{$^2$National Optical Astronomy Observatory, Tucson, AZ, 85718 USA}
\affil{$^3$Steward Observatory, University of Arizona, Tucson, AZ, 85718 USA}
\affil{$^4$Texas A\&M University, College Station, TX, 77843 USA}
\affil{$^5$Department of Astronomy, University of Washington, Seattle, WA, 98195, USA}
\affil{$^6$Cerro Tololo Inter-American Observatory, Casilla 603, La Serena, Chile}
\affil{$^7$Harvard Smithsonian Center for Astrophysics, Harvard University, Cambridge, MA, 02138 USA}
\email{daveb@slac.stanford.edu}

\begin{abstract}
The science goals for ground-based large-area surveys, such as the Dark Energy Survey, Pan-STARRS, and the Large Synoptic Survey Telescope,
require calibration of broadband photometry that is stable in time and uniform over the sky to precisions of a per cent or better.
This performance will need to be achieved with data taken over the course of many years, and often in less than ideal conditions.
This paper describes a strategy to achieve precise internal calibration of imaging survey data taken in less than ``photometric'' conditions,
and reports results of an observational study of the techniques needed to implement this strategy.
We find that images of celestial fields used in this case study with stellar densities $\sim 1$/arcmin$^2$ 
and taken through cloudless skies can be calibrated with relative precision $\sim 0.5\%$ (reproducibility).
We report measurements of spatial structure functions of cloud absorption observed over a range of atmospheric conditions,
and find it possible to achieve photometric measurements that are reproducible to 1\% in images that were taken through cloud layers
that transmit as little as 25\% of the incident optical flux (1.5 magnitudes of extinction).
We find, however, that photometric precision below 1\% is impeded by the thinnest detectable cloud layers.
We comment on implications of these results for the observing strategies of future surveys.
\end{abstract}


\section{ Introduction}
\label{Sec:intro}
 
A digital camera on a modern ground-based astronomical telescope will count a fraction of the photons produced by a
celestial source that reach the top of the atmosphere.
For broad-band observations the digital count (ADU) associated with a source
is proportional to the integral of the optical flux $F_\nu(\lambda)$ from the
source weighted by the observational bandpass, $S_b(x,y,alt,az,t,\lambda)$,
\begin{equation}
\label{eqn:cnts}
             ADU_b^{meas} = A \! \int_0^{\Delta T} \!\!\!\! dt \int_0^\infty \!\!\!\!
                             {F_\nu(\lambda) S_b(x,y,alt,az,t,\lambda) \lambda^{-1} d\lambda},
\end{equation}
where $A$ is the area of the telescope pupil and $\Delta T$ is the duration of the exposure.
The units of flux $F_{\nu}(\lambda)$ are ergs cm$^{-2}$ s$^{-1}$ Hz$^{-1}$,
and the factor $\lambda^{-1}d\lambda$ counts the number of photons per unit energy at a given wavelength.
(Strictly, this should be $(h\lambda)^{-1}d\lambda = - (h\nu)^{-1}d\nu$, but the units can be chosen to absorb
the factor of Planck's constant $h$ into the definition of the instrumental system response.)
The coordinates $(x,y)$ are those of the source image in the focal plane of the camera,
$(alt,az)$ are the altitude and azimuth of the telescope pointing,
and $t$ is the time (and date) of the observation. 

Assuming the atmospheric and instrumental properties are uncorrelated, the optical passband can be separated into two functions,  
\begin{equation}
\label{eqn:optb}
         S_b(x,y,alt,az,t,\lambda) = S_b^{inst}(x,y,t,\lambda) \times S^{atm}(alt,az,t,\lambda),
\end{equation}
where $S^{atm}$ is the optical transmittance (dimensionless) from the top of the atmosphere to the input pupil of the telescope,
and $S_b^{inst}$ is the instrumental system response (ADU/photon) from photons through the input pupil of the telescope to ADU counts in the camera.
This instrumental ``throughput'' includes the reflectance of the mirrors, transmission of the refractive optics and optical filters, efficiency 
of the camera sensors, and the gain of the electronics used to read out the detectors.
The relative spatial variation in $S_b^{inst}$ is usually measured with some combination of artificial illumination
with diffuse light from a screen mounted in the dome of the telescope housing
and star flats assembled from dithered exposures of fields with high densities of stars \citep{magnier04, regnault09}.
If care is taken to manage effects of thermal and mechanical changes of the telescope,
then it is possible to obtain a relatively stable instrumental response throughout a night of observing.
Careful analysis can reduce errors in the flat fielding across small fields of view to sub-percent accuracies \citep{landolt92, stetson05}.

But the atmospheric transmittance can vary considerably more rapidly and by significantly greater amounts \citep{stubbs07}. 
Precise photometry is traditionally achieved only during times when atmospheric conditions are stable and clear of clouds.
A night is generally regarded as ``photometric'' if multiple observations of the same target yield sufficiently 
small variations in measured magnitudes.
Only at the best sites do these conditions exist for significant fractions of the observing calendar. 

The science goals for future ground-based all-sky surveys, such as the Dark Energy Survey (DES) \citep{flaugher07}                         ,
Pan-STARRS \citep{kaiser02}, and the Large Synoptic Survey Telescope (LSST) \citep{ivezic08}
pose stringent requirements on the stability and uniformity of photometric measurements.
These surveys seek relative calibration of photometry that is stable in time and uniform over the sky to precisions of a per cent or better.
This performance will need to be achieved with measurements made from multiple images taken over the course of many years.
And to maximize efficiency, these surveys will observe in less than ideal conditions.
In this paper we demonstrate a method to precisely calibrate data taken in non-photometric conditions.

\subsection{Atmospheric Extinction and the Nature of Clouds}

Processes that attenuate light as it propagates through the atmosphere include absorption and scattering (Rayleigh)
by molecular constituents (O$_{2}$, O$_{3}$, water vapor, and trace elements),
scattering (Mie) by airborne macroscopic particulate aerosols with physical dimensions comparable to the wavelength of visible light,
and shadowing by ice crystals and water droplets in clouds.
These process all have characteristic dependences on wavelength and airmass, and atmospheric transmittance can be separated into three terms,
\begin{eqnarray}
\label{eqn:Sgray}
         S^{atm}(alt,&&az,t, \lambda) = S^{gray}(alt,az,t)          \times {}   \nonumber\\
                    && {}  \times S^{molecular}(alt,az,t, \lambda)  \times {}   \nonumber\\
                    && {}  \times S^{Mie}(alt,az,t, \lambda).
\end{eqnarray} 

In a previous paper \citep{burke10} we reported a technique to remove effects of attenuation by atmospheric molecules and aerosols in 
calibration of ground-based observations at optical and near-infrared (NIR) wavelengths.
The technique reported there takes advantage of state-of-the-art models of atmospheric optical radiation transport and
readily-available codes to accurately compute atmospheric extinction over a wide range of observing conditions.
Spectra of a small catalog of bright ``probe'' stars are taken
as they progress across the sky and back-light the atmosphere during the night.
The signatures of various atmospheric constituents in these spectra are used to extract the makeup of the atmosphere across the sky. 
This technique was shown to provide excellent reconstruction of the wavelength-dependent portion of atmospheric transmission.

In addition to determination of the wavelength-dependent atmospheric extinction,
calibration of wide-field imaging survey data requires recognition and correction for
absorption and scattering of light by water droplets and ice crystals in clouds.
These particulates are large compared to the wavelength of visible light, so they produce shadows that
are wavelength independent ($S^{gray}$ in Eq. \ref{eqn:Sgray}).
Cloud structure can be intricate with significant spatial variations across a single field of view,
and can change in the time interval between exposures \citep{fliflet06, koren08}.
In the worst case, the loss of light will be too severe to allow useful data to be taken.
But it will be important to be able to correct data for thin cloud cover that may not be apparent to the naked eye,
and it will certainly be of great benefit to be able to take useful data over an extended range of atmospheric conditions. 
There are several important features of future dedicated surveys that should enable this to be done.
These surveys will use telescopes with fields of view several degrees across,
and they will be sensitive enough to capture several tens of thousands of stars in each image.
These surveys also will observe these stars multiple times, and so will be sensitive to changes in observing conditions.

\subsection{Strategy for ``All-Weather'' Calibration and Outline of this Paper}

Here we extend our previous work to include photometric corrections for cloud cover.
We seek a calibration of each image in a data set relative to a reference condition of the sky encountered during the survey.
This reference condition may be the most photometric night, but it need not be;
it could be a frequently encountered condition that optimizes statistical sampling of celestial objects. 
Absolute calibration of this reference condition is a later step.

Our goal is a robust calibration procedure that will achieve highly reproducible results from survey data taken over
a wide range of conditions rather than optimum results from tightly selected subsamples of the data.
Our strategy is to construct a catalog of stars that appear as point-sources in the survey images themselves.
The number of such sources, and the precision with which their magnitudes are measured, must be sufficient 
to create an accurate (up to an overall constant) map $S^{gray}(alt,az,t)$ of the gray extinction at all points on each image.
The required density and precision of measurements of calibration sources is determined by the spatial structure of the absorption of light
by cloud cover and the photometric requirements of the survey.
The goal of this study is to provide a more quantitative picture of these conditions and requirements.

The calibration procedure begins with a standard photometric reduction of images carried out using bias,
dark, and flat-field images taken on a nightly cadence.
The technique discussed in our previous paper is used to determine wavelength-dependent atmospheric extinction for each image; 
and the observed instrumental magnitudes of sources on each image are corrected to the top of the atmosphere with the assumption that the 
spectral energy distribution (SED) $f_{\nu}(\lambda)$ of the object is independent of wavelength ({\it i.e.} ``flat''). 
An internal standard magnitude system is defined, and linear color corrections are then applied to transform the flat SED magnitudes
to in-band top-of-the-atmosphere magnitudes.
The final step is a global calibration analysis based on the successful SDSS \"Ubercal technique \citep{padman08}.
This analysis determines the wavelength-independent extinction for every image in the survey sample,
and completes the definition of the internal calibration.
This paper discusses each of these in the context of analysis of data acquired during an observing run that we carried out at Cerro Tololo
in northern Chile.

\section{Observing Campaign and Reduction of Data}
\label{sec:campaign}

An observing campaign was done for this study during a 3-night period in July 2009
at the Cerro Tololo InterAmerican Observatory (CTIO) in northern Chile.
Data were taken on each night with the SMARTS 1.5m Cassegrain telescope with the Ritchey-Chr\'etien spectrograph (RCSPEC),
and with the Blanco 4m telescope with the MOSAIC II wide-field imager.

\subsection{The Spectroscopic Data and Analysis}

Targets observed with the spectrograph are listed in Table \ref{table:spectargets}.
These were all used in our previous studies,  
and the data were acquired and reduced in the same manner as reported in the earlier paper.
These data were used to extract extinction parameters for the wavelength-dependent processes introduced above.

\begin{table} 
\caption{Stars Observed with SMARTS RCSPEC.
         \label{table:spectargets}}
\begin{center} 
\begin{tabular}{lccl} 
\tableline 
\noalign{\smallskip}
  Target     &      RA (J2000)    &      DEC (J2000)      \\  
\noalign{\smallskip}
\tableline
\noalign{\smallskip} 
CD-329927    &   14 11 46  &   -33 03 14   \\
LTT7379      &   18 36 26  &   -44 18 37   \\
HD189910     &   20 03 18  &   -25 08 45   \\
HD207474     &   21 49 10  &   -02 02 24   \\
LTT9239      &   22 52 41  &   -20 35 53   \\
F110         &   23 19 28  &   -05 09 56   \\
CD-35534     &   01 32 04  &   -34 29 15   \\
\noalign{\smallskip}
\tableline 
\end{tabular} 
\end{center}
\end{table}

Data were taken with RCSPEC during parts of each of the three nights of the run.
The instrumental setup employed a low-dispersion grating (\#11) at 12.5$\degr$ tilt blazed at 8000\AA.
All observing was done in first order with an OG530 blocking filter to eliminate second-order light at wavelengths below 10500\AA.
This resulted in a dispersion of 5.4\AA/pixel on the CCD, FWHM of 16.4\AA, and resolution R $\approx$ 400 at 6500\AA.
The slit on RCSPEC does not rotate to track the parallactic angle of refraction,
so all spectra were taken with a 10$\arcsec$ full aperture.
Target stars were first visually centered with the slit closed to 1$\arcsec$, then the slit was opened to acquire spectra.

Reduction of spectra was carried out as in our previous work.
Briefly, bias frames and dome flats were taken daily and used in reductions of the 2-d CCD images of the slit.
Quartz lamp flats and neon lamp calibration exposures were taken with each new pointing of the telescope.
Reductions of 2-d images to 1-d spectra were performed with IRAF software\footnote{IRAF software is distributed by the
National Optical Astronomy Observatory, which is operated by the Association of Universities for Research in Astronomy (AURA)
under cooperative agreement with the National Science Foundation.},
and included overscan and bias removal, sky subtraction, and wavelength calibration. 
Initial correction of the spectra for instrumental signature and flux calibration was done using standard targets Fiege 110 and LTT 9239
without telluric correction.

The reduced spectra were fit with model stellar spectra \citep{kurucz93} and templates of atmospheric absorption
computed with the MODTRAN code \citep{MODTRAN}.
These fits yield coefficients that define the column heights of various atmospheric components relative to the 1976 U.S. Standard Atmosphere mix 
used to compute the templates.
The fitting model for atmospheric transmittance of light used in this analysis is summarized with the formula:
\begin{eqnarray}
\label{eqn:fitform}
S^{model}&&(alt,az,t;\lambda) = S^{gray}(alt,az,t)                                                             \times {} \nonumber\\ 
    & & {}\times   (1.0 - C_{mol} (BP(t)/BP_0)     A_{mols}(z(alt);\lambda))                                   \times {} \nonumber\\
    & & {}\times   (1.0 - \sqrt{C_{mol}BP(t)/BP_0} A_{mola}(z(alt);\lambda))                                   \times {} \nonumber\\
    & & {}\times   (1.0 - C_{O3}                   A_{O3}  (z(alt);\lambda))                                  \times {} \nonumber\\
    & & {}\times   (1.0 - C_{H2O}(alt,az,t)        A_{H2O} (z(alt);\lambda))                                   \times {} \nonumber\\
    & & {}\times    e^{\left(-z(alt) \cdot \tau_{aerosol}(alt,az,t;\lambda)\right)}.
\end{eqnarray}
\noindent The attenuation coefficients $A_{i}$ are computed as one (1.0) minus the transmission MODTRAN templates.
Refinement of the instrumental signature is also included in these fits as described in our earlier publication.

For purposes of this work, the important components are molecular scattering and absorption ($A_{mol}$), ozone ($A_{O3}$),
and the aerosol (Mie) optical depth represented by the formula
\begin{equation}
\label{eqn:aertau}
   \tau_{aerosol}(alt,az,t; \lambda) = (\tau_{0} + \tau_{1} EW + \tau_{2} NS) \left(\frac{\lambda}{\lambda_0}\right)^\alpha,
\end{equation}
where $\lambda_0$ = 6750\AA~is chosen for convenience in the middle of the wavelength range of the observations,
and $EW = cos(alt)sin(az)$ and $NS = cos(alt)cos(az)$ are projections of the telescope pointing respectively
in the east-west and north-south directions.
Molecular scattering is given by the barometric pressure (the value $BP_0 = 782$ mb is used for reference on Cerro Tololo),
and the parameters $C_{O3}$, $\alpha$, and $\tau_{i}$ were derived for each night of observing.
Absorption by water vapor ($A_{H2O}$) is negligible in the $r$ band used primarily in the analysis reported here.
The results of the fits to the data taken during the run are given in Table \ref{table:specfits}.
These are used to correct all imaging data and to define the atmospheric component
of the standard passbands (Subsection \ref{subsection: Stdbands} below).
The atmospheric extinctions ranged from a few per cent in $i$-band images taken near zenith
to $\sim 20\%$ for $g$-band images taken at the largest zenith angles.
From our previous work we expect measurements of $r$-band extinction to be accurate to $\sim 0.003-0.004$ magnitudes
over the range of colors of stars and atmospheric depth encountered in this study.

\begin{table}
\caption{Summary of Spectral Calibration Fit Results
         \label{table:specfits}}
\begin{center}
\begin{tabular}{lcccccl}
\tableline
\noalign{\smallskip}
  Date         & $C_{O3}$   &  $\tau_0(\%)$  &  $\tau_{1}(\%)$   &  $\tau_{2}(\%)$       &  $\alpha$  \\
\noalign{\smallskip}
\tableline
\noalign{\smallskip}
   July 4    &    0.96    &       1.29   &    0.05              &   -0.01               &  -1.66   \\
   July 5    &    0.77    &       2.63   &   -0.09              &   -0.01               &  -2.27   \\
   July 6    &    0.87    &       1.29   &    0.01              &    0.03               &  -1.45   \\ 
\noalign{\smallskip}
\tableline
\end{tabular}
\end{center}
\end{table}

\subsection{The Imaging Data and Reductions}
\label{section:reductions}

We want to isolate the effects of the spatial structure of cloud densities on our ``all weather'' calibration strategy from
other potential sources of instability or non-uniformity.
In addition to making direct measurement of extinction by atmospheric molecules and aerosols
(and avoiding optical bands with substantial absorption of light by water vapor),
we have chosen an observing scheme that minimizes potential variations in the measurements of 
light from individual sources due to variations in the instrumental response of the telescope and camera.

\begin{table*} 
\caption{Fields Observed with Blanco/MOSAIC II.
         \label{table:targets}}
\begin{center} 
\begin{tabular}{lcccccl} 
\tableline 
\noalign{\smallskip}
  Field     &  Nickname &      RA            &      DEC              & Galactic Latitude         &  Calibration      \\
            &           &     (J2000)        &      (J2000)          & ($\deg$)                  &  Stars            \\
\noalign{\smallskip}
\tableline
\noalign{\smallskip} 
CD-329927                 & CD-32       &   14 11 46         &   -33 03 14           &     39           &    1645              \\
SDSS J210330.85-002446.4  & J2103       &   21 03 31         &   -00 24 46           &    -29           &    1480              \\
SDSS J221458.36-002511.9  & J2214       &   22 14 58         &   -00 25 12           &    -44           &     154              \\
SDSS J233040.47+010047.4  & J2330       &   23 30 40         &    01 00 50           &    -56           &     323              \\
\noalign{\smallskip}
\tableline 
\end{tabular} 
\end{center}
\end{table*}

\begin{figure}
\plotone{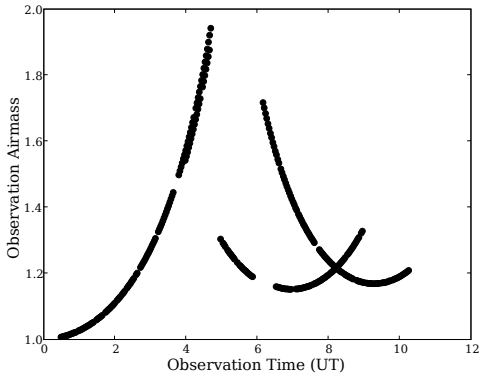}
\caption{The airmasses of the imaging observations used in this study. Three fields observed over three nights
are folded onto the single UT axis of the figure. Data for J2214 on July 5 are not shown as they are not used in the analysis.
          \label{fig:airmass}}
\end{figure}

Imaging data were taken with the Blanco telescope and the MOSAIC II imager with the CTIO $griz$ filters (c6017-c6020).
The targeted  $36' \times 36'$ fields are given in Table \ref{table:targets}.
These fields were chosen because they are relatively well known, were conveniently visible during the run, and because they cover a range of
Galactic latitudes and hence, stellar densities.
Each field was tracked for several hours by the equatorially mounted Blanco telescope
so motion of the celestial scene on the focal plane of the MOSAIC II instrument was minimized.
A summary of the observing campaign is given in Table \ref{table:observing},
and Figure \ref{fig:airmass} shows the airmasses at which data were taken.
The table includes a characterization of the atmospheric condition during each observing period;
this is not a rigorous determination of the photometric quality of the data, but rather a simple summary from visual appearance.
A more quantitative definition of photometric quality is given in Section \ref{sec:photometric} below.

\begin{table*} 
\caption{Summary of Blanco/MOSAIC II Observing Campaign.
         \label{table:observing}}
\begin{center} 
\begin{tabular}{lccccccl} 
\tableline 
\noalign{\smallskip}
  Date   & MOSAIC II Ops      &  Field       & General       &   Observable      &  Images      &     Calibratable  \\
  (2009) &       (UT)         &              & Conditions    &     Time (\%)     &  All/$r$-band  &      Images (\%)  \\
\noalign{\smallskip}
\tableline
\noalign{\smallskip}
July 4   &  06:30 - 09:00  &   J2103      & Partly Cloudy &    66             &    70/40     &         82        \\
July 5   &  04:00 - 04:40  &   CD-32      &  Cloudy       &    20             &    21/12     &         70        \\
July 5   &  05:40 - 07:10  &   J2214      & Very Cloudy   &    NA             &    40/24     &         NA        \\
July 6   &  00:30 - 04:40  &   CD-32      &   Clear       &    100            &   117/67     &         100       \\
July 6   &  04:50 - 05:50  &   J2103      &   Clear       &    100            &    28/12     &         100       \\
July 6   &  06:10 - 10:15  &   J2330      &   Clear       &    100            &   112/62     &         100       \\
\noalign{\smallskip}
\tableline
\end{tabular} 
\end{center}
\end{table*}

Images were taken in pairs of 15-second exposures with nominal gain setting of two e$^-$ per ADU count, and the pixels were read out unbinned.
The time interval between exposures was approximately two minutes.
The specific pattern $rr-gg-rr-ii-rr-zz-rr$ was repeated as a set
to provide a high density of $r$-band images and color measurements for each detected object.
Short 15-second exposures were used in this study in order to capture spatial structure in cloud opacity as it will appear in LSST images;
it can be expected that spatial structure will be less intricate when averaged over longer exposures such as those that will be used by DES.

Bias and flats were taken daily in each band, and the IRAF MSCRED package was used to correct the science images
obtained on each corresponding night.
Each image was fitted with a World Coordinate System (WCS) solution using the IRAF MSCRED and MSCFINDER packages
with object lists taken from the USNO-B astrometric catalog.
Fitting residuals in each axis were typically 0.08 $\arcsec$ (rms), or approximately one-third of the pixel dimension and easily sufficient for
source matching.
The WCS mapping also allows the images and photometry to be corrected for geometric distortion. 
The WCS solution was used to convert the pixel coordinates of each object reported by DoPHOT \citep{schechter93} to coordinates on the sky, 
and the image then $sinc$ resampled with the IRAF MSCIMAGE task to an image with equal-area projection.
Each object then has coordinates on the USNO-B system, and the photometry is corrected for optical distortion of pixel areas.
No further correction was applied at this stage of the analysis for non-uniformity of the instrument response.

Object identification and instrumental photometry for each identified source were done following a procedure that uses a version of
DoPHOT modified by one of us \citep{saha11}.
The procedure renormalizes all PSF-fitted magnitudes to the system of 20 pixel (5.4 $\arcsec$) radius apertures,
and de-trends the data for variations of the PSF across the field of view.
The reference aperture is large enough that the measurements are robust against the seeing variations encountered during our observations.
DoPHOT reports error estimates for individual objects based on shot noise from source and background counts modulated by the
chi-squared from fit residuals for individual objects.
The procedure propagates these errors, and adds in quadrature additional uncertainties incurred in applying the aperture corrections.
The resulting estimated errors $\sigma^{phot}$ have been validated as being realistic,
and our analysis below also finds good agreement in these data.

Instrumental magnitudes were defined in bands $b = gri$ for each detected source $i$ on each image $j$ by the formula,
\begin{equation}
\label{eqn:maginst}
     m_b^{inst}(i,j) \equiv -2.5\log_{10}\left( ADU_b \over \Delta T \right) + 30.75.
\end{equation}
The measured digital counts $ADU_b$ were corrected and normalized as described above. 
The arbitrarily chosen zero point corresponds to magnitude differences V - $r \approx$ 2 between the V-band used in the on-line MOSAIC II
exposure time calculator and the $r$-band defined here.

The wavelength-dependent portion of atmospheric transmission was computed using the appropriate coefficients from the fits to the spectroscopic data,
({\it cf.} Eqs. \ref{eqn:Sgray} and \ref{eqn:fitform}),
\begin{equation}
\label{eqn:schrom}
S^{chromatic} \equiv S^{molecular}(alt,az,t,\lambda) \!\! \times \!\! S^{Mie}(alt,az,t,\lambda).
\end{equation}
The instrumental passbands $S_b^{inst}$ (Eq. \ref{eqn:optb}) were computed using filter transmission curves\footnote{Website www.ctio.noao/instruments/FILTERS}
and detector efficiencies\footnote{Website www.ctio.noao.edu/mosaic} found on the CTIO website;
these were taken to be independent of location in the focal plane.
A corrected top-of-the-atmosphere ``flat-SED'' source magnitude was then computed by correcting the raw instrumental magnitude for both
wavelength-dependent atmospheric extinction and the instrumental response of the appropriate band,
\begin{widetext}
\begin{equation}
\label{eqn:mTOA}
m_b^{TOA} \equiv   -2.5 \log_{10}\left( \frac{ ADU_b }
      {\Delta T \int_0^\infty  S^{chromatic}(alt,az,t,\lambda) S^{inst}_{b}(\lambda) \lambda^{-1} d\lambda } \right) + 30.75.
\end{equation}
\end{widetext} 
The integral in Equation \ref{eqn:mTOA} was computed numerically for each source on each image as a sum over 1 nm wide wavelength bins.
The resulting corrected magnitudes were then used to construct object catalogs for the different fields observed during the run.

\subsection{Object Catalogs}
\label{subsection:objcats}

In what follows, we consider only those objects that DoPHOT unambiguously classified as stars (Type = 1).
The DoPHOT software is particularly good at separating stars from extended objects (Schechter {\it op.cit.}).
And as shown below, the colors of our final sample track the well-known stellar locus with little indication of contamination by non-stellar objects.

An object catalog was created for each field by associating sources seen on repeated images at the same celestial coordinates to a single object.
Subsets of twenty images of each field were chosen as ``build'' samples,
and stellar sources with $m_r^{TOA} < 24.5$ (S/N $>$ 1) were selected from each of these images.
A nested sorting routine then identified nearest neighbors and next-to-nearest neighbors for pairs of sources found on different images.
A friends-of-friends algorithm was used to collect these pairs into objects.
Pairs were considered to come from the same object if they are within 1 $\arcsec$ of each other on the sky;  
stellar densities in the sample fields are sufficiently low that with this choice there is negligible contamination from random overlap of true objects.
The resulting number of objects found in each field is given in Table \ref{table:targets}.
The full catalog for the study was then compiled by associating sources found on the remaining images to objects found
in the ``build'' samples.

\begin{figure}
\plotone{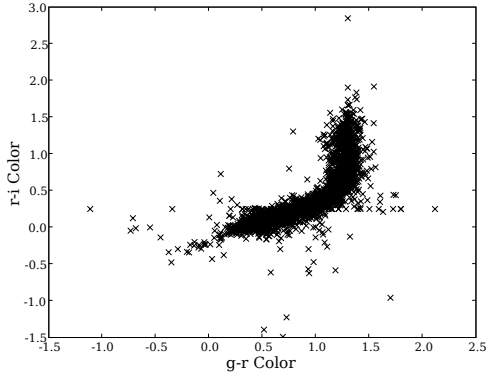}
\caption{Color-color (g-r vs. r-i) of all objects computed from initial estimates of top-of-the-atmosphere magnitudes $m_b^{TOA}$. 
          \label{fig:Color_Color}}
\end{figure}

A first estimate of the magnitude of each object in each band was made by finding the brightest observed $m_b^{TOA}$ and computing an average value
$\overline{m_b^{TOA}}$ from all observations of that object that were within 0.06 mag ($\sim 3 \sigma^{phot}$) of the brightest.
The distribution of colors $g-r \equiv \overline{m_g^{TOA}} - \overline{m_r^{TOA}}$ and $r-i \equiv \overline{m_r^{TOA}} - \overline{m_i^{TOA}}$
computed from these magnitudes is shown in Figure \ref{fig:Color_Color}.
The distribution is consistent with the well-known main-sequence stellar locus, and indicates a clean separation of stellar point sources in the data.
The analysis then concentrated on the numerous $r$-band data by selecting objects that satisfied the following criteria:
\begin{itemize}
\item The estimated object magnitude in the $r$-band satisfies $22.0> \overline{m_r^{TOA}} > 17.0$;
this corresponds to signal-to-noise ratios of approximately 10-300 for the exposure time, seeing, and sky conditions of our sample.
\item The object was observed in more than 10 $r$-band images with photometric measurement error $\sigma^{phot} < 0.10$.
\item The object is spatially separated by more than 5$\arcsec$ from any other object in the catalog.
\end{itemize}
All objects that passed these criteria were randomly separated into ``calibration'' objects (with 70\% probability)
and ``test'' objects (with 30\% probability).
As implied by the terminology, calibration objects were used to determine calibration constants for each image, and test objects
were used to test the accuracy of the interpolation of these constants to random positions on each image.

A first estimate of $S^{gray}(alt,az,t)$ at the location of each calibration object on each image was computed as the difference of the estimated 
$\overline{m_r^{TOA}}$ and the $m_r^{TOA}(alt,az,t)$ observed in the image.
A first estimate of the average cloud opacity across the field of view on each image was then computed as the average of the estimated $S^{gray}$ of all
calibration objects seen on the image.
An image was considered ``calibratable'' if,
\begin{itemize}
\item There were at least 100 calibration objects visible in the image.
\item The estimated mean gray extinction in the $r$-band was less than 1.5 mag.
\end{itemize}

A summary of the observing conditions, fraction of observable time, and fraction of calibratable images for subsets of the data 
is included in Table \ref{table:observing}.
Essentially all images taken in clear conditions could be calibrated, and even in rather poor conditions the fraction of images that could
be calibrated with good precision remains high.
But, as discussed below, the precision of the calibration falters as the photon statistics worsen and stars drop out of the calibration sample
with deteriorating observing conditions.
The field J2214 was observed in very cloudy conditions that yielded extremely poor results as the number of calibration stars fell
dramatically and photometric reduction errors became large.
No further analysis was done of data on this target field.

\subsection{Standard Passbands and Magnitudes}
\label{subsection: Stdbands}

\begin{figure}
\plotone{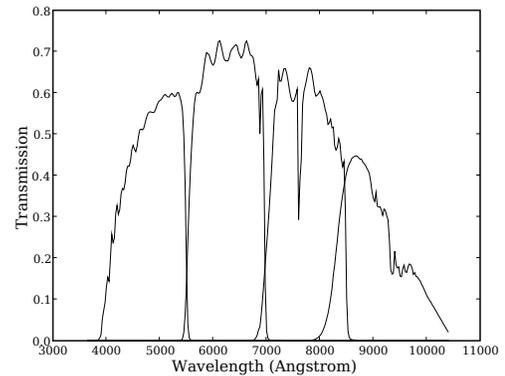}
\caption{Standard passbands $S_b^{std}$ used for this analysis. 
          \label{fig:Standard_Passbands}}
\end{figure}

To obtain a well-defined in-band magnitude for each object, we transform each measured $m_r^{TOA}$ to an equivalent
magnitude that would have been seen through a fixed ``standard'' passband \cite{fukugita96},
\begin{equation}
\label{eqn:stdpasses} 
S_b^{std}(\lambda) \equiv S_b^{inst,std}(\lambda) \times S^{chromatic,std}(\lambda).
\end{equation}
The instrumental passbands $S_b^{inst,std}$ were those computed from quantum efficiency and filter transmission data downloaded from the
Blanco/MOSAIC II website.
The atmospheric transmission $S^{chromatic,std}$  was computed with MODTRAN for an airmass of 1.3 using the spectroscopic data from the third night of the run.
The resulting $S_b^{std}$ passbands shown in Figure \ref{fig:Standard_Passbands} are typical of the observational passbands (corrected to airmass of 1.3) 
for much of the data used in this paper.
These form an internal photometric system suitable for our purposes here;
we make no attempt to determine transformations to any other published system.

The difference between a magnitude measured through an observational passband and what would be measured through a standard passband is \citep{ivezic07}, 
\begin{widetext}
\begin{equation}
\label{eqn:mstd}
  m_b^{std} - m_b^{TOA} \equiv \Delta m_b^{std}
       = 2.5\, \log\left( \frac{\int_0^\infty S_b^{std}(\lambda) \lambda^{-1} d\lambda}{\int_0^\infty S_b^{inst}(\lambda)S^{chrom}(\lambda) \lambda^{-1} d\lambda}
      \frac{\int_0^\infty f_\nu(\lambda) S_b^{inst}(\lambda)S^{chrom}(\lambda) \lambda^{-1} d\lambda}{\int_0^\infty f_\nu(\lambda) S_b^{std}(\lambda) \lambda^{-1} d\lambda}     \right),
\end{equation}
\end{widetext}
where the source SED $f_\nu(\lambda)$ is defined by
\begin{equation}
                 F_\nu(\lambda) = F_0 f_\nu(\lambda),
\end{equation}
with $f_\nu(\lambda)=1$ at some chosen reference wavelength.
The value of $F_0$ does not enter the analysis,
and a useful choice for the reference wavelength is the weighted mean of the standard $r$ passband.
The normalizations of the standard passbands,
\begin{equation}
\label{eqn:stdnorm}
        S_b^{norm} \equiv \int_0^\infty S_b^{std} \lambda^{-1} d\lambda,
\end{equation}
are 0.150, 0.151, 0.107, and 0.060 for $b=griz$ respectively, 
and the corresponding weighted mean wavelengths $\lambda_b$ = 4804.4\AA, 6232.5\AA, 7700.5\AA, and 8978.0\AA.

\begin{figure}
\plotone{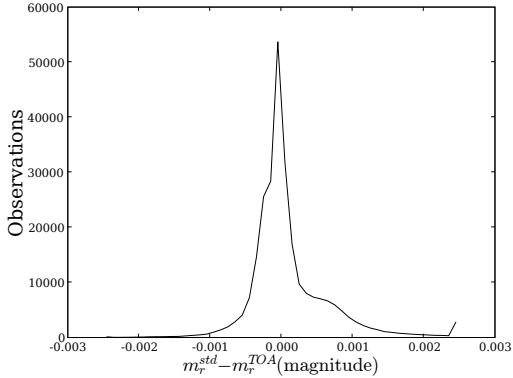}
\caption{Transformations of calibration observations from measured top-of-the-atmosphere magnitudes to the standard $r$-band.
         All calibration observations are plotted, and the structure reflects the various combinations of airmass and
         atmospheric conditions encountered during the campaign. Overflow and underflow counts are shown in the outer bins of the plot. 
          \label{fig:Standard_Xform}}
\end{figure}

The $g$ and $i$ bands were used to make a linear correction to the slope of the object SED across the $r$ band.
\begin{equation}
\label{eqn:fnu}
   f_\nu(\lambda) \approx  1.0 + \frac{F_i - F_g}{(\lambda_i - \lambda_g)} \frac{(\lambda - \lambda_r)}{F_r},
\end{equation}
where the flux values were computed as,
\begin{equation}
\label{fluxes}
F_b = 10^{-0.4 (m_b^{TOA}-30.75)}.
\end{equation} 
The distribution of $\Delta m_r^{std}$ is shown in Figure \ref{fig:Standard_Xform} for the observations of each calibration (and test) object.
As can be seen from Eq. \ref{eqn:mstd} if a particular observation is made through a passband $S_r^{chrom}$ that is identical to $S_r^{std}$, then
$\Delta m_r^{std} \equiv 0$ for that observation.
Generally, the choice of a standard passband that is typical of the actual observing conditions results in relatively modest values
of $\Delta m_r^{std}$ for most observations of stars in the sample.

\section{Calibration of Source and Object Catalogs}
\label{sec:calibration}

Photometric calibration can be separated into three steps:
\begin{itemize}
\item Relative calibration: internal normalization of measurements in a given bandpass relative to all
other measurements made in the same bandpass across the sky and at different times.  This is the only part of the calibration
that we address in this work. 
\item Absolute calibration of colors: determination of the ratios of flux normalizations of the standard passbands.
\item Absolute calibration of flux: definition of an overall physical scale. 
\end{itemize}
This can be made numerically explicit by writing the calibration errors in a linear decomposition,
\begin{equation}
\label{eqn:photoSysErr}
  m_b^{std} = m_b^{true} + \sigma^{phot} + \delta_b(x,y,alt,az,t,S_b(\lambda),SED) + \Delta_b,
\end{equation}
where $m_b^{true}$ is the true in-band magnitude of the object at the top of the atmosphere.
The photometric errors $\sigma^{phot}$ ({\it cf.} Sec. \ref{section:reductions}) include statistical fluctuations in source and background counts,
random errors in corrections for seeing-dependent aperture losses
and other uncertainties in the assignment of ADU counts to source flux measurements.
The $\Delta_b$ and $\delta_b$ terms represent errors in the calibration of ADU counts to physical units.
The $\Delta_b$ are fixed offsets of the zero-points of the
photometric bands from the absolute normalization of the physical scale that are determined in the second and third calibration steps.
The $\delta_b$ are relative errors of the internal zeropoint around $\Delta_b$ that depend on position in
the field of view ($x,y$), position on the sky ($alt,az$), date and time of the observation,
the observational bandpass $S_b(\lambda)$, and the source SED.
The average of $\delta_b$ over the survey is zero by
construction, but a value must be determined for each source location in each image.
In this work, we primarily address determination of that part of $\delta_r(x,y,alt,az,t,S_r,SED)$ caused by variation in
transmission of light through the atmosphere, in particular variation due to gray cloud cover $S^{gray}(alt,az,t)$.

\subsection{Global Calibration}

After reduction of each image, correction for chromatic atmospheric extinction, and transformation to the standard passbands,
a global self-calibration procedure was used to minimize the dispersion of residual errors in all observations of the calibration objects.
This process is based on techniques originated for imaging surveys \citep{glazebrook94, macdonald04};
the specific implementation used here is based on the ``\"Ubercal'' procedure developed for SDSS \citep{padman08}
and applied to data taken in photometric conditions free of clouds by several authors \citep{wittman11, schlafly12}.
This technique minimizes residual errors in both the image reductions and chromatic extinction corrections,
and we extend it here to correct for the gray opacity of cloud cover over a range of non-photometric conditions.

``Calibration patches'' were defined on the camera focal plane that are small compared to the full field of view,
but large enough to contain sufficient numbers of calibration stars to achieve good precision.
For example, $\sim 100$ calibration stars with photometric errors $\sigma^{phot} < 0.1$ would provide sub-percent relative calibrations
if the cloud structure is sufficiently simple to allow a single average to be taken over each patch.
Images taken by future surveys are expected to contain $\sim 1$ such star per square arcmin,
and an important objective of our study is to determine if this is sufficient to reach the goals of these surveys.
Several of the fields we chose for this study (Table \ref{table:targets}) are at sufficiently low Galactic latitudes
to provide similar densities of stars in 15 second exposures with the Blanco.
So we partition the MOSAIC II $36' \times 36'$ focal plane into sixteen $9' \times 9'$ patches.

Future surveys will ``dither'' pointings from epoch to epoch to control systematic errors.
The calibration is joined smoothly across patch boundaries as
stars fall on different patches on different epochs across the sky. 
We implement this feature in our study by defining two different patterns of patches - one pattern for even-numbered images,
and the other for odd-numbered images.
The patterns are internally dithered by shifting the boundaries of the central four patches by $\pm 400$ pixels
relative to a perfectly symmetrical division of the MOSAIC II focal plane into sixteen patches of equal size.
This means that the outer twelve patches alternate in area from image to image,
and generally means that back-to-back pairs of images that make up our observing cadence are analyzed with complimentary patterns.
Several values of the size of the dither were tried, and it was found that steps of $\pm 20\%$ of the size of the patch work well.

We define coordinates $(x,y)$ that span the focal plane of the MOSAIC II imager, and $(x^p,y^p)$ as local coordinates that span a single patch $p$.
The units of both are pixel counts with origin at the center of the imager or patch respectively, and normalized to lie between -1 and +1.
We note that the MOSAIC II layout\footnote{Website www.ctio.noao.edu/noao/content/mosaic-ii-ccd-imager}
corresponds to the horizontal ($x$) coordinate increasing to celestial North (declination) and the vertical ($y$) coordinate
increasing to celestial East (right ascension).
The calibration zero point relative errors are written in terms of these variables as,
\begin{eqnarray}
\label{eqn:GCMerrormodel}
\delta_r(x,y,alt,az,&&t,S_r,SED)  \rightarrow  \delta_r(x,y,x^p,y^p,j) {} \nonumber  \\
                    &&  {}  \equiv \delta_r^{inst}(x,y) + \delta_{(p,j)}^{gray}(x^p,y^p),  {}
\end{eqnarray}
where the time variable has been subsumed into the image index $j$, and the telescope $(alt,az)$ has been similarly subsumed into the patch index $p$.
No color correction is made after transformation of each measured magnitude to the standard $r$-passband defined in Section \ref{subsection: Stdbands}. 
The $\delta_r^{inst}$ is a correction for non-uniformity of the $r$-band instrumental response that remains
after the flat-field and geometric optical projection corrections are applied during the image reduction process.
The instrumental correction is assumed to be constant during the three nights that data were collected for this study. 
But each patch $p$ on each image $j$ is characterized by a unique wavelength-independent correction
$\delta^{gray} = -2.5{\rm log}_{10}(S^{gray})$ ({\it cf.} Eqs. \ref{eqn:Sgray} and \ref{eqn:mTOA}) 
that is predominantly due to absorption of light by cloud cover.

The global calibration procedure minimizes the relative error $\delta_r$
in the photometric zero-point for each patch $p$ on each $r$-band image $j$ of the accumulated survey by minimizing
\begin{equation}
\label{eqn:uberchisq}
\chi^{2} = \sum_{(i,j)} {\frac{\left( m_r^{std}(i,j) - \left( m_r^{GC}(i) + \delta_r(x,y,x^p,y^p,j) \right) \right)^2}
                                           {\sigma^{phot}(i,j)^2 + \left( \sigma_0 \right)^2 }}, 
\end{equation} 
where the summation is over all calibration objects $i$ in all images $j$;
the coordinates $(x,y)$ and $(x^p,y^p)$ are those of object $i$ in image $j$.
The fitted parameters are the estimates $m_r^{GC}$ of the in-band magnitudes of each calibration object,
and the errors in the zero points $\delta_r$.
The parameter $\sigma_0$ is introduced to control possible underestimates of the errors of the brightest objects.
It is used only in the fitting step of the analysis, and separate fits with values $\sigma_0 =$ 0.001 and 0.003 were tested.
The larger value is set by the expected signal-to-noise of the brightest calibration objects ($m_r = 17.0$).
The differences in the parameters fitted with the two values are found to be statistically negligible,
and all results presented below were derived from fits made with $\sigma_0 = 0.001$. 

We use low-order polynomials to model the corrections $\delta^{inst}$ and $\delta^{gray}$, 
\begin{equation}
\label{eqn:illumform}
\delta_r^{inst}(x,y) \equiv \delta_xx + \delta_yy + \delta_{x2}x^2 + \delta_{y2}y^2 + \delta_{xy}xy,
\end{equation}
and
\begin{equation}
\label{eqn:grayform}
\delta_{(p,j)}^{gray}(x^p,y^p) \equiv \sum_{n=0}^{k}\sum_{m=0}^{k-n} s_{(n,m)}^{gray}(x^p)^n(y^p)^m.
\end{equation}
The geometric correction is taken to be zero at the coordinate origin (the center of the MOSAIC II focal plane array),
so there are five terms fitted to the entire data sample.
Separate fits were done with the order of the polynomials in $\delta_{(p,j)}^{gray}$ set to $k =$ 1, 2, and 3;
the higher order fits include six ($k$ = 2) or ten ($k$ = 3) parameters for each of the sixteen patches on each image,
which yield totals of 21014 or 32726 parameters respectively. 
No explicit requirement is made that the $\delta_{(p,j)}^{gray}$ solutions be continuous or smooth across patch boundaries,
but the internal dither of even and odd images implicitly enforces compatible solutions in neighboring patches.
Complete analyses were done with the second and third-order fits to test the sensitivity of our conclusions 
to the number of free parameters in the model.
We discuss below optimization of the complexity of the fit model, and sensitivity to over-fitting the information content of the data.

It is useful to define a measured gray extinction for observation $j$ of calibration object $i$ as
({\it cf.} Eqs \ref{eqn:photoSysErr}, \ref{eqn:GCMerrormodel}, \ref{eqn:uberchisq})
\begin{equation}
\label{eqn:grayext}
E^{gray}(x,y,i,j) \equiv m_r^{std}(i,j) - \delta_r^{inst}(x,y) - m_r^{GC}(i).
\end{equation}
This parameter, which depends on the global fit only through quantities that are determined simultaneously for all images,
can be compared with the gray correction fitted at the location of the calibration stars image-by-image (Eq. \ref{eqn:grayform}).
It is the basis for the analysis of the spatial structure functions of the gray extinction due to clouds that is presented below.

\subsection{Results of the Global Calibration Fit}
\label{sec:thefit}

The $\chi^2$ per degree of freedom (DOF) was minimized using the SciPy\footnote{Website www.scipy.org}
conjugate gradient routine optimize.fmin.cg
with convergence accepted when the change in $\chi^2$/DOF per iteration was less than $10^{-5}$.
The second (DOF = 165041) and third (DOF = 153329) order fits converged to $\chi^2$/DOF = 1.886 and 1.754 respectively.

\begin{figure}
\epsscale{1.0}
\plotone{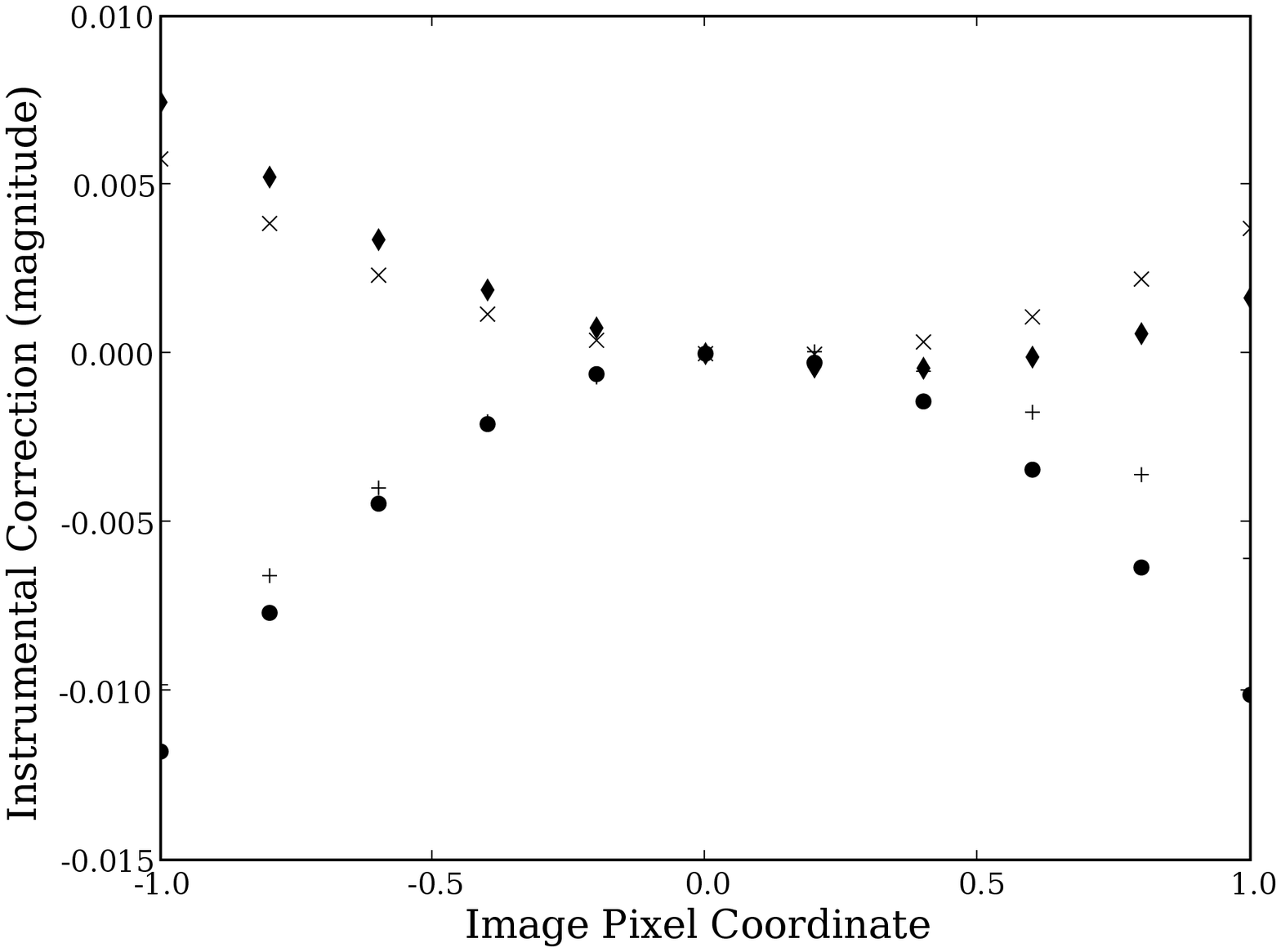}
\plotone{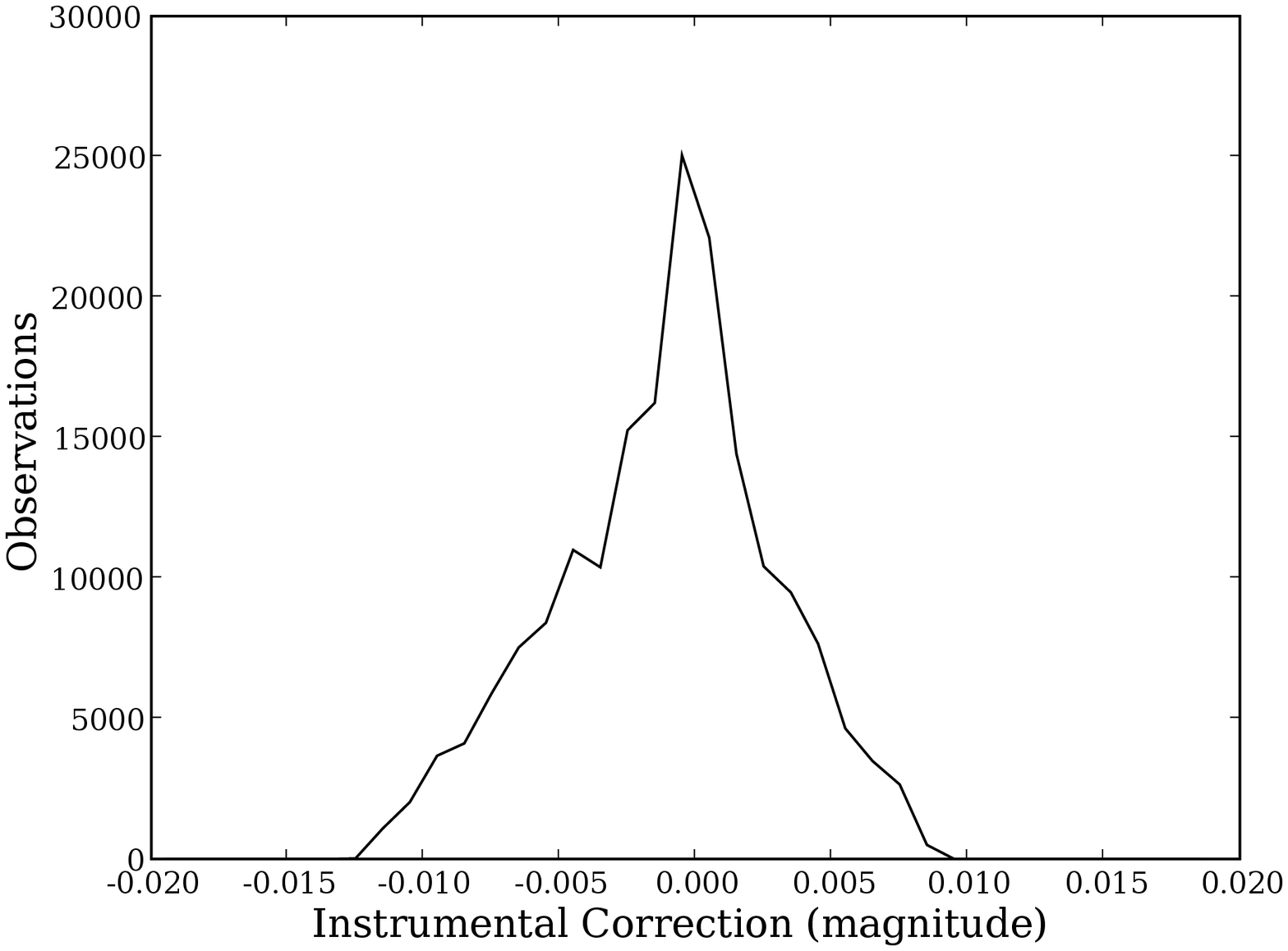}
\epsscale{1.0}
\caption{Fitted instrumental corrections across the full MOSAIC II focal plane: (top) Points along x at y=0 (x), along y at x=0 (+), and 
diagonally at y=x ($\bullet$) and y=-x ($\blacklozenge$), and (bottom) distribution of instrumental corrections applied to the data.
The observing strategy limited the motion of objects on the focal plane to those created by telescope pointing and tracking errors.    
          \label{fig:Illumination_Corrections}}
\end{figure}

\begin{figure}
\epsscale{1.0}
\plotone{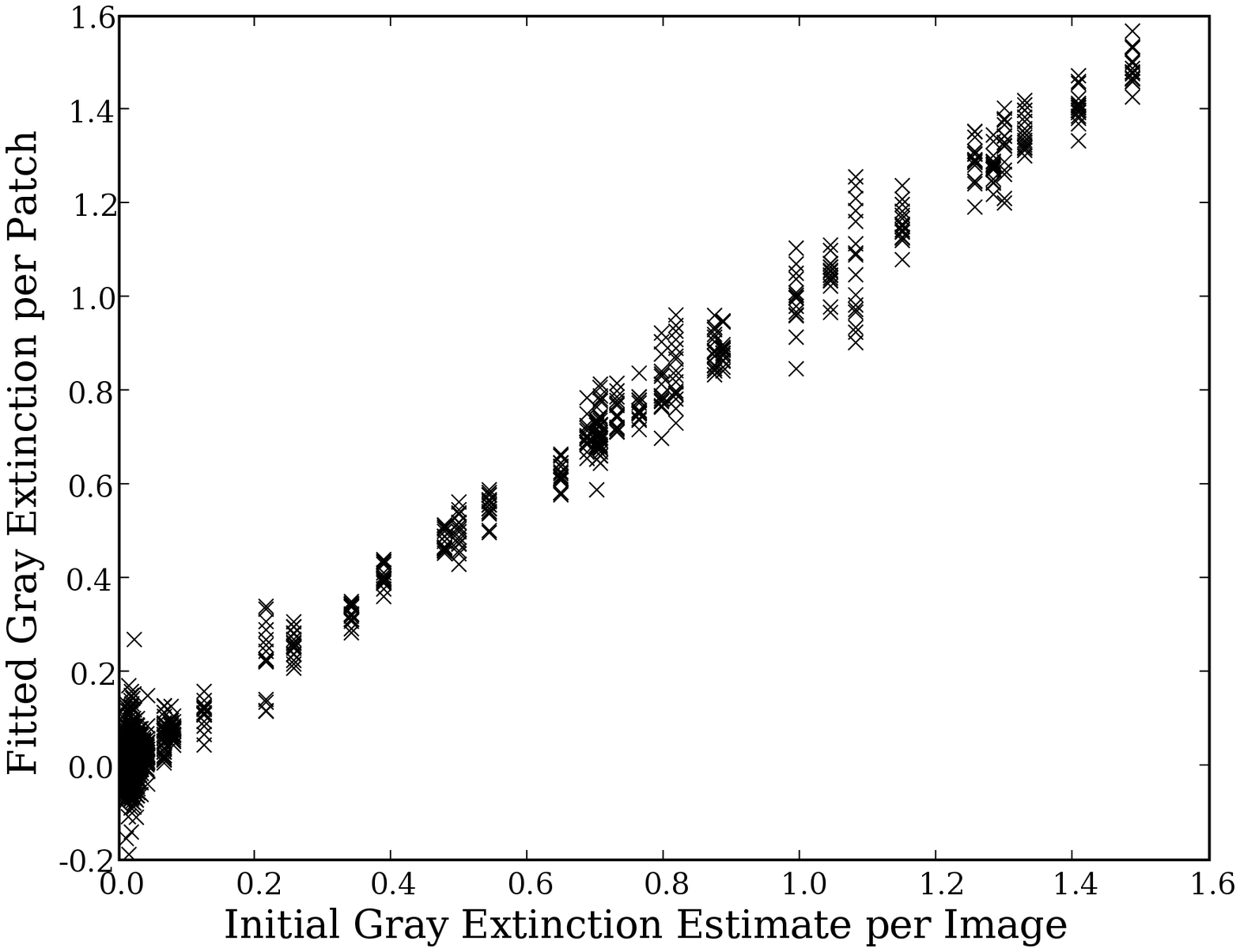}
\plotone{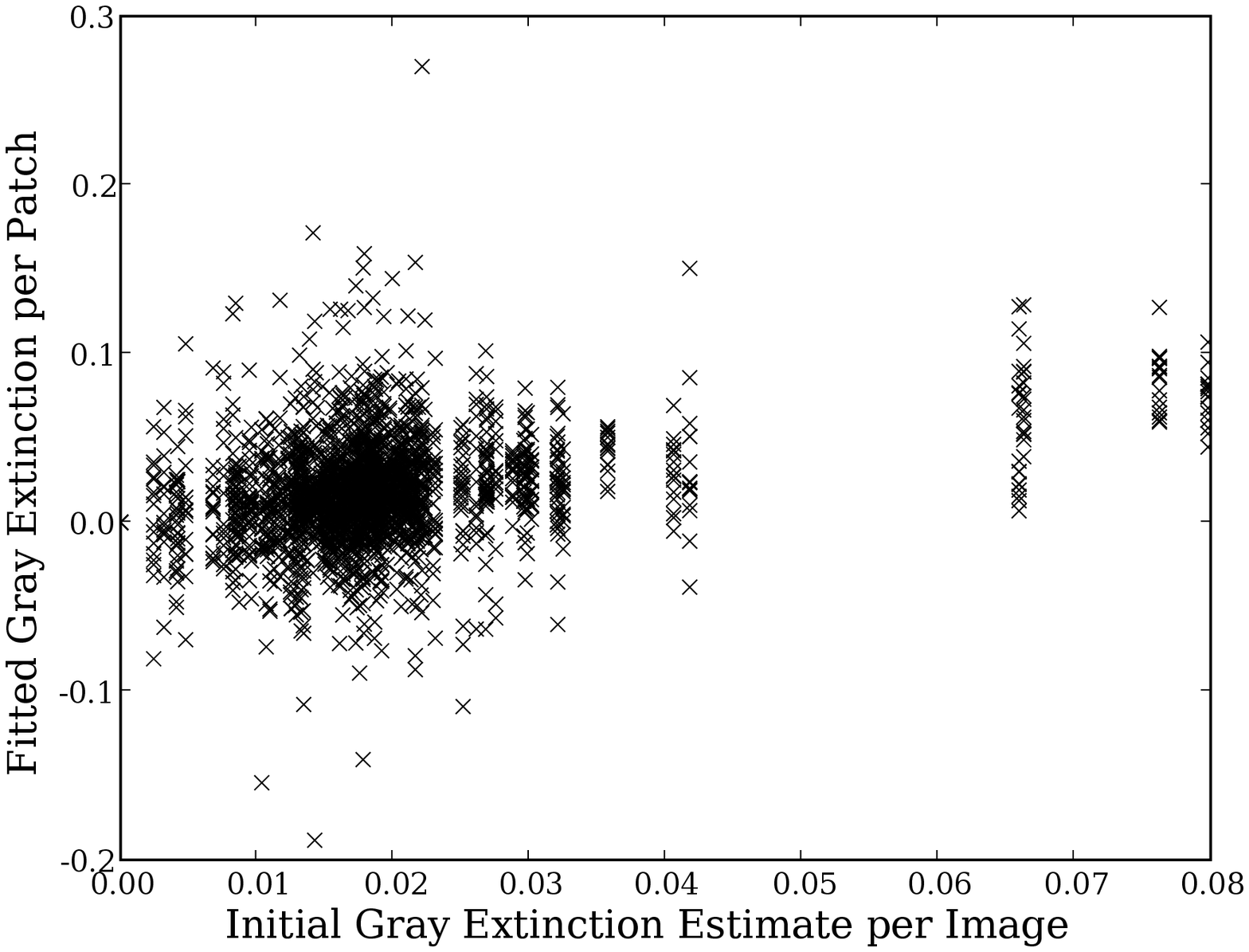}
\epsscale{1.0}
\caption{Fitted extinction (constant term) for each calibration patch compared with initial estimate of extinction for each image.
All data (top) considered calibratable in Table \ref{table:observing}; and shown (bottom) with expanded horizontal scale at low values of gray extinction. 
          \label{fig:Initial_Extinction}}
\end{figure}

\begin{figure}
\epsscale{0.50}
\plotone{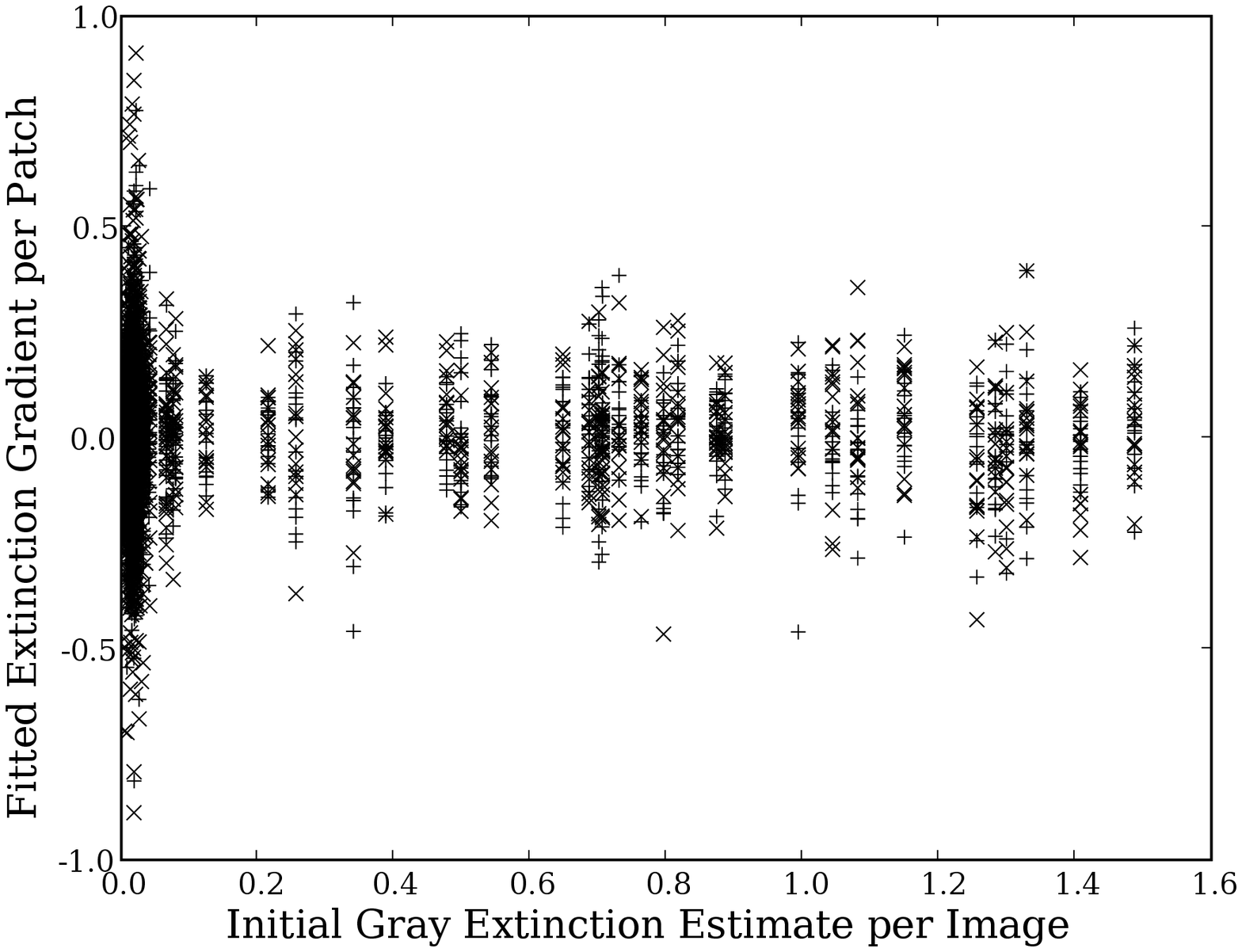}
\plotone{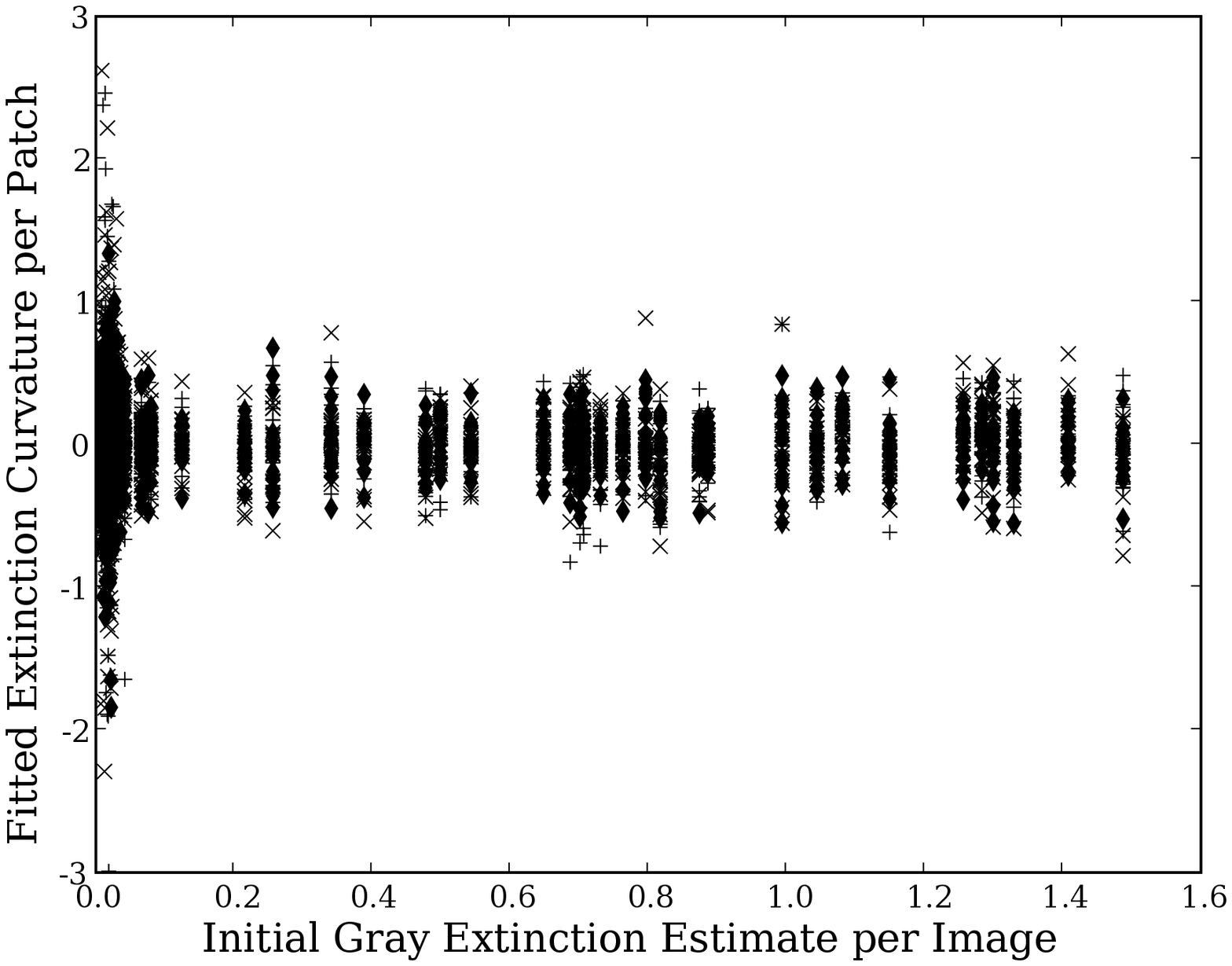}
\plotone{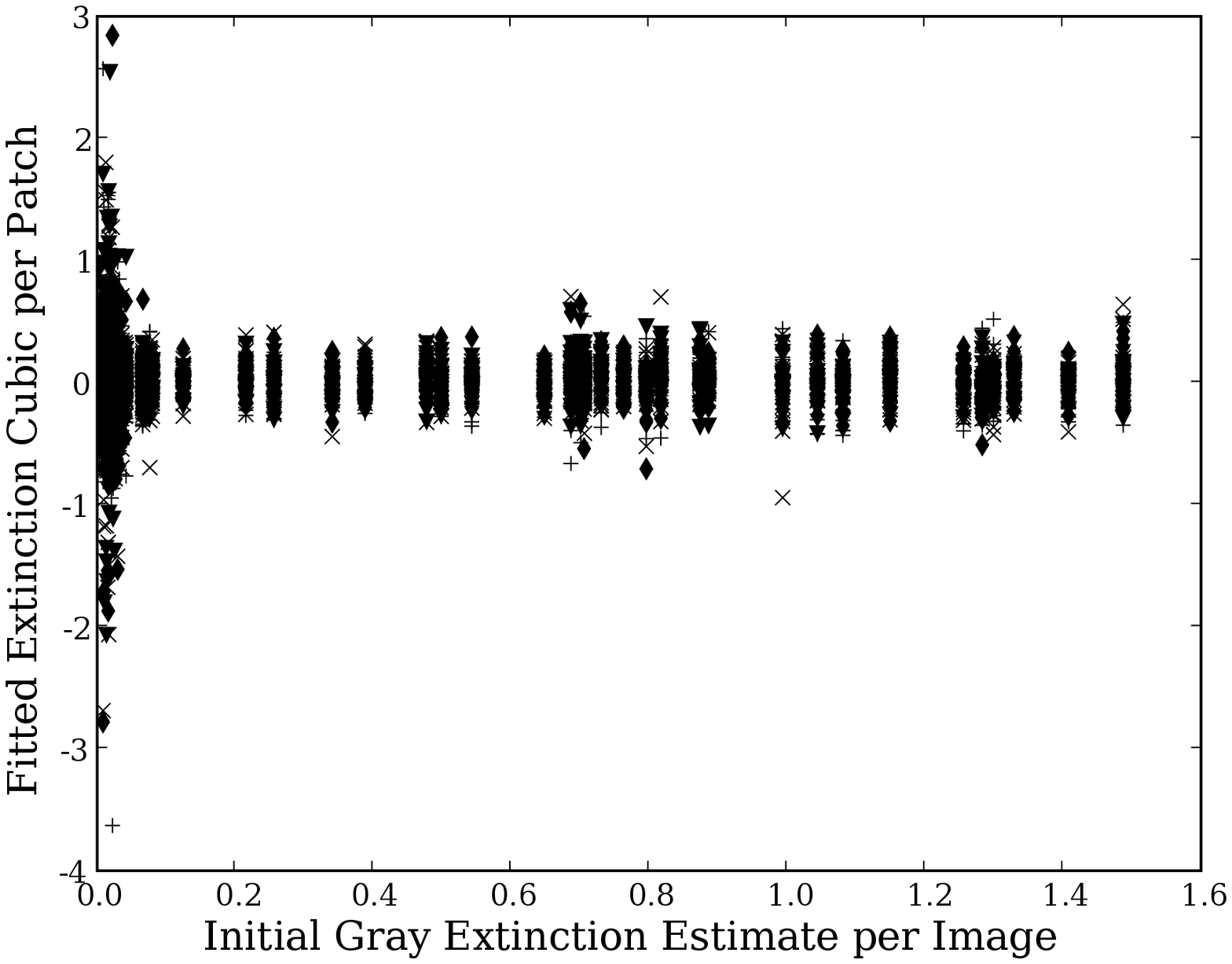}
\epsscale{1.0}
\caption{The gray extinction parameters fit to the third-order model: (top left) linear terms, (top right) quadratic terms, and (bottom) cubic terms.   
          \label{fig:M10_GC_Fits}}
\end{figure}

The fitted calibration parameters are presented in Figures \ref{fig:Illumination_Corrections},
\ref{fig:Initial_Extinction}, and \ref{fig:M10_GC_Fits}.
The instrumental correction that is used for all images is shown in Figure \ref{fig:Illumination_Corrections};
the distribution of the correction for the observed data is shown as a histogram in the same figure.
With few exceptions the correction is less than 0.010 mag.
The telescope pointing for a given field changed by no more than $\pm 400$ pixels ($\pm 0.1$ in the image coordinate space), 
so variations in this correction for a given calibration object were $\le 0.001$ mag.
This is included in the analysis, but has negligible effect on our conclusions.

The atmospheric gray model coefficients are plotted against the initial estimate of the average extinction for each image
({\it cf.} Section \ref{subsection:objcats}) in Figures \ref{fig:Initial_Extinction} and \ref{fig:M10_GC_Fits}.
The constant terms for the sixteen patches on each image are plotted in Figure \ref{fig:Initial_Extinction};
seen on the full scale of extinction, the mean of the fitted patches tracks well the initial estimate of the average over the image.
The figure also includes a plot on an expanded scale to highlight the distribution of images with the least gray extinction.
A tight core of images is evident (the figure is saturated) that we identify as containing those taken through cloudless photometric sky.
As discussed below, the absolute zero-point of the calibration is not constrained by the fitting algorithm; 
its value is set by the procedure used to initialize the fitting process.  
As expected, the higher-order terms are approximately centered on zero, and exhibit the same (saturated) photometric core of observations.

\begin{figure}
\epsscale{1.0}
\plotone{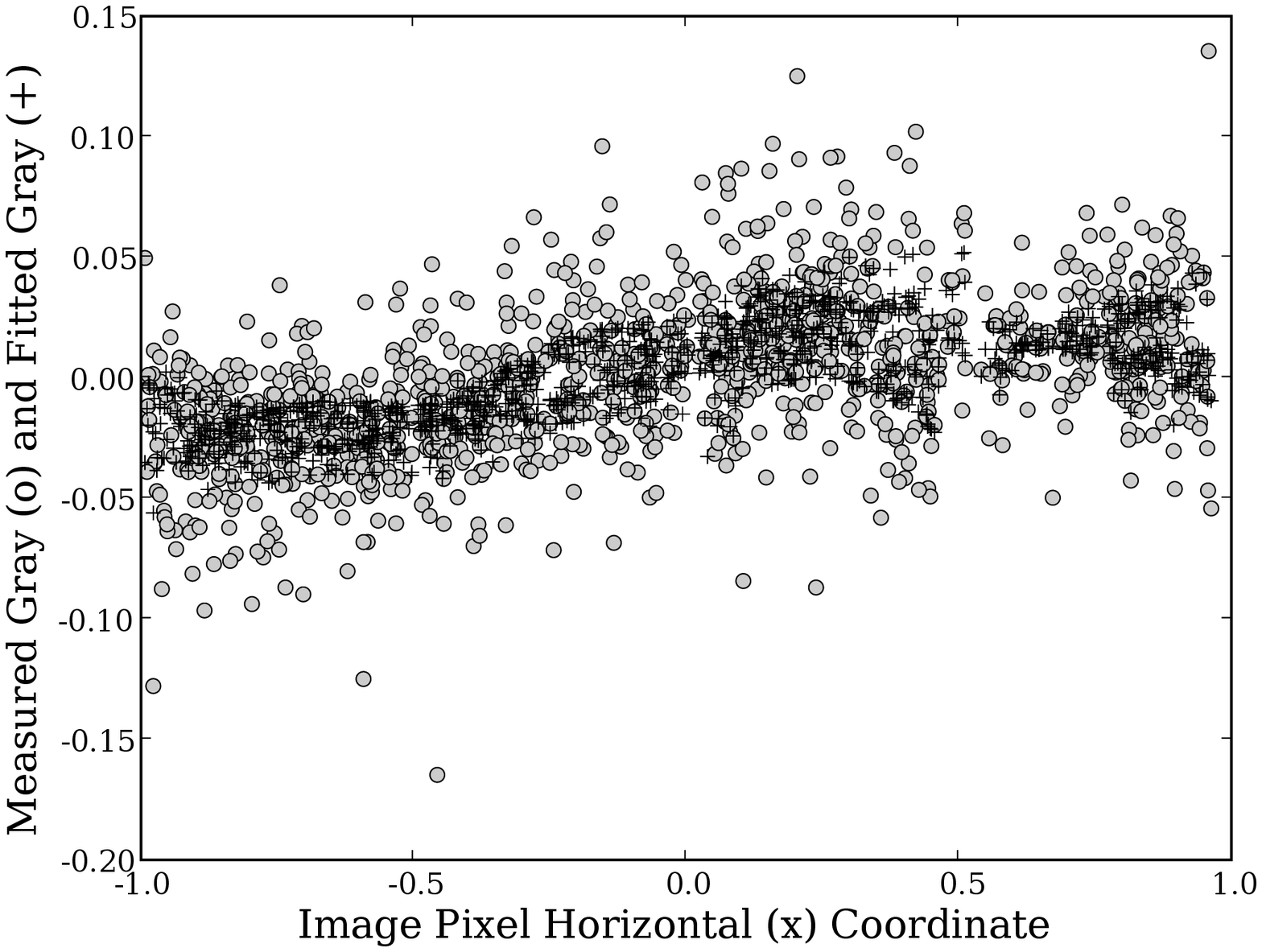}
\plotone{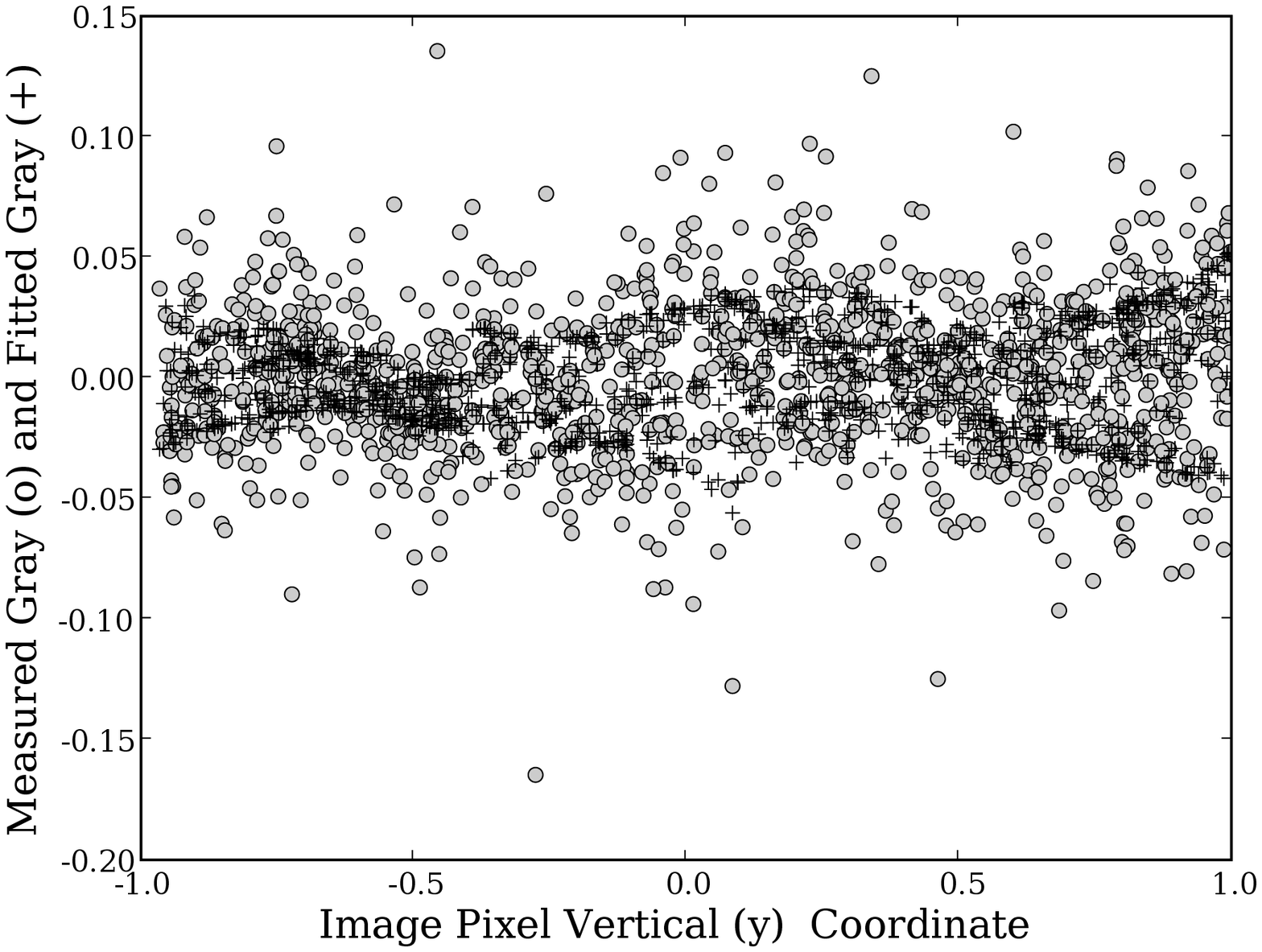}
\epsscale{1.0}
\caption{Gray extinction on image 1132 of J2103 observed on July 4.
        Shown are the measured extinction ($\circ$) and model fit (+) projected onto the horizontal (NS) and vertical (EW) image coordinates for
        the second-order model.    
        The mean over the full image of the measured extinction (0.480) has been subtracted from all plotted values.
          \label{fig:Sample_Fits_0}}
\end{figure}

\begin{figure}
\epsscale{1.0}
\plotone{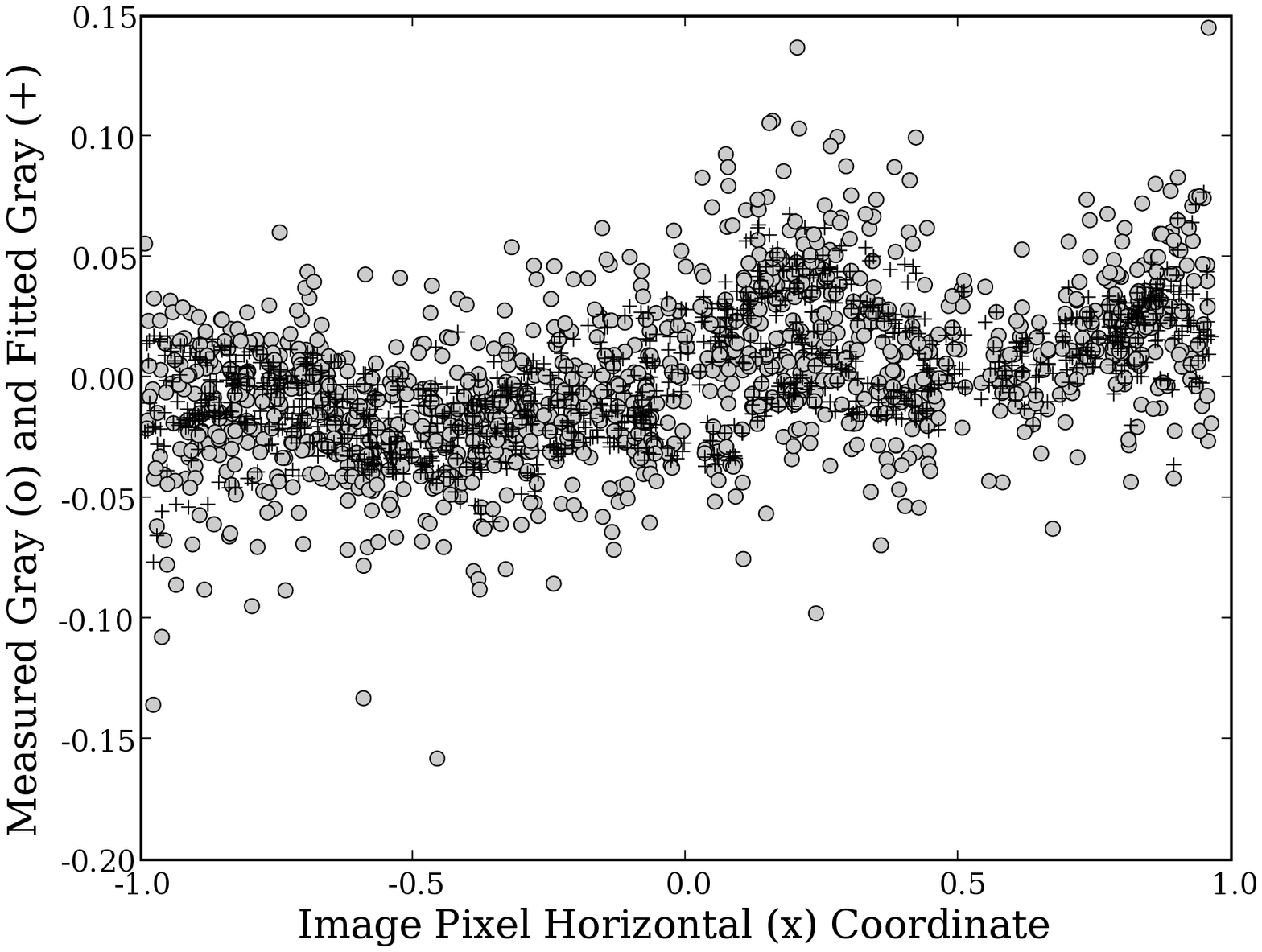}
\plotone{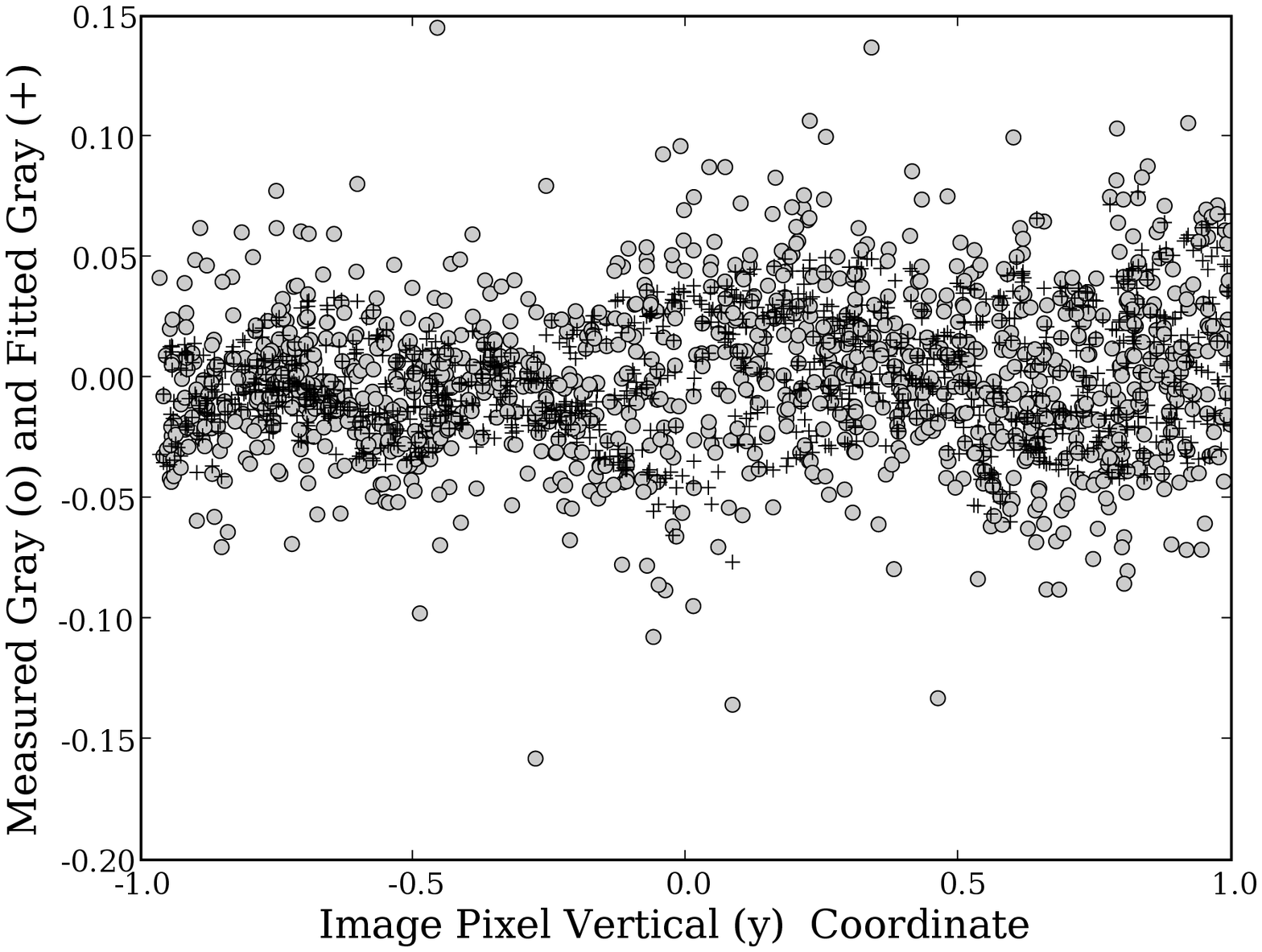}
\epsscale{1.0}
\caption{Gray extinction on image 1132 of J2103 observed on July 4.
        Shown are the measured extinction ($\circ$) and model fit (+) projected onto the horizontal (NS) and vertical (EW) image coordinates for
        the third-order model.    
        The mean over the full image of the measured extinction (0.480) has been subtracted from all plotted values.
          \label{fig:Sample_Fits_1}}
\end{figure}

\begin{figure}
\epsscale{1.0}
\plotone{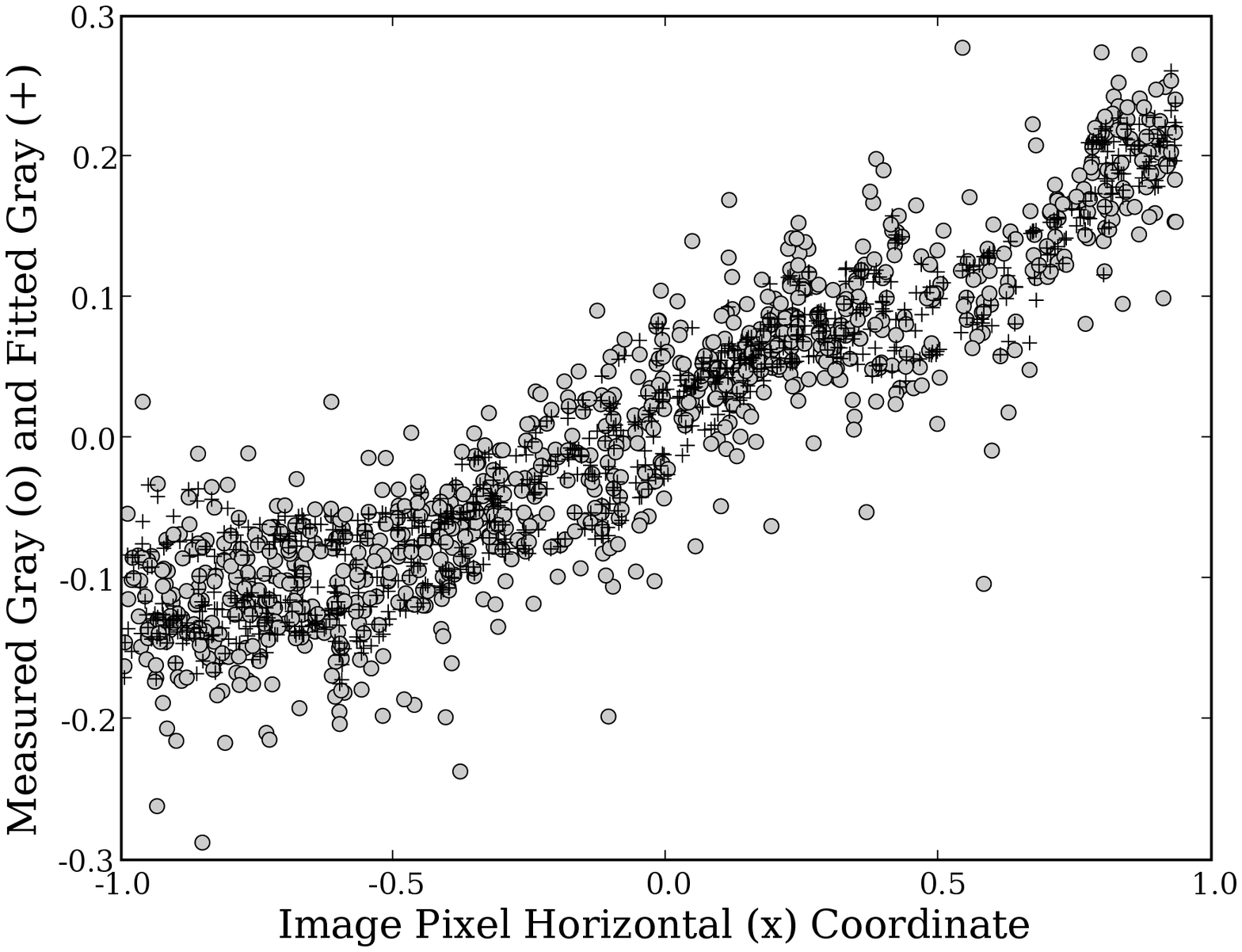}
\plotone{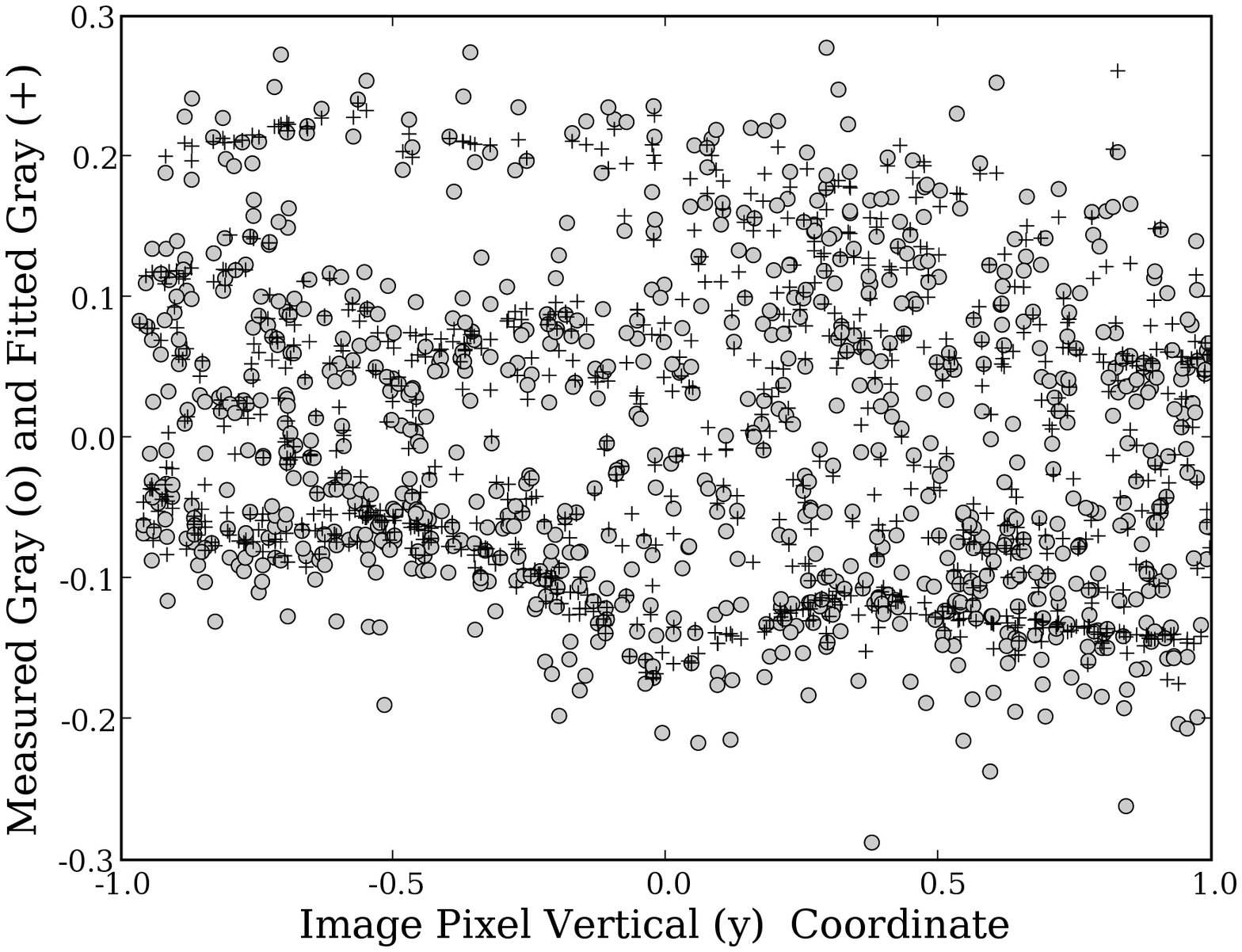}
\epsscale{1.0}
\caption{Gray extinction on image 1107 of J2103 observed on the partly cloudy night of July 4.
        Shown are the measured extinction ($\circ$) and third-order model fit (+).    
        The mean over the full image of the measured extinction (1.089) has been subtracted from all plotted values.
          \label{fig:Sample_Fits_2}}
\end{figure}

\begin{figure}
\epsscale{1.0}
\plotone{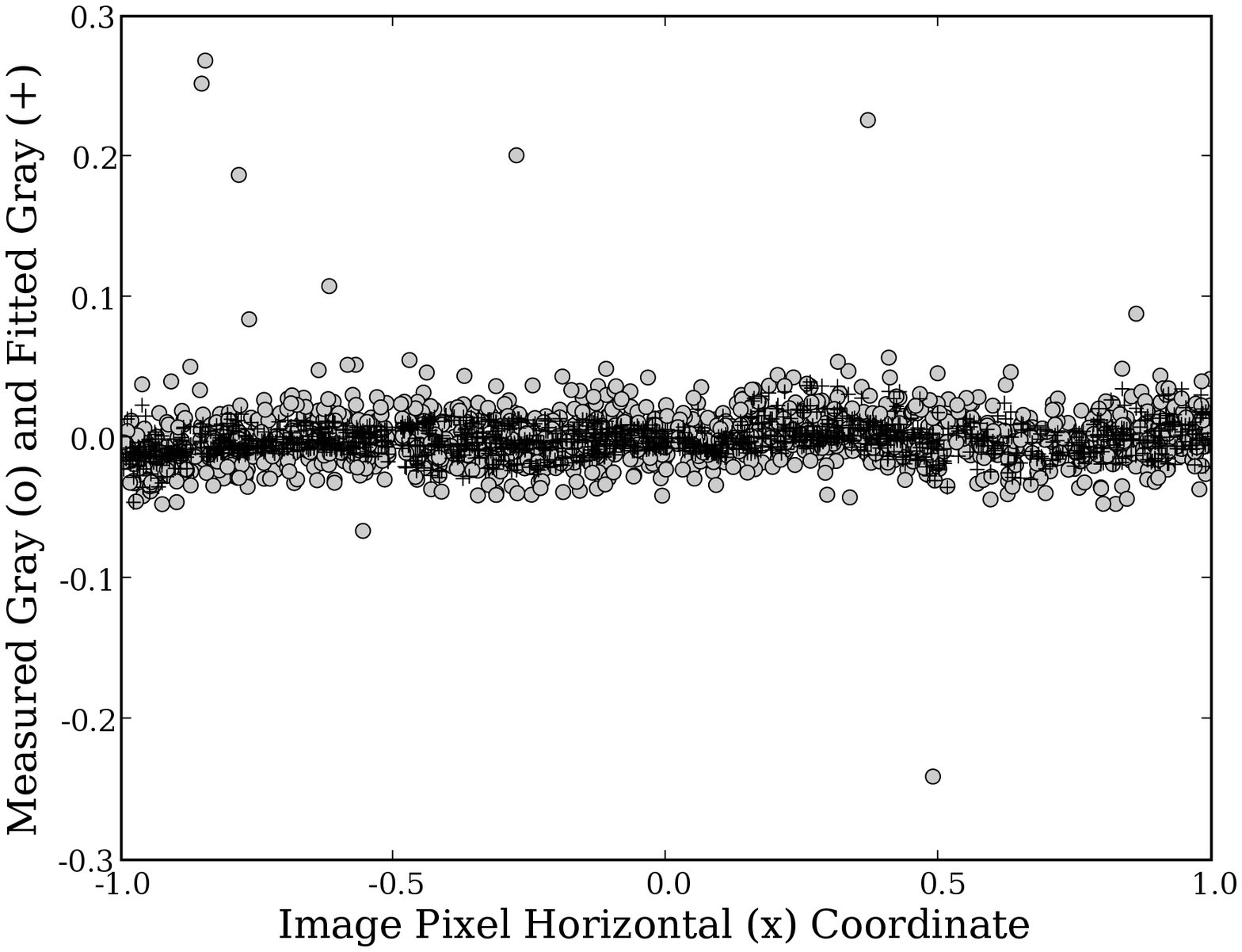}
\plotone{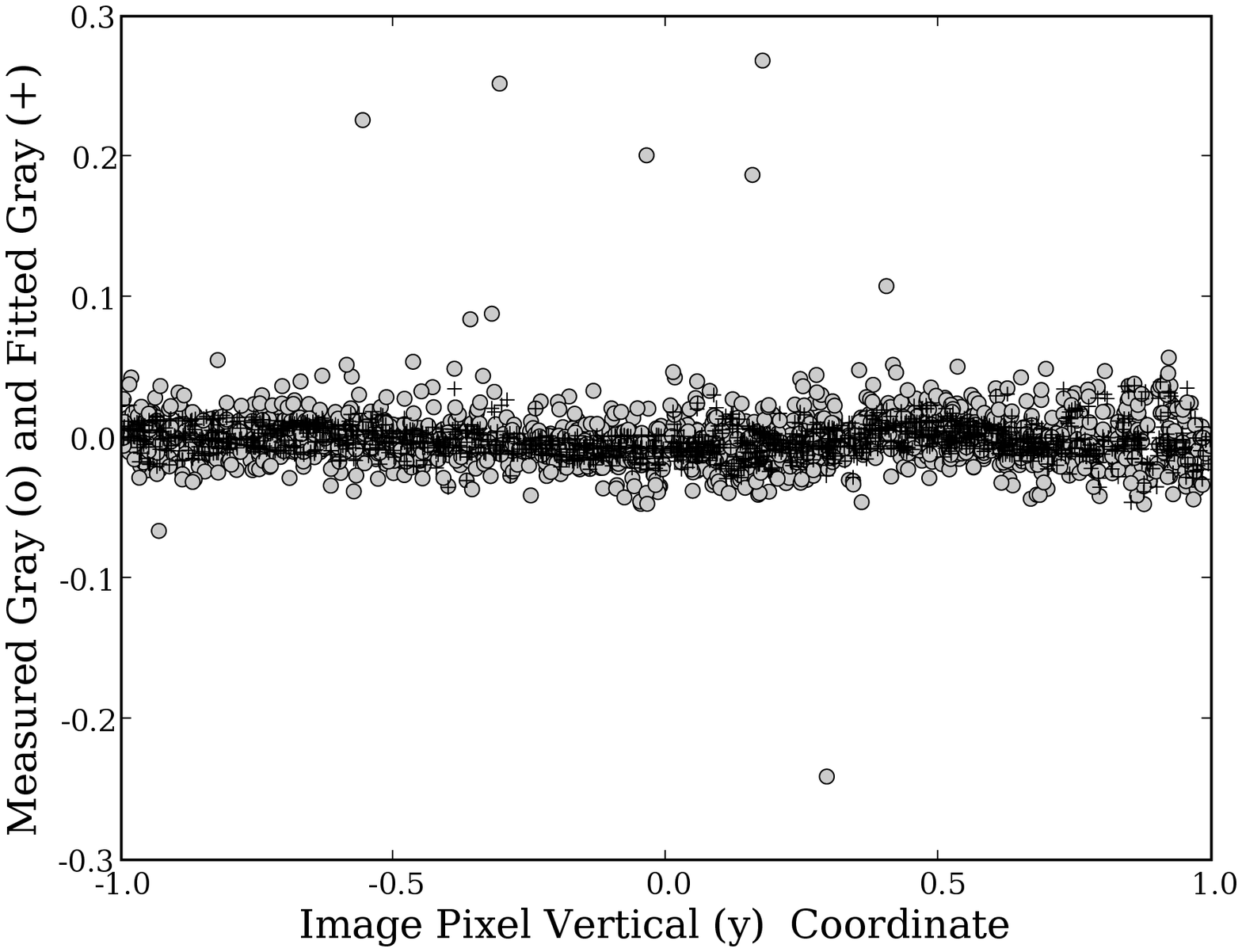}
\epsscale{1.0}
\caption{Gray extinction on image 3070 of CD-329927 observed in photometric conditions on July 6.
        Shown are the measured extinction ($\circ$) and third-order model fit (+).    
        The mean over the full image of the measured extinction (0.020) has been subtracted from all plotted values.
        The data are plotted on the same full scale as those in Figure \ref{fig:Sample_Fits_2}.
          \label{fig:Sample_Fits_3}}
\end{figure}

Fits to samples of images selected to cover ranges of mean cloud thickness are shown in Figures \ref{fig:Sample_Fits_0},
\ref{fig:Sample_Fits_1}, \ref{fig:Sample_Fits_2}, and \ref{fig:Sample_Fits_3}.
The measured (Eq. \ref{eqn:grayform}) and fitted (Eq. \ref{eqn:grayext}) gray values for each calibration object 
in the image are plotted against the $x$ and $y$ locations of the object in the image coordinate system.
The mean of the measured gray values (Eq. \ref{eqn:grayext}) of all calibration objects on the image (given in the figure caption)
has been subtracted from the both the measured and fitted values to avoid loss of resolution on the vertical scale.
The variation in the fitted values at a fixed $x$ or $y$ is due to the sampling of the extinction
by calibration objects in the corresponding orthogonal coordinate.
Figures \ref{fig:Sample_Fits_0} and \ref{fig:Sample_Fits_1} compare results from the second and third-order fits to the same data,
while Figures \ref{fig:Sample_Fits_2} and \ref{fig:Sample_Fits_3} show third-order fits to extreme cases of cloud structure.

All the fits are well behaved across the field of view, and thanks to the dither of the grouping of calibration
stars between even and odd numbered images, they are smoothly connected across patch boundaries.
(Patch boundaries occur at approximately half-integer intervals in the image coordinate space,
and the dither corresponds to $\pm0.1$ in the image coordinate space.) 
Though it takes a moment to adjust the eye, distinctive features can be discerned in the structure of the cloud optical depth.
For example, image 1132 of J2103 observed through an average of 0.480 mag of gray extinction on the night of July 4
shows two regions of approximately constant gray extinction when projected
onto the horizontal coordinate (i.e. $-1.0 < x < -0.4$ and $0.2 < x < 1.0$); these project to two bands along the vertical coordinate.
Similar structure can be seen especially in image 1107 taken through 1.089 mag of gray extinction. 
Image 3070 of CD-329927 was taken through little, if any, cloud cover in what are characterized as photometric conditions.
The scatter of the measurements in this image is dominated by statistical errors, 
but both the measured and fitted values show small systematic variations across the field of view that persist from image to image.
These variations contribute to the ``zero-point structure function'' discussed below.

\begin{figure}
\epsscale{1.0}
\plotone{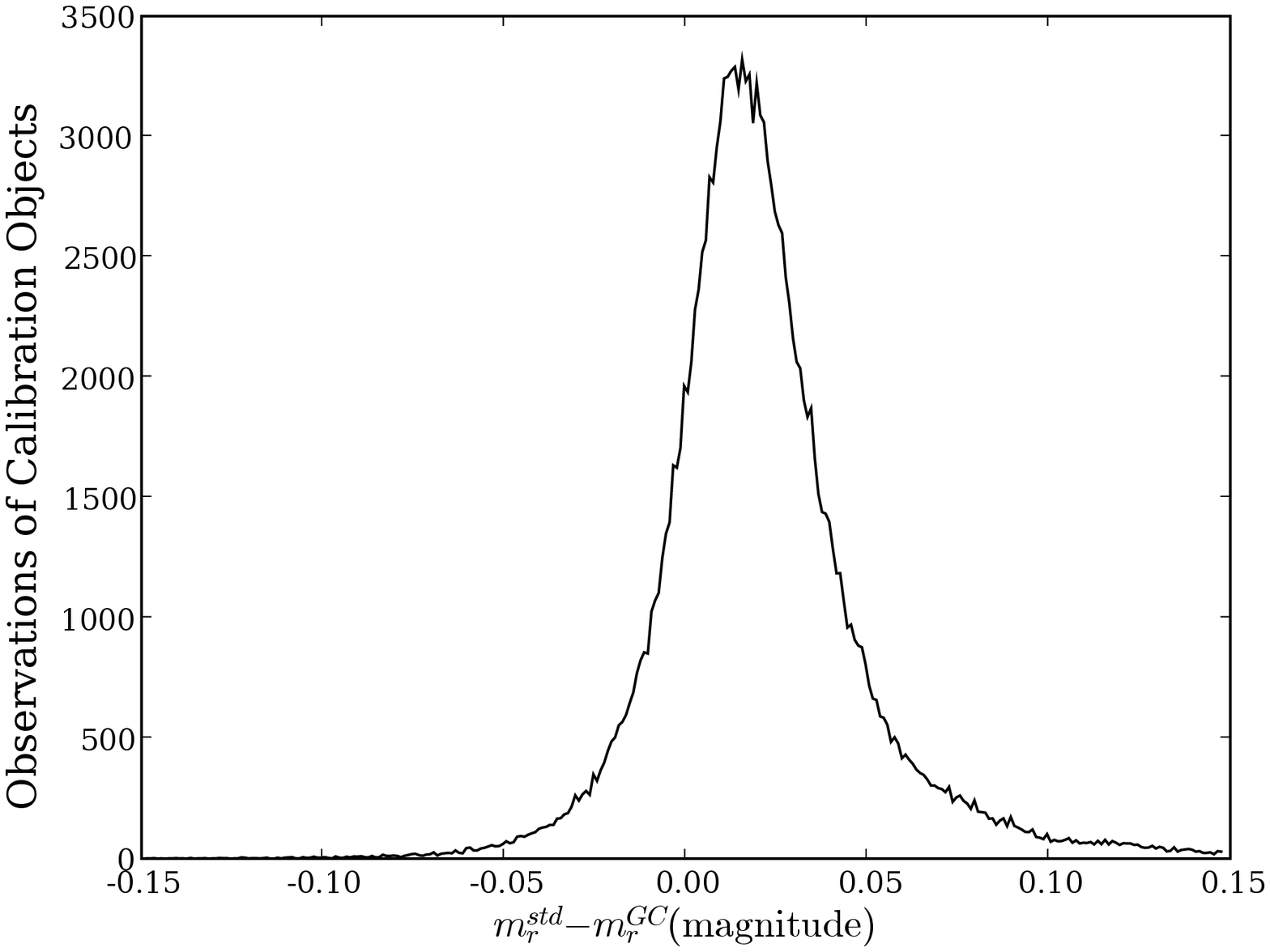}
\plotone{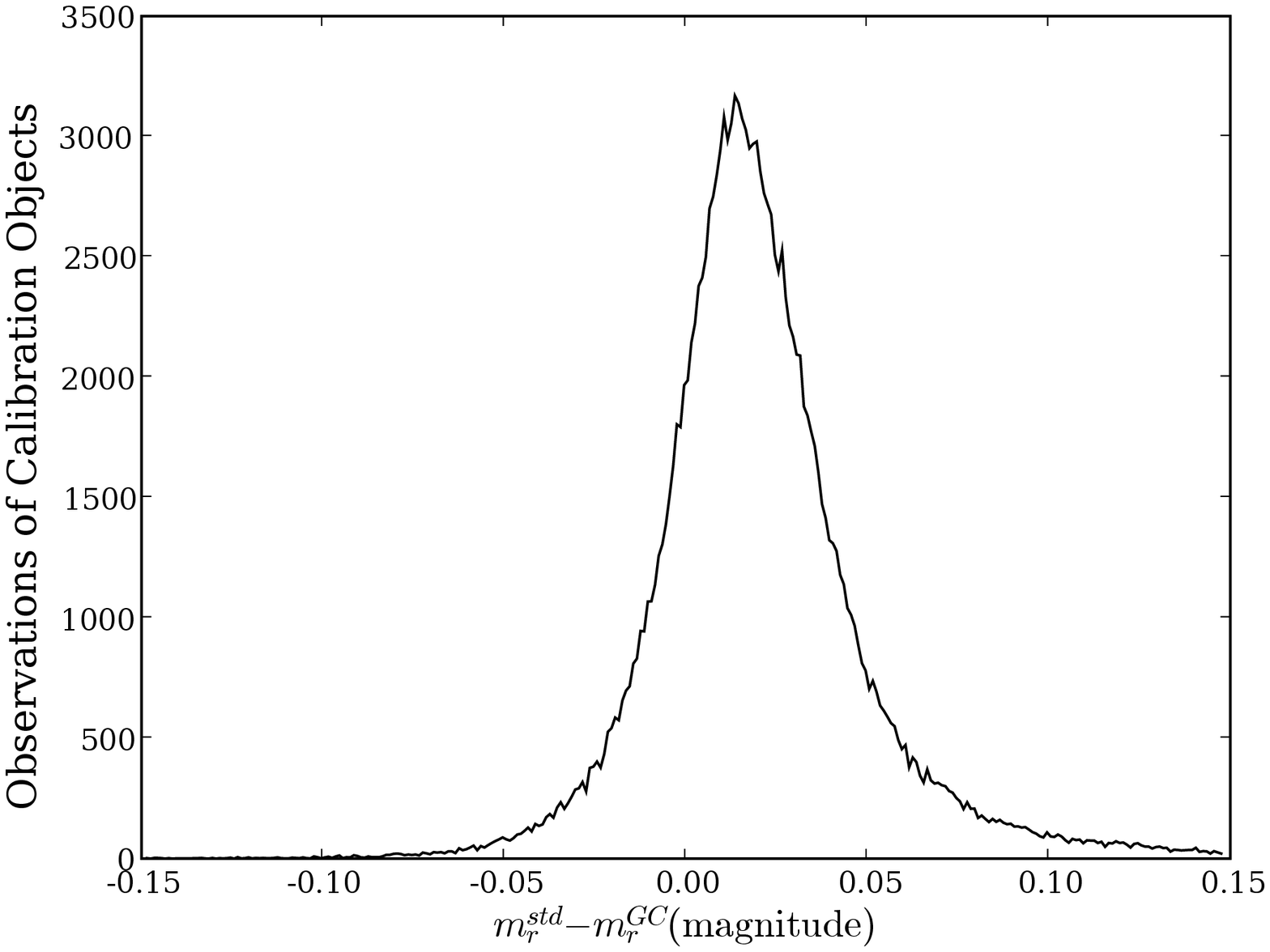}
\caption{Distribution of fitted gray extinctions for individual measurements with (top) the second-order model,
   and (bottom) the third-order model. Note the expanded scale; underflow (51 counts) and overflow (29103 counts) bins are not plotted.  
          \label{fig:GC_Corrections}}
\epsscale{1.0}
\end{figure}

Finally in this section, we comment on the origin and meaning of the calibration zero-points.
Expanded distributions of the $E^{gray}$ values (Eq. \ref{eqn:grayext}) for the two model fits are shown in Figure \ref{fig:GC_Corrections}.
It can be seen that the fitted $m_r^{GC}$ are systematically brighter than the observed magnitude $m_r^{std}$
with a most probable offset of about 0.02 mag.
The technique used to make initial estimates of the flux of the calibration objects and the average gray extinction in each image
produces a systematic underestimate of the object magnitude and overestimate of the extinction;
the results depend on the number of observations that are made and on the range of magnitudes used to compute the average extinction.
The fitting procedure is insensitive to a common offset of all observed magnitudes within a given field and band,
and does not provide any mathematical constraint between celestial fields that do not spatially overlap. 
So only to the extent that the instrumental and chromatic atmospheric extinction calibrations are absolute can we expect
the $r$-band zero-points to be the same in the three fields used in this study.
But we can make meaningful comparisons of the repeated observations of each object in a particular field and passband.

\subsection{Information Content of the Calibration Model and Density of Calibration Data}
\label{sec:fitbehaviour}

We want to determine the order of the model used to fit the gray extinction that maximizes the information captured in the model.
But to avoid introducing unphysical behaviour in the solution, we do not want to severely over-fit the calibration data. 
Statistical evaluation of the information in the atmospheric model (e.g. the Akaike Information Criterion 
corrected for sample size) indicates that the model captures more information with the addition of the cubic terms. 
But we find that the physical parameters deduced from the full analysis (presented below) are equivalent for the second and third-order models,
and conclude that, for these data, there is no gain from including still higher-order parameterizations.

\begin{figure}
\plotone{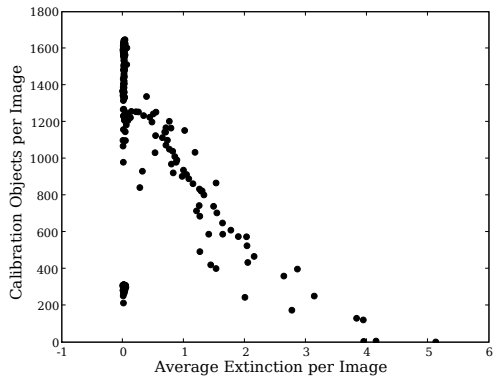}
\caption{The number of calibration objects per image as a function of the 
   initial estimate of the average extinction across the field of view.  The plot includes all images whether calibratable or not,
   and the cluster of images in the lower left corner are those taken of J2330 in clear conditions.
          \label{fig:ncalstars}}
\epsscale{1.0}
\end{figure}

The density of calibration stars limits the detail that can be successfully built into the calibration model.
As noted above, in cloudy conditions the calibration becomes poorer as the flux of light from calibration stars is reduced,
and also as the density of useful calibration stars becomes smaller.
Figure \ref{fig:ncalstars} shows the count of calibration stars for all observations as a function of the cloud absorption.
As expected the number of stars that pass the cuts applied in the selection of the calibration and test samples decreases considerably with absorption,
but this is not the only effect we need to take into account.

\begin{figure}
\epsscale{0.5}
\plotone{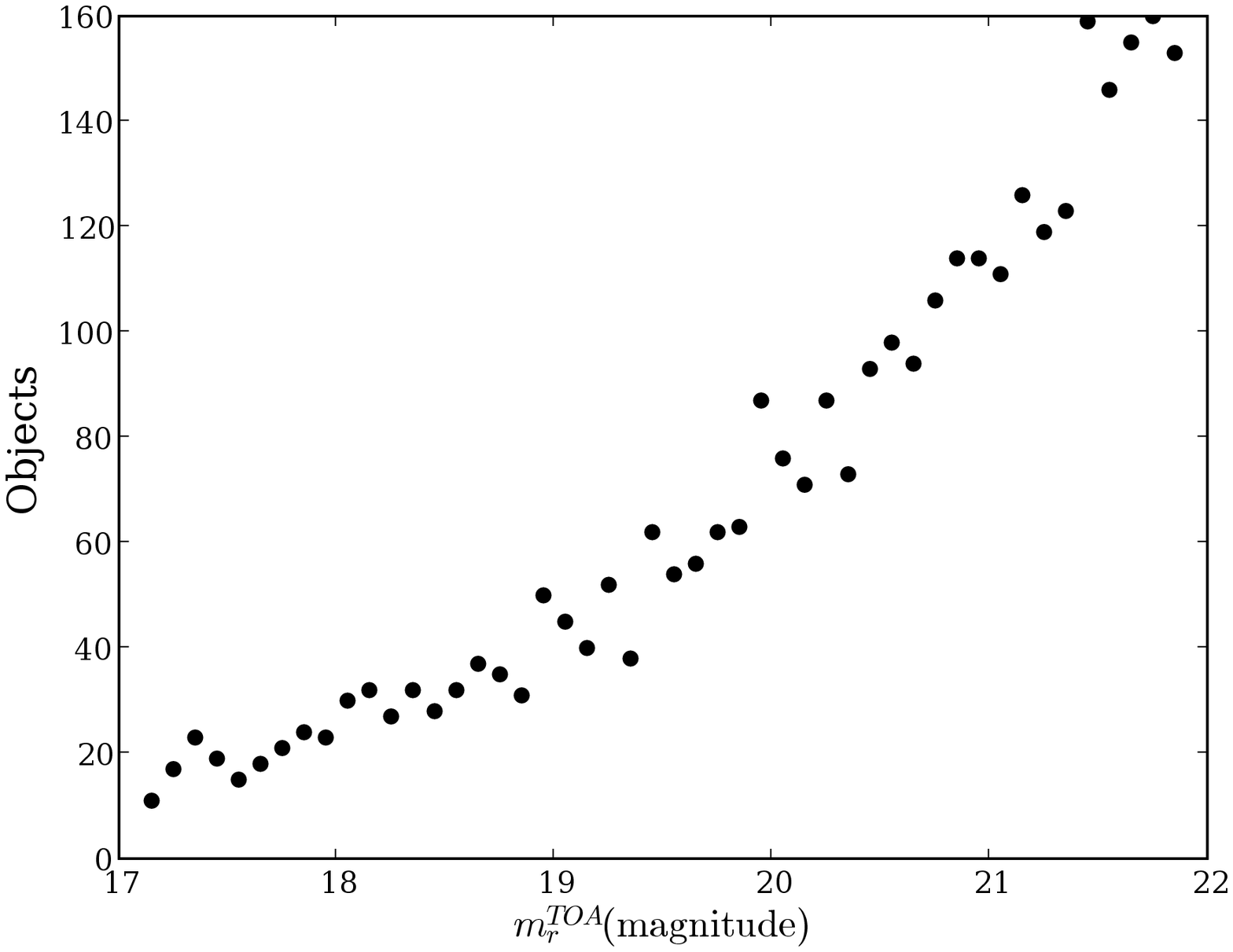}
\plotone{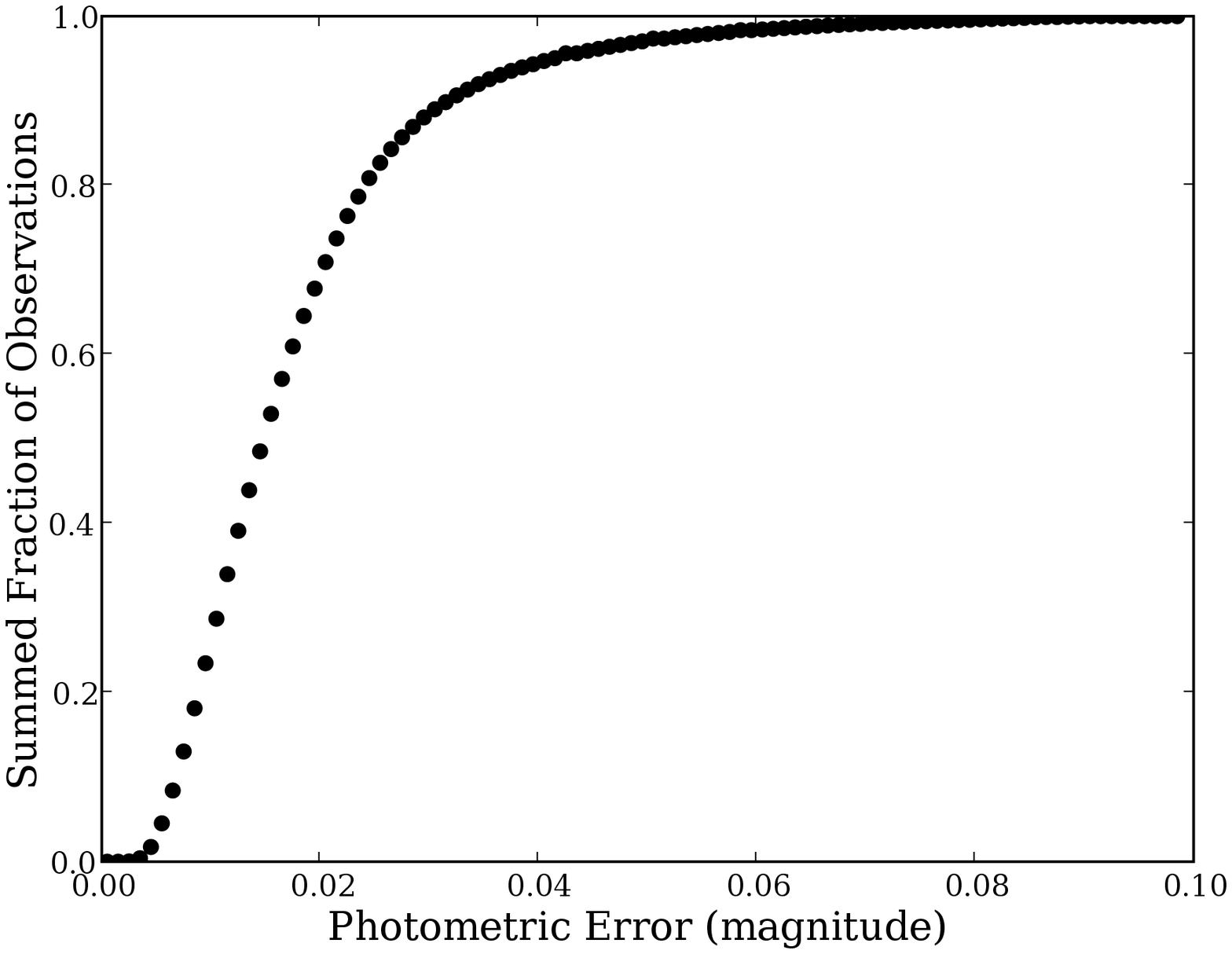}
\plotone{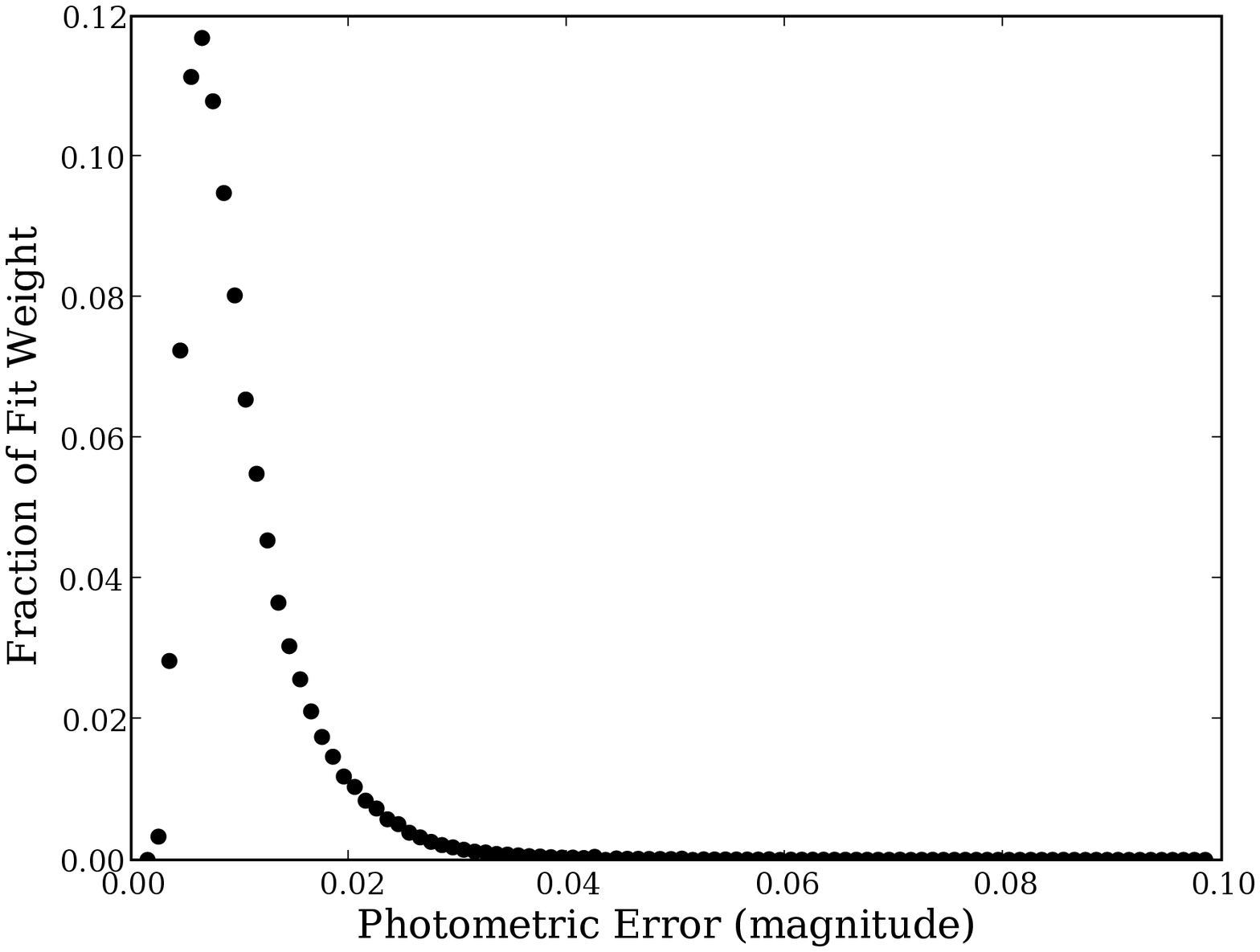}
\plotone{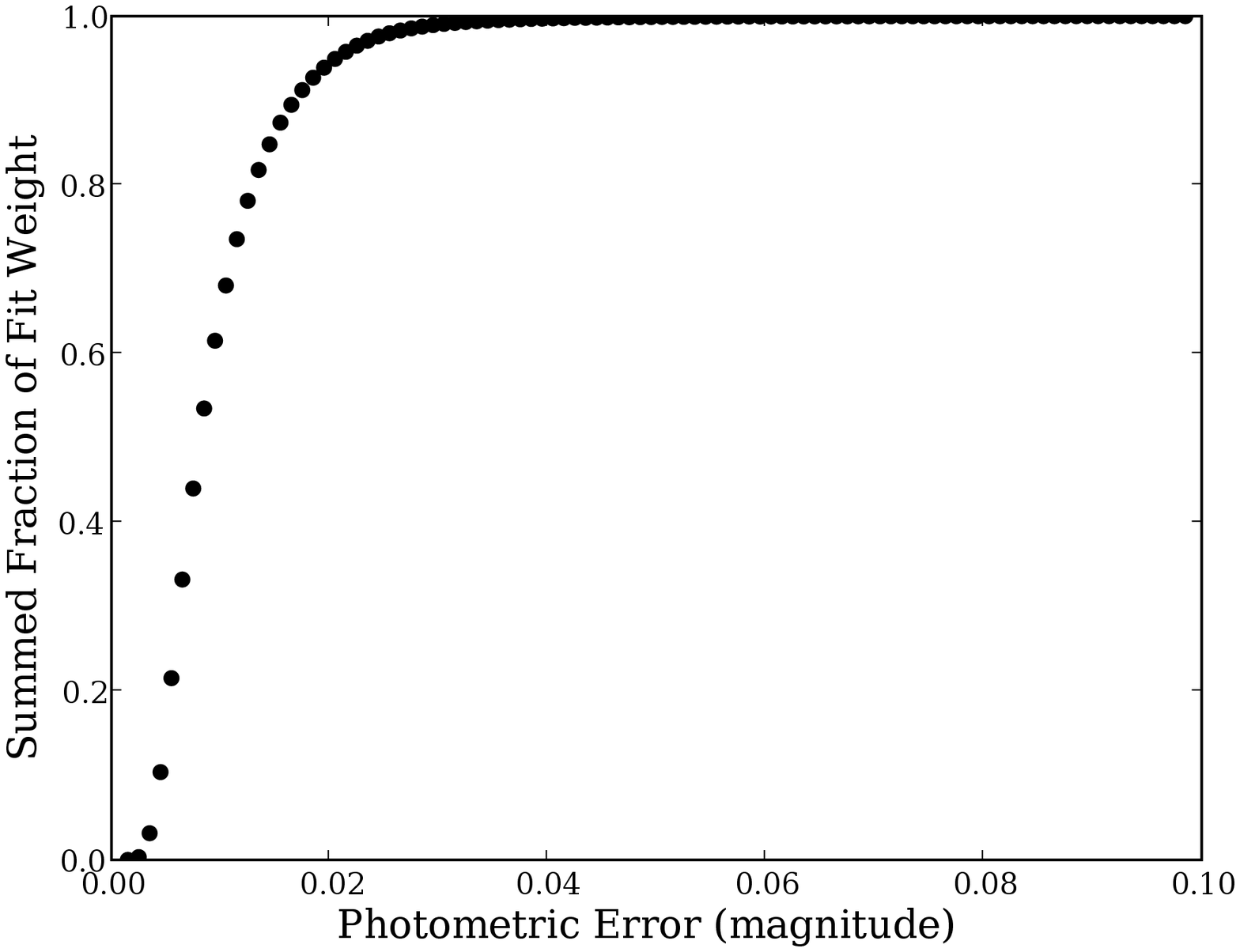}
\caption{Distribution (top left) of magnitudes of calibration objects;
         cumulative distribution (top right) of photometric errors of observations of calibration objects;
         distribution (bottom left) of weights in the global calibration fit;
     and cumulative distribution (bottom right) of calibration weights.  
          \label{fig:calmags}}
\epsscale{1.0}
\end{figure}

The $\chi^2$ statistic depends on the distribution of magnitudes of the calibration stars,
and the effective density of calibration points will not be simply the numerical density of calibration stars. 
We can define an effective weight to the global calibration $\chi^2$ from stars of a given photometric error,
\begin{equation}
\label{eqn:chiwgt}
 f_{\chi}(\sigma^{phot}) \equiv \frac{N(\sigma^{phot})}{ \big( \sigma^{phot} \big)^2},
\end{equation} 
where $N(\sigma^{phot})$ is the number of observations of calibration objects with photometric error $\sigma^{phot}$ within a given data sample.
Shown in Figure \ref{fig:calmags} is $f_{\chi}$ and its cumulative distribution computed for the full sample of calibration objects.
It can be seen that, for these data, $60\%$ of the $\chi^2$ comes from observations with $\sigma^{phot} < 0.01$,
while as also shown in the figure, this corresponds to about 30\% of the observations of calibration objects. 
So the effective density of calibration points is less than that estimated from the number counts alone.

\begin{figure}
\epsscale{0.50}
\plotone{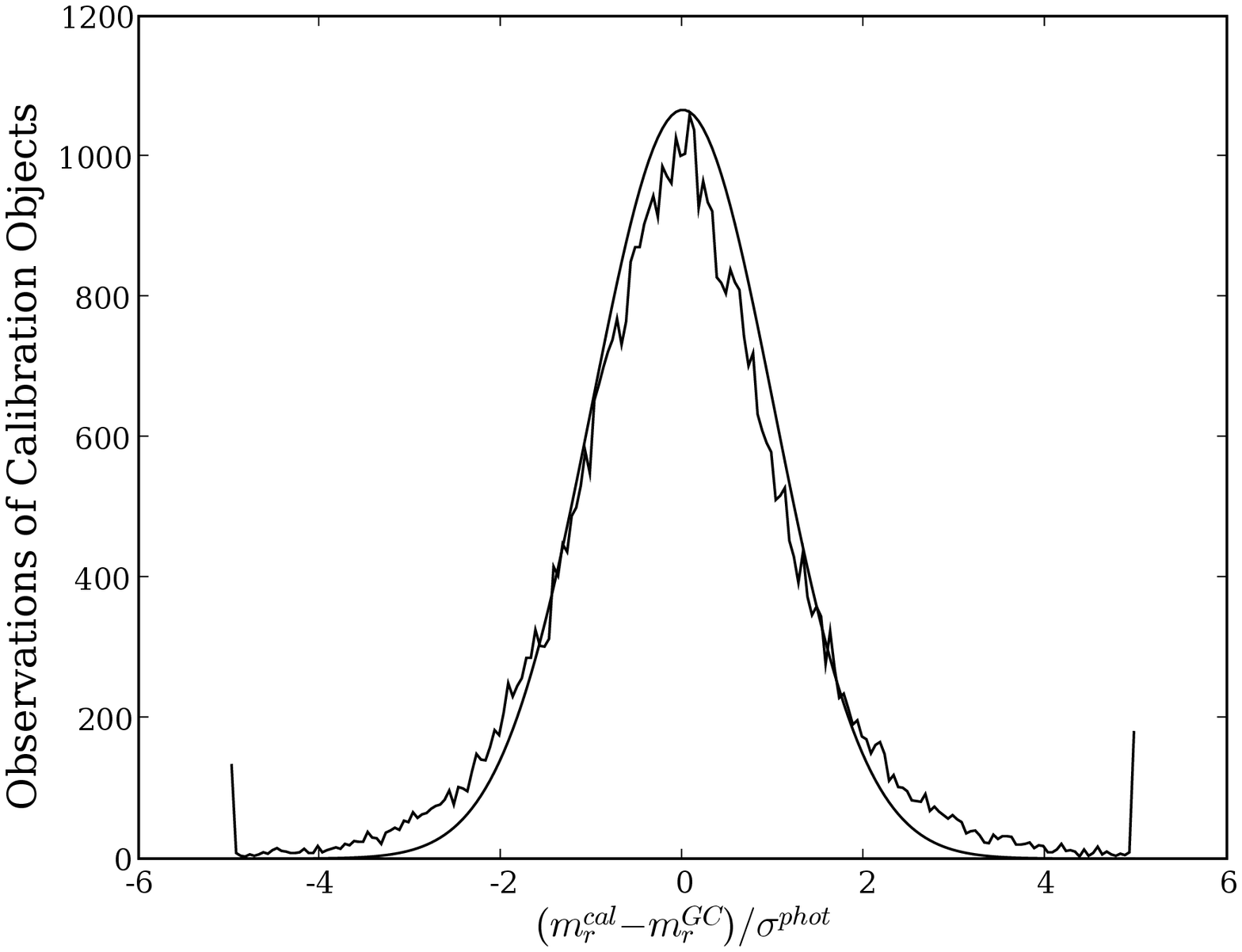}
\plotone{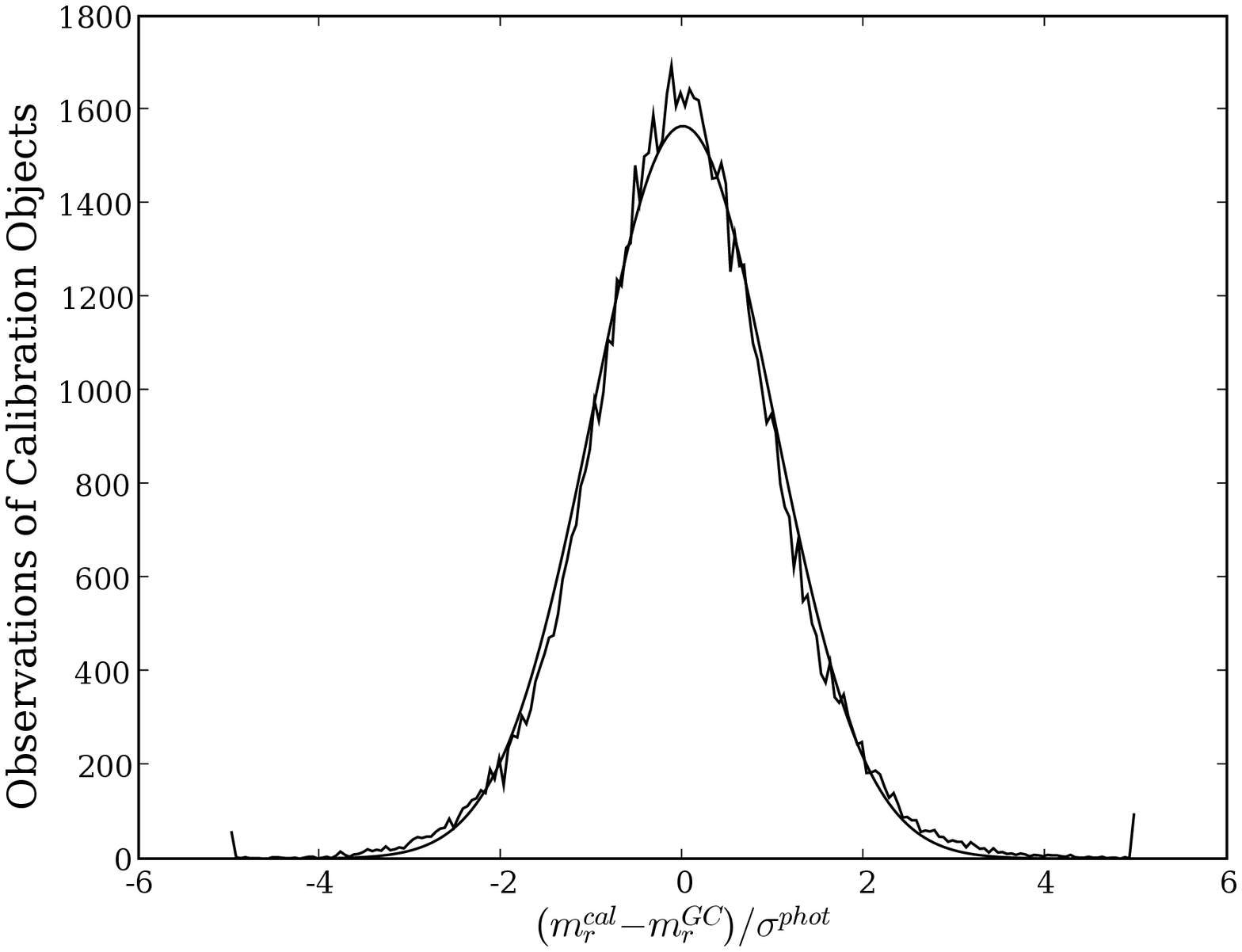}
\plotone{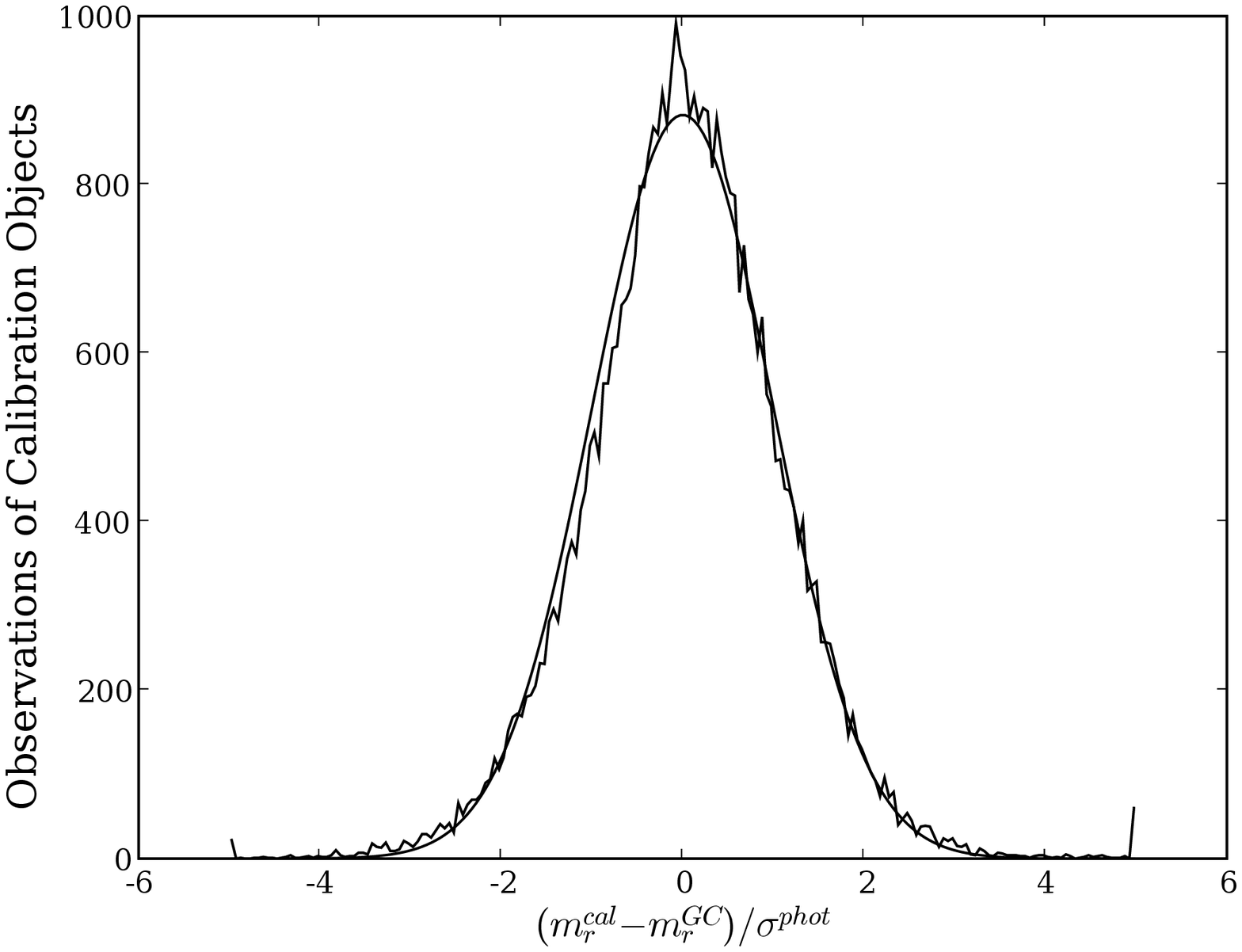}
\plotone{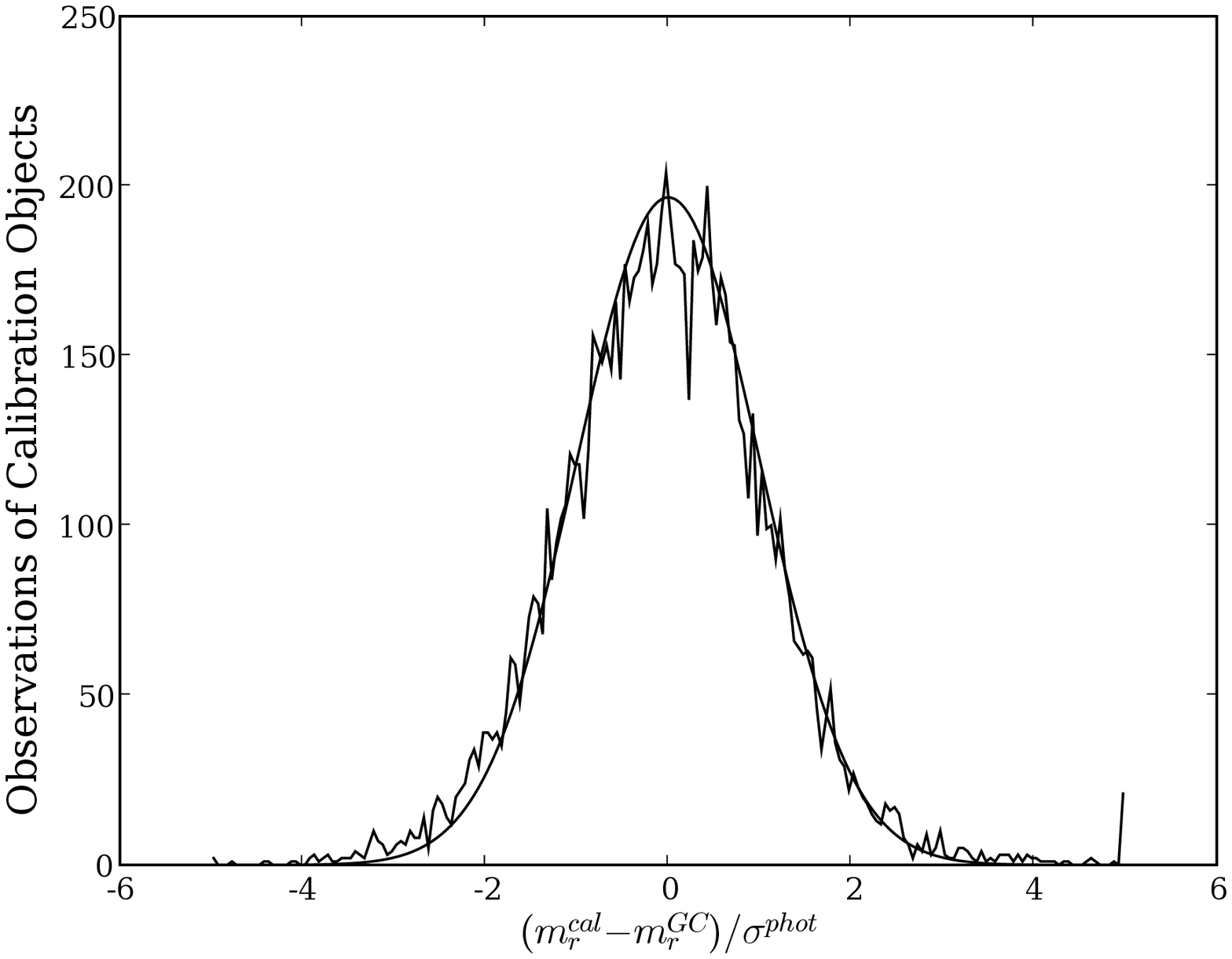}
\caption{Distributions of ``pulls'' of the calibration second-order model fit for observations with photometric 
         error $\sigma^{phot}$ in ranges, (top left) 0.00 - 0.01, (top right) 0.01 - 0.02, (bottom left) 0.02 - 0.04, and (bottom right) 0.04 - 0.10.
         Overflow and underflow counts are shown at $\pm 5$ on the horizontal axis.
         Overlaid on the data are Gaussian distributions with s.d. = 1.0 and normalized to the total number of observations.     
          \label{fig:M6_calpulls}}
\epsscale{1.0}
\end{figure}

\begin{figure}
\epsscale{0.50}
\plotone{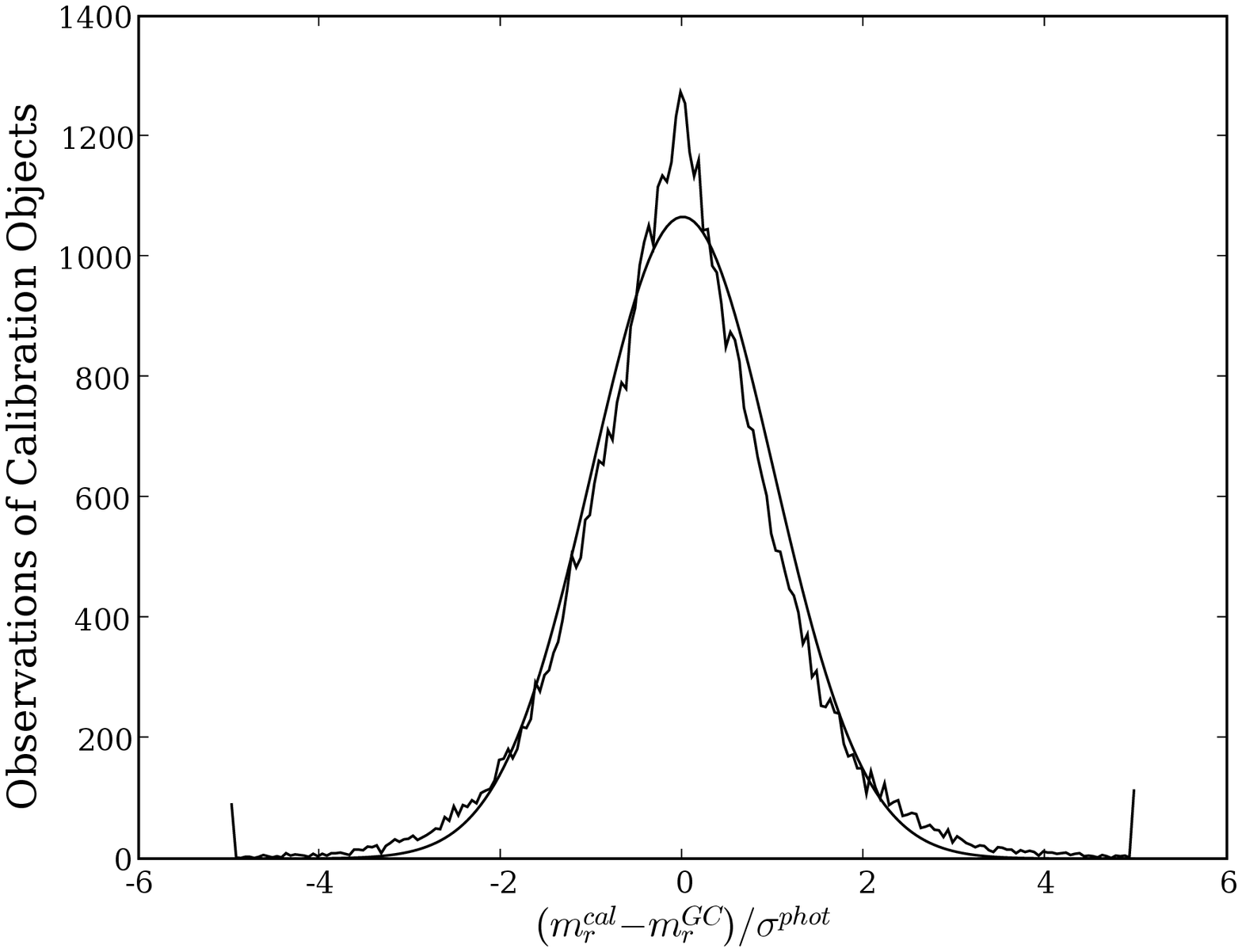}
\plotone{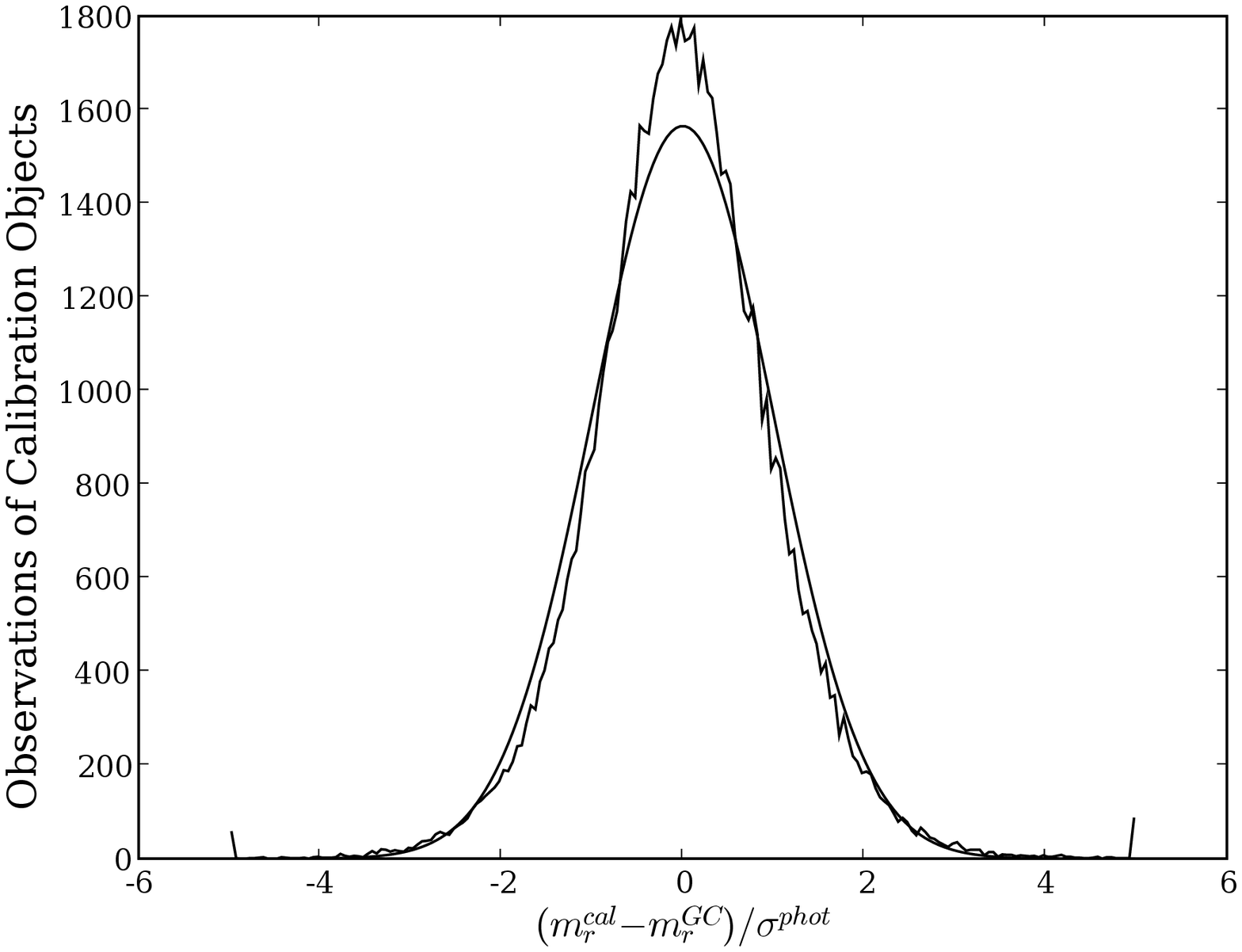}
\plotone{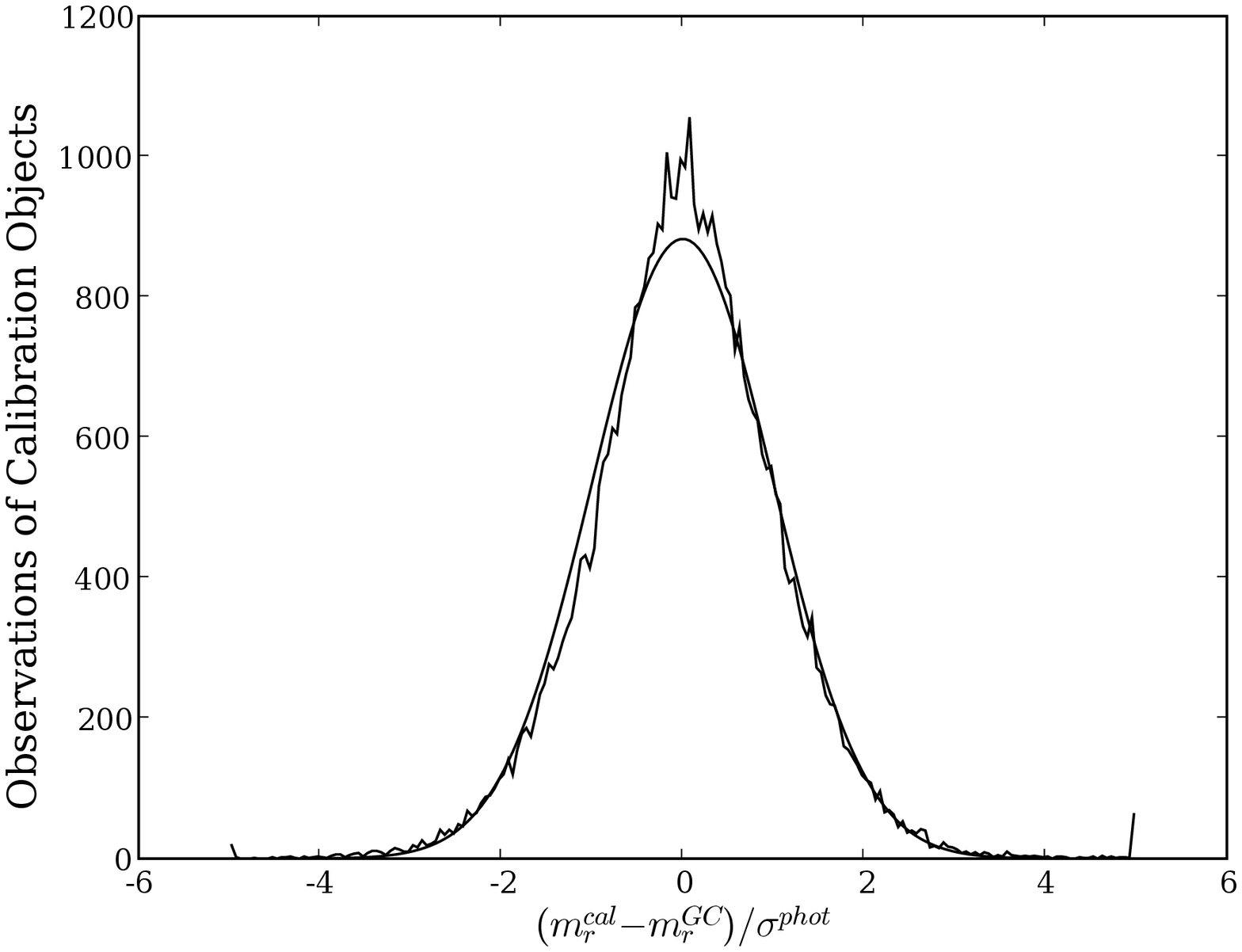}
\plotone{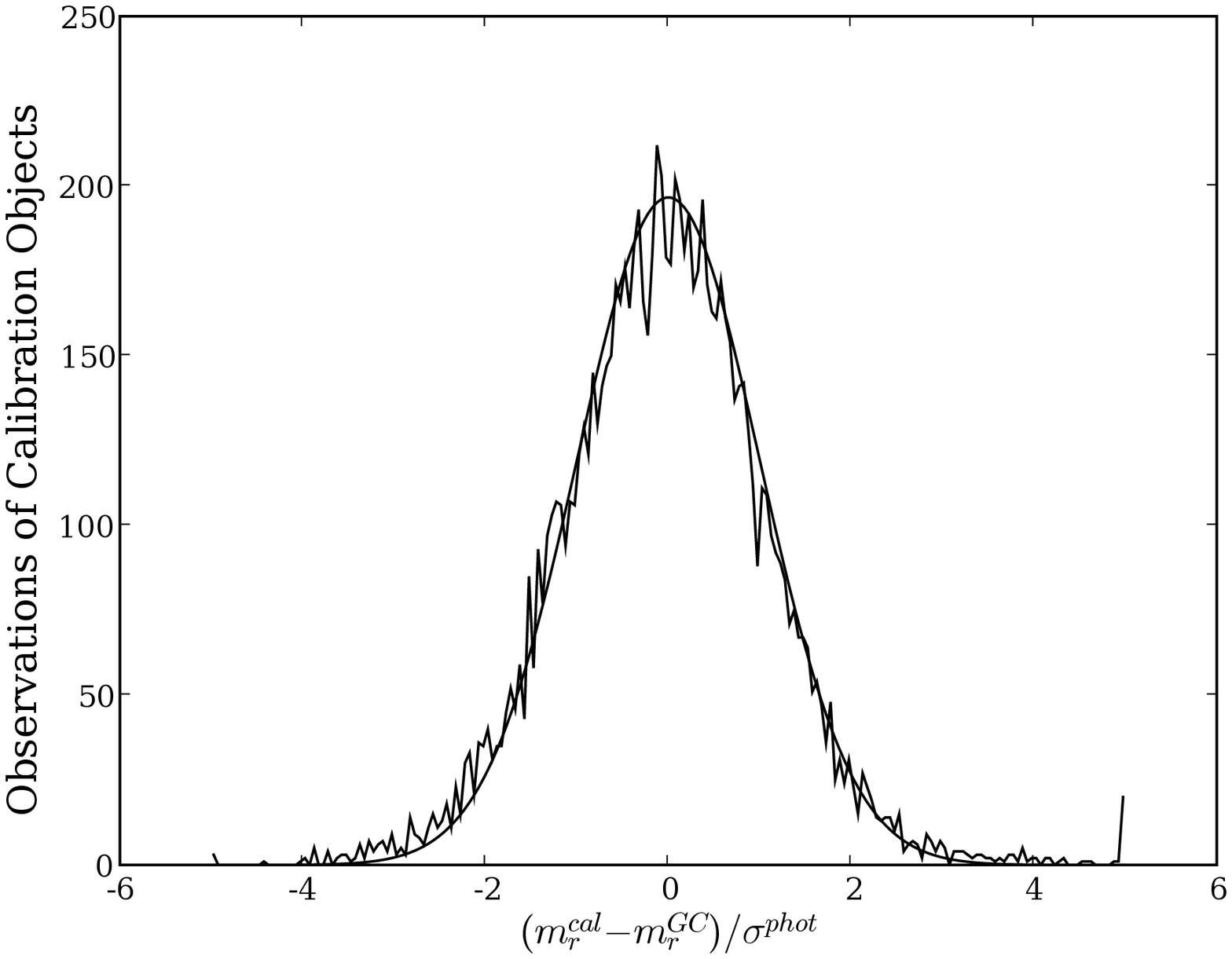}
\caption{Distributions of ``pulls'' of the calibration third-order model fit for observations with photometric 
         error $\sigma^{phot}$ in ranges, (top left) 0.00 - 0.01, (top right) 0.01 - 0.02, (bottom left) 0.02 - 0.04, and (bottom right) 0.04 - 0.10.
         Overflow and underflow counts are shown at $\pm 5$ on the horizontal axis.
         Overlaid on the data are Gaussian distributions with s.d. = 1.0 and normalized to the total number of observations.     
          \label{fig:M10_calpulls}}
\epsscale{1.0}
\end{figure}

Further insight to the information in the model can be gained by examining the distribution of residual differences between
magnitudes of the calibration objects determined from the global fits and the magnitudes found in individually calibrated observations.
The measured magnitudes of all calibration and test objects $i$ on all images $j$ were corrected using the results of the global fit to 
yield the calibrated observations $m_r^{cal}(i,j)$,
\begin{equation}
\label{eqn:mGC}
m_r^{cal}(i,j) = m_r^{std}(i,j) -  \delta_r(x,y,x^p,y^p,j).
\end{equation}
For calibration objects, these can then compared to $m_r^{GC}$ extracted from the fit to compute the residual of each observation,
\begin{equation}
\label{eqn:caldms}
\Delta_m^{cal}(i,j) \equiv m_r^{cal}(i,j) - m_r^{GC}(i).
\end{equation}
Distributions of the residuals normalized by their photometric measurement errors (``pulls'' $\equiv  \Delta_m^{cal}(i,j)/\sigma^{phot}(i,j)$)
are shown in Figures \ref{fig:M6_calpulls} and \ref{fig:M10_calpulls}. 
Shown overlaid on the data are Gaussian distributions with standard deviation equal one and normalized to the total number of observations in each plot;
these would be expected to reproduce the distribution of pulls if the fitting model is true and complete and the photometric errors are Gaussian.
It can be seen that the assignments of errors in the photometric reductions are generally quite good.
But the distributions for the third-order fit are generally more narrow than statistics would predict.
This indicates some level of over-fitting to the data.
Conversely the distributions for the second-order fit,
particularly the pulls to the observations with $\sigma^{phot} < 0.01$ that dominate the fit,
indicate that there remains information in the data not fully captured in the model.
We complete a full analysis and present results for both models in order to test the sensitivity of our conclusions to the order of the model.

\subsection{Photometric Calibration Precision}
\label{sec:results}

The primary measure of success for the work here is the internal precision (reproducability) of the photometric calibration of the data.
We want the repeatability of measurements of magnitudes of a celestial object to be limited only by the shot noise in
the photon counts of the source and background sky.

\begin{figure}
\epsscale{1.0}
\plotone{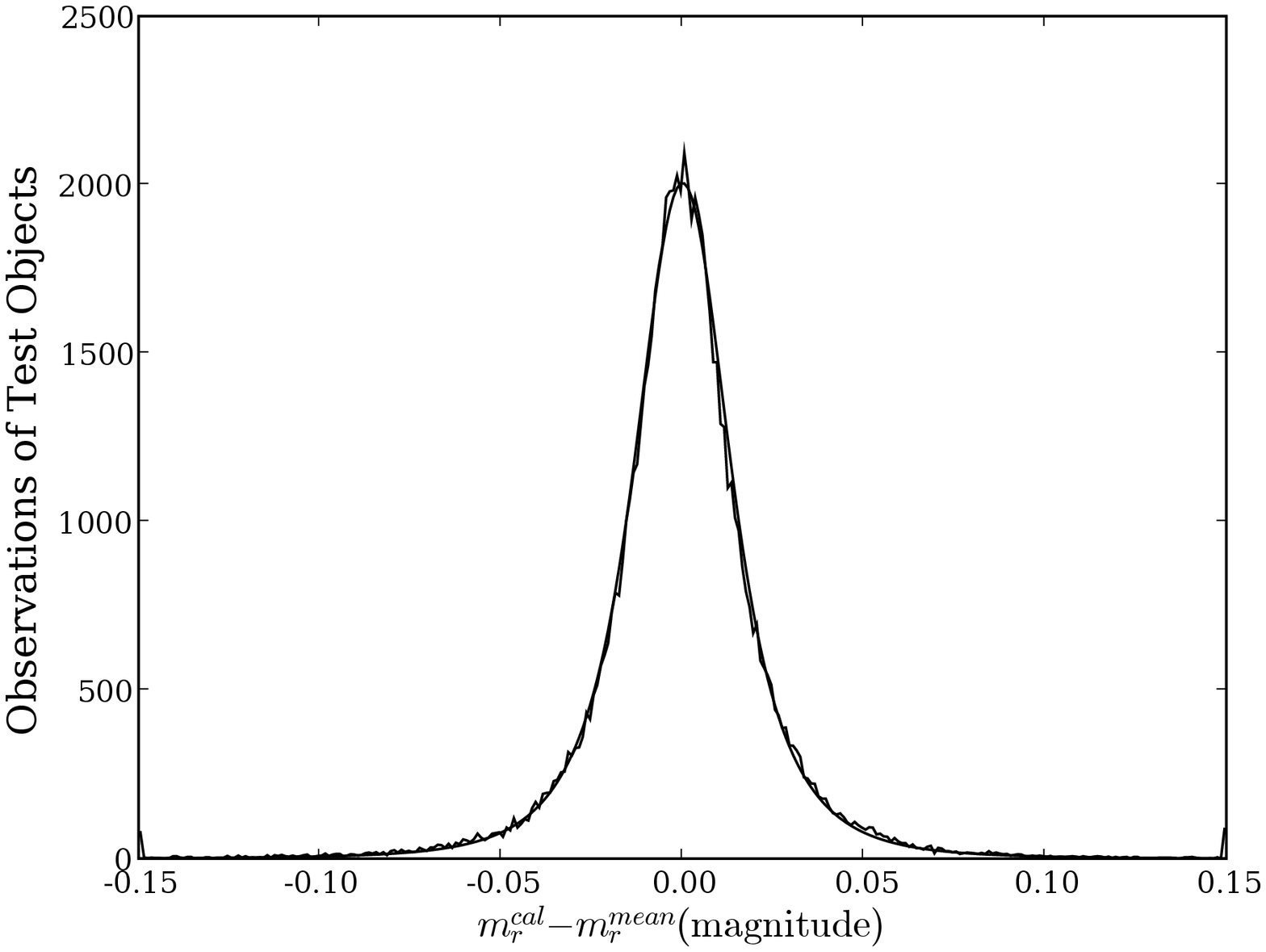}
\plotone{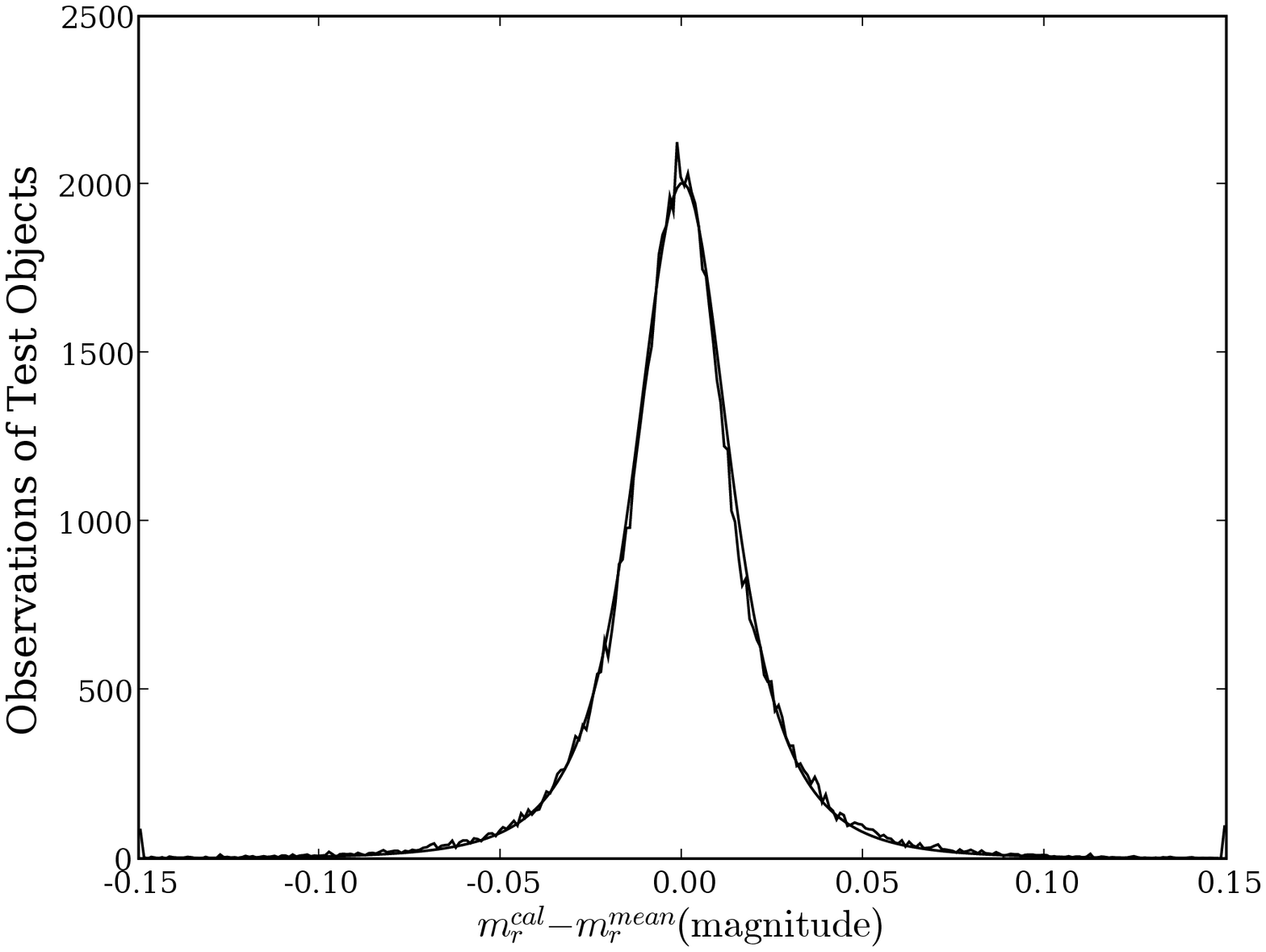}
\caption{Dispersion of all calibrated test observations.
   The horizontal axis is the difference between individual test observations and the mean magnitude of all observations of the same star.
   Over or under flow counts are accumulated in bins at the extreme ends of the horizontal range.
   The curve is computed from the estimated photometric extraction errors and a fitted calibration error (see text).
   Result of (top) the second-order model analysis, and (bottom) the third-order model analysis.  
          \label{fig:GC_Test_0}}
\epsscale{1.0}
\end{figure}

The precision of the calibration procedure cannot be estimated from the residuals of the observations of calibration objects 
as they are biased by the fit itself.
So we use the test objects that were randomly separated from the calibration objects, but in all other respects were chosen and processed in the same way.
The observations of test objects randomly sample the error in the calibration model and provide an unbiased estimate of the calibration precision.  
The error-weighted mean of the calibrated magnitudes $m_r^{cal}(i,j)$ of each test object $i$ was computed,
\begin{equation}
m_r^{mean}(i) = \frac{\sum_{j} m_r^{cal}(i,j)\sigma^{phot}(i,j)^{-2}}{\sum_{j} \sigma^{phot}(i,j)^{-2}},
\end{equation} 
and the residuals of individual measurements of test objects then defined in a way analogous to that for the calibration objects,
\begin{equation}
\Delta_m^{test}(i,j) \equiv m_r^{cal}(i,j) - m_r^{mean}(i).
\end{equation} 
The distribution of $\Delta_m^{test}$ is shown in Figure \ref{fig:GC_Test_0} for all observations of all test objects in the fields under study.
The solid curve shown in the plot is a model computed by representing each observation by a unit Gaussian,
\begin{eqnarray}
\label{eqn:calmdl}
N^{mdl}&&(\Delta_m(i,j)) \!\! =  \!\! \frac{\Delta_{bin}}{2.5066 \sigma^{mdl}(i,j)}   \times  {}   \nonumber\\
       &&   {}  \times {\rm exp}\bigg ( \!\!-0.5\Big(\frac{\Delta_m(i,j)}{\sigma^{mdl}(i,j)}\Big)^2\bigg),
\end{eqnarray}
where the parameter $\Delta_{bin}$ is the bin size in $\Delta_m$ used to histogram the data,
and $\sigma^{mdl}$ is a model value for the full statistical error in the measurement.
The data do not support a full evaluation of the covariance of the errors in the different parameters derived by the calibration fit.
So we assume them to be independent of each other and to be Gaussian distributed with standard deviation $\sigma^{cal}$ which we take to 
be a single constant for a given sample of the data.
The total error in the calibrated magnitude extracted from an observation of a test object is then,
\begin{equation}
\label{eqn:mdlsig}
\sigma^{mdl}(i,j)^2 = \sigma^{phot}(i,j)^2 + (\sigma^{cal})^2.
\end{equation}

\begin{figure}
\epsscale{1.0}
\plotone{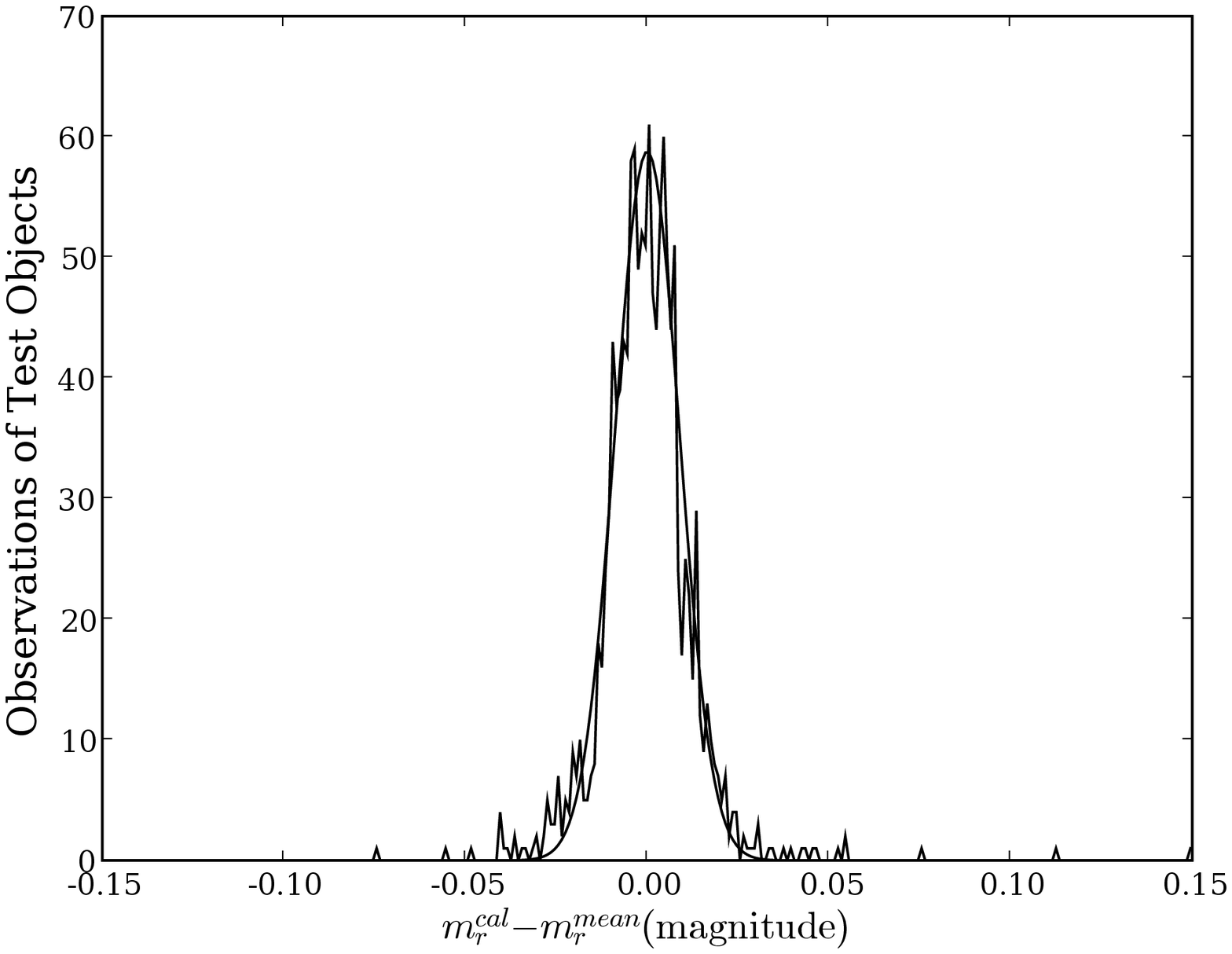}
\plotone{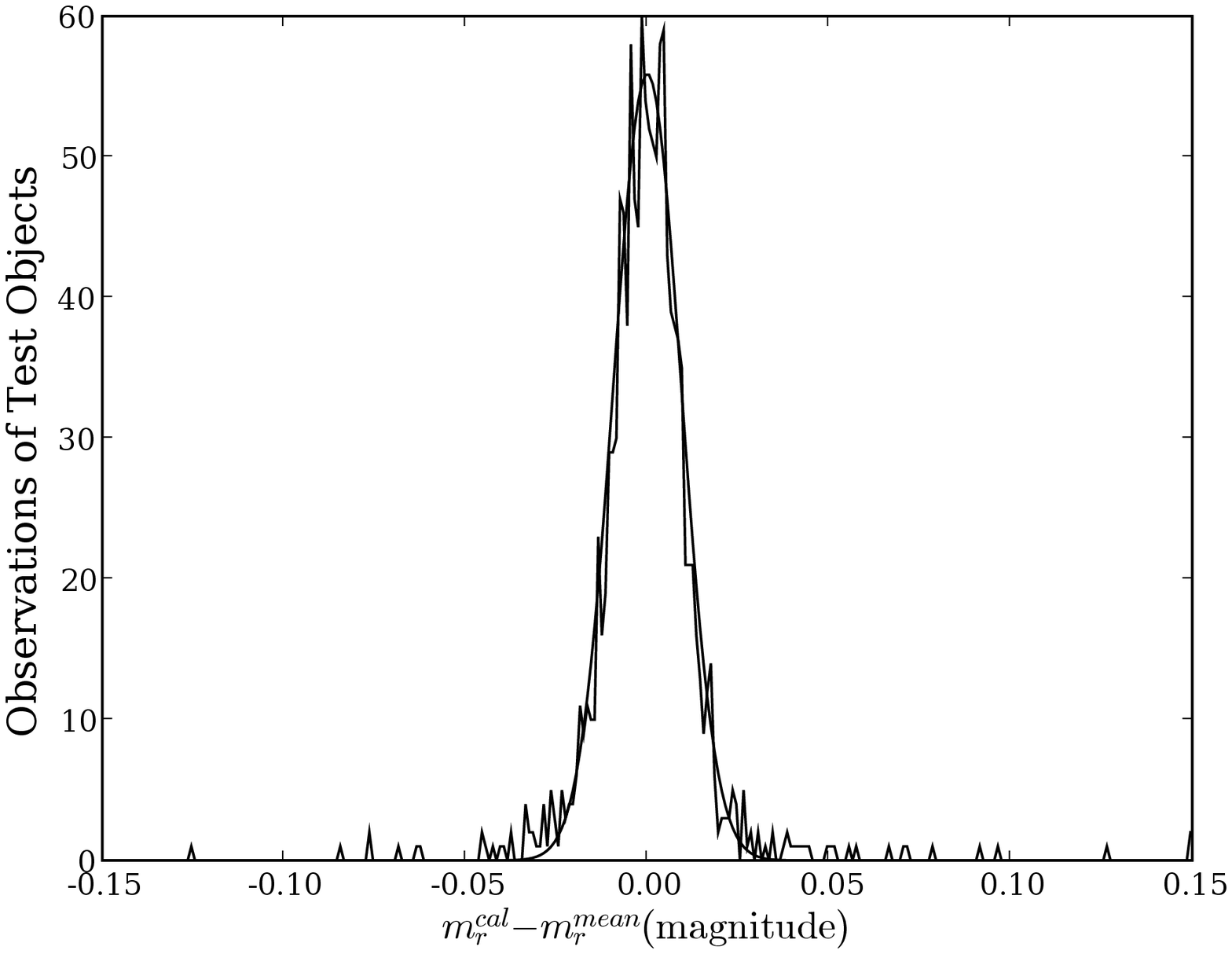}
\caption{Dispersion of calibrated test observations with $\sigma^{phot} < 0.005$.
   The horizontal axis is the difference between individual test observations and the mean magnitude of all observations of the same star.
   Over or under flow counts are accumulated in bins at the extreme ends of the horizontal range.
   The curve is computed from the estimated photometric extraction errors and a fitted calibration error (see text).
   Result of (top) the second-order model analysis, and (bottom) the third-order model analysis.        
          \label{fig:GC_Test_2}}
\epsscale{1.0}
\end{figure}

The value of $\sigma^{cal}$ was varied and a best fit ($\chi^2$ with root-$N^{mdl}$ errors) made to distributions of
residuals for various subsets of the test observations.
The fits are made to the histogram counts using bins with $N^{mdl} > 5$ and $\Delta_m < 0.1$.
The curve in Figure \ref{fig:GC_Test_0} is the sum of $N^{mdl}$ over all observations of all test objects computed with the value of the 
calibration error $\sigma^{cal}$ that yields the best fit to the data.
We also show in Figure \ref{fig:GC_Test_2} the equivalent plot for the subset of observations with $\sigma^{phot} < 0.005$.
The model yields a good fit in all cases, so we conclude that our simple ansatz
for the calibration error is sufficient to characterize the data.

The values of $\sigma^{cal}$ extracted from samples of the data with various views are summarized in
Tables \ref{table:models}, \ref{table:fitresults}, and \ref{table:calgray}.
Table \ref{table:models} gives results of fits with two calibration models to data with differing photometric errors, 
and includes numerical values for an intermediate sample chosen with $\sigma^{phot} < 0.010$ not shown in the figures.
Given in Table \ref{table:fitresults} are values of $\sigma^{cal}$ extracted from time-ordered subsets of the data (Table \ref{table:observing}).
This table includes the mean estimated photometric reduction error for each of the various observational samples,
and also includes the fraction of the images in each sample that were taken in conditions that were judged to be ``photometric'' as described
in Section \ref{sec:photometric} below.
Table \ref{table:calgray} provides values of $\sigma^{cal}$ extracted from subsets of the data partitioned according to the
gray extinction (Equation \ref{eqn:grayext}) averaged over the calibration objects on each image.
The entries in the last column of this table are computed by subtraction in quadrature of the precision of the calibration in the ``photometric'' sample
from the those determined from the test objects in ``non-photometric'' images.
We do not use data from J2330 in constructing Table \ref{table:calgray} as this field at high Galactic latitude contains
an extremely low density of calibration stars.

We can make a rough accounting of the contributions to the standard deviation $\sigma^{cal} \approx 0.005$ magnitude
reported in Table \ref{table:calgray} for observations made in ``photometric'' conditions.
Errors $\sigma^{phot}$ in measurements of the fluxes from stars that effectively contribute to the
calibration at a point on a given image are largely statistical;   
from the values of the zero-point structure functions (ZPSF) presented below  
we estimate these to be $\sim 0.003$ magnitude for fields at lower Galactic latitudes.
Our observing and analysis procedures minimize contributions from uncorrected variations in seeing, instrumental response,
and wavelength-dependent extinction; these are $\le 0.001$ magnitude each. 
Finally, there are contributions from {\it error} in the estimates of the $\sigma^{phot}$ of the test star magnitudes used to extract $\sigma^{cal}$.
The view of the data in Table \ref{table:models} indicates that multiplicative errors are small,
but additive errors could be several millimags though our resolution is not sufficient to make a good estimate;
in this sense the calibration itself could be slightly better than five millimags. 
We note that additive errors do not substantially contribute to the entries in the last column of Table \ref{table:calgray}.

\begin{table} 
\caption{Comparison of Calibration Models.
         \label{table:models}}
\begin{center} 
\begin{tabular}{lccc} 
\tableline 
\noalign{\smallskip}
 Sample     &      Mean                  & \multicolumn{2}{c}{Test Object}    \\
            &    Photometric             & \multicolumn{2}{c}{Precision $\sigma^{cal}$ (mag)}           \\
            &      Error (mag)           & \multicolumn{2}{c}{Model Order}           \\
            &                            &   $2^{nd}$  &  $3^{rd}$                      \\
\noalign{\smallskip}
\tableline
\noalign{\smallskip}
All Images             &  0.021          &   0.007    &  0.007                        \\
$\sigma^{phot} <$ 0.010  &  0.007          &   0.008    &  0.008                        \\ 
$\sigma^{phot} <$ 0.005  &  0.004          &   0.009    &  0.008                        \\    
\noalign{\smallskip}
\tableline 
\end{tabular} 
\end{center}
\end{table}

\begin{table} 
\caption{History of Calibration Results.
         \label{table:fitresults}}
\begin{center} 
\begin{tabular}{lcccccl} 
\tableline 
\noalign{\smallskip}
  Date       &    Field     & Mean             &   Fraction            &    Test Object            \\
 (2009)      &              & Photometric      &     of Images         &    Precision              \\
             &              & Error (mag)      &   ``Photometric''     &     $\sigma^{cal}$ (mag)    \\
\noalign{\smallskip}
\tableline
\noalign{\smallskip}
July 4       &   J2103      &    0.023         &       0.24            &        0.006       \\
July 5       &   CD-32      &    0.043         &       0.00            &        0.009       \\
July 6       &   CD-32      &    0.018         &       0.97            &        0.005       \\
July 6       &   J2103      &    0.018         &       0.75            &        0.008       \\
July 6       &   J2330      &    0.018         &       0.90            &        0.012       \\   
\noalign{\smallskip}
\tableline 
\end{tabular} 
\end{center}
\end{table}

\begin{table} 
\caption{Gray Extinction and Calibration Error.
         \label{table:calgray}}
\begin{center} 
\begin{tabular}{lccc} 
\tableline 
\noalign{\smallskip}
 Mean                   & Test Object              &   Gray Calibration        \\
Gray Extinction         &   Precision              &     Error                 \\
                        &  $\sigma^{cal}$ (mag)      &       $\sigma$ (mag)      \\
\noalign{\smallskip}
\tableline
\noalign{\smallskip}
``Photometric''         &  0.005                    &    NA                     \\
0.02 - 0.10             &  0.007                    &   0.005                   \\ 
0.10 - 0.80             &  0.008                    &   0.006                   \\ 
0.80 - 1.50             &  0.009                    &   0.007                   \\  
\noalign{\smallskip}
\tableline 
\end{tabular} 
\end{center}
\end{table}

We make a number of immediate observations from the results presented in this section:
\begin{itemize}
\item The second-order and third-order models yield essentially equivalent values for $\sigma^{cal}$.
So we conclude that these models nearly optimally capture the information in the data, and in particular that
the densities of calibration stars in these fields do not support further definition of the gray structure.
\item Calibrations of the CD-329927 field are consistent with the simple expectation that the precision 
becomes worse as the photometric reduction errors of the calibration sample increase.
\item The loss of statistical power as the density of calibration stars is reduced is apparent in the case of J2330 which is at high Galactic latitude.
The number of stars in the J2330 field that contribute to the global calibration is $\sim 0.1$ per square arcmin,
while $\sim 0.4-0.5$ stars per square arcmin contribute to the calibration of the CD-329927 and J2103 fields.
(See discussion in Section \ref{sec:fitbehaviour}.)
\item Calibrations under a variety of conditions are seen to be sub-percent in nearly all cases;
the availability of even a subset of ``photometric'' images provides tight constraint on calibration of entire samples.
\item From the view of Table \ref{table:calgray} it is seen that the calibration precision degrades rapidly with the onset of even thin cloud layers,
and then grows more slowly as the cloud layer thickens. We next examine this behaviour more quantitatively. 
\end{itemize}

\section{Photometric Conditions and Gray Structure Functions}
\label{sec:photometric}

The complexity of cloud structure at short spatial scales will determine how finely the gray extinction must be sampled by
calibration stars if we are to achieve good photometric precision in non-photometric conditions.
So we want to use our data to study the spatial structure of gray extinction by clouds,
and to determine how it affects the precision of our calibration process. 
For this we use the values of $E^{gray}(i,j)$ defined in Equation \ref{eqn:grayext}.

As discussed in Section \ref{sec:thefit} above, there is an undetermined $r$-band zero point for each field
that appears as a fixed bias of $E^{gray}$.
Such a fixed bias will not affect measurements of structure functions, 
but spatial correlations in the errors in the fitted magnitudes $m_r^{GC}$ can.
Other sources of systematic error in measurements of structure functions are caused by drift of the telescope pointing
between observations of a given field coupled with errors in the correction for instrumental response,
and variations in the chromatic atmospheric extinction corrections during a night.
The instrumental and chromatic atmospheric threats are combinations of two or more errors,
and our observing strategy was designed to keep them small, but we seek a method to estimate the full error in any case.  
Fortunately it turns out that we can identify photometric subsamples of the observations that can be used
to determine zero-points and to measure correlations in $E^{gray}(i,j)$ that are not directly created by cloud cover.

\begin{figure}
\epsscale{1.0}
\plotone{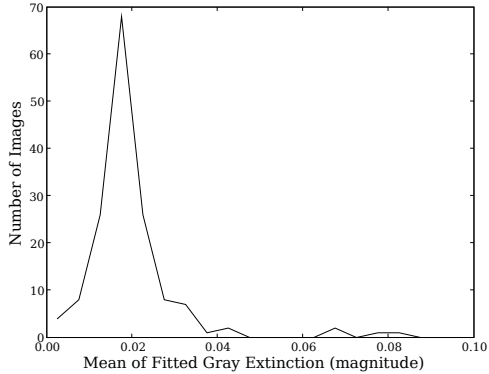}
\caption{Gray zero point distribution; see also Figure \ref{fig:Initial_Extinction}.   
          \label{fig:ZPN2}}
\epsscale{1.0}
\end{figure}

We compute $\overline{E^{gray}(j)}$, the gray extinction averaged over the calibration objects $i$ in image $j$,
and collect histograms of this average for all images.
The second-order and third-order fits yield essentially identical distributions, and we plot the distribution for the third-order 
model on a greatly expanded scale in Figure \ref{fig:ZPN2}  ({\it cf.} Figures \ref{fig:Initial_Extinction} and \ref{fig:GC_Corrections}).
We interpret the peak in the distribution to come from observations taken through minimal cloud cover, 
so the position of this peak defines the zero-point for our photometric scale.  
More completely, we extract this peak separately for each field, and define a relative zero point $ZPT$ for each sample.
We note that statistical errors on the $ZPT$ are $\sim \sigma^{phot}/\sqrt N_{obj} \sim 0.002$ for the number of objects $N_{obj}$ in each image.
We consider images with  $\vert \overline{E^{gray}(j)} - ZPT \vert < 0.02$ to have been taken through ``photometric'' sky.
The fractions of ``photometric'' images in each historical data sample are given in Table \ref{table:fitresults},
and the calibration precision for the combined ``photometric'' sample is given on the first line of Table \ref{table:calgray}.

\subsection{Gray Structure Functions}
\label{sec:cloudstructure}

We define the measured gray structure function ($GSF$) using the calibration objects on each image in a given sample as,
\begin{eqnarray}
\label{eqn:gsf}
&&GSF^{meas}(d^{sky}) \equiv \frac{1}{N_{pair}(d^{sky})} \times  {}  \nonumber \\ 
   && {}   \times \sum_{(i,i') \in j} (E^{gray}(x,y,i,j) - E^{gray}(x',y',i',j))^2,
\end{eqnarray}
where the summation extends over all pairs of objects $(i,i')$ in the image $j$
that are separated by the distance $d^{sky} (\arcmin) = |\vec x - \vec x'|$ on the sky.
The measured structure function is binned in $d^{sky}$ and averaged over all calibration images $j$ in a chosen sample of data.
Note that the structure function defined this way $GSF \rightarrow 0$ for completely correlated gray patterns (or as $d_{sky} \rightarrow 0$),
and $GSF \rightarrow 2 \Big [ \Big <(E^{gray})^2 \Big> - \Big <E^{gray} \Big >^2 \Big ]$ for completely uncorrelated patterns. 

As introduced in the previous section, the measured $GSF^{meas}$ may include spatial correlations due to systematic errors in the fitted
magnitudes of the calibration objects, drift of the telescope pointing coupled with errors in the measured instrument response,
or changes in undetected spatial variation of atmospheric extinction during an observing sequence (e.g. changing spatial profiles of aerosols).
We keep track of these potential sources of error in what follows. 
Specifically we write ({\it cf.} Eqns \ref{eqn:photoSysErr}, \ref{eqn:GCMerrormodel}, and \ref{eqn:grayext}) 
\begin{eqnarray}
\label{eqn:calgrayerr}
E^{gray}(x,y,&&i,j) =  \delta^{gray}(x,y,i,j) \!  + \! (m_r^{true}(i) - m_r^{GC}(i)) \!  +  {}  \nonumber \\
 && {}   + \epsilon_{r}^{inst}(x,y,j) + \epsilon_{r}^{chrom}(x,y,j) + \sigma^{phot},
\end{eqnarray}
where $\epsilon_{r}^{inst}$ and $\epsilon_{r}^{chrom}$ are possible residual errors arising from the instrumental and atmospheric throughput in the $r$-band.
Here, as before, $\sigma^{phot}$ represents contributions from random photometric errors.

The measured structure function can not be negative, and even the random photometric errors will produce a positive offset.
We compute an estimate $R$ of this offset with the assumption that the photometric errors are Gaussian random variables,
\begin{equation}
\label{eqn:ranD}
R(\sigma,\sigma') =  \int_{-\infty}^{\infty} \int_{-\infty}^{\infty} G(\sigma; \chi)G(\sigma'; \chi')(\chi - \chi')^2 d\chi d\chi',
\end{equation}
where $G(\sigma; \chi)$ is a Gaussian with unit area and width $\sigma$ (c.f. Eqn \ref{eqn:calmdl}).  
We evaluate this integral and create a look-up table for the range of widths $\sigma$ and $\sigma' < 0.10$
used in the selection of calibration objects.
The error in the measured structure function is then computed in parallel with the calculation of the $GSF$ calculation
using the values of $\sigma^{phot}$ encountered in the data.
\begin{eqnarray}
\label{eqn:ranDcomp}
GSF^{ran}&&(d^{sky}) \equiv  \frac{1}{N_{pair}(d^{sky})}  \times   {}   \nonumber \\ 
   &&  {}   \sum_{(i,i') \in j} R(\sigma^{phot}(i,j),\sigma^{phot}(i',j)).
\end{eqnarray}
This background is binned in $d^{sky}$, averaged over all images in the sample, and subtracted from the measured structure function,
\begin{equation}
\label{eqn:gsfcor} 
GSF^{meas-ran}(d^{sky}) \equiv GSF^{meas}(d^{sky}) - GSF^{ran}(d^{sky}).
\end{equation}

Our assumption that images in our photometric samples are taken through a nearly cloudless sky corresponds to assuming
that $\delta^{gray} \approx 0$ for those observations.
So we use those images to estimate the contributions to $GSF(d^{sky})$ of the remaining errors.
We define the measured zero point structure function $ZPSF^{meas-ran}(d^{sky})$ to be the $GSF^{meas-ran}$
evaluated for images in the photometric samples, and take it as an estimator for the $ZPSF$,
\begin{eqnarray}
\label{eqn:ZPSFa}
ZPS&&F(d^{sky}) \equiv   \frac{1}{N_{pair}(d^{sky})}  \sum_{(i,i') \in j}                    {} \nonumber\\ 
   & & {} \bigg(  ( (m_r^{true}(i) - m_r^{GC}(i)) - (m_r^{true}(i') - m_r^{GC}(i')) )    +  {} \nonumber\\
   & & {} + ( \epsilon_r^{inst}(x,y,i) - \epsilon_r^{inst}(x',y',i') )                -  {} \nonumber\\
   & & {} + ( \epsilon_r^{chrom}(x,y,i) - \epsilon_r^{chrom}(x',y',i') )               -  {} \nonumber\\
   & & {} + ( \sigma^{phot}(i) - \sigma^{phot}(i') )                    \bigg)^2      +  {} \nonumber\\
   & & {} -  \frac{1}{N_{pair}(d^{sky})}  \!\! \sum_{(i,i') \in j} \!\!  R(\sigma^{phot}(i),\sigma^{phot}(i')).
\end{eqnarray} 
If the random photometric errors $\sigma^{phot}$ are uncorrelated with the remaining errors,
and their contribution properly represented by $R(\sigma^{phot}(i),\sigma^{phot}(i'))$, then  
\begin{eqnarray}
\label{eqn:ZPSFb}
ZPS&&F(d^{sky})  =    \frac{1}{N_{pair}(d^{sky})}  \sum_{(i,i') \in j}                       {} \nonumber\\
   & & {}    \bigg( ( (m_r^{true}(i) - m_r^{GC}(i)) - (m_r^{true}(i') - m_r^{GC}(i')) )  +  {} \nonumber\\
   & & {} + ( \epsilon_r^{inst}(x,y,i) - \epsilon_r^{inst}(x',y',i') )                +  {} \nonumber\\
   & & {} + ( \epsilon_r^{chrom}(x,y,i) - \epsilon_r^{chrom}(x',y',i') )              \bigg)^2.
\end{eqnarray}
This is the residual correlation introduced by the analysis procedure that we need to subtract from the 
$GSF^{meas-ran}$ to obtain the true gray structure in the data.
We note that the $ZPSF$ computed with this procedure will not include correlations between true gray extinction and errors in the calibration fit,
but we expect those to be small.

\begin{figure}
\epsscale{1.0}
\plotone{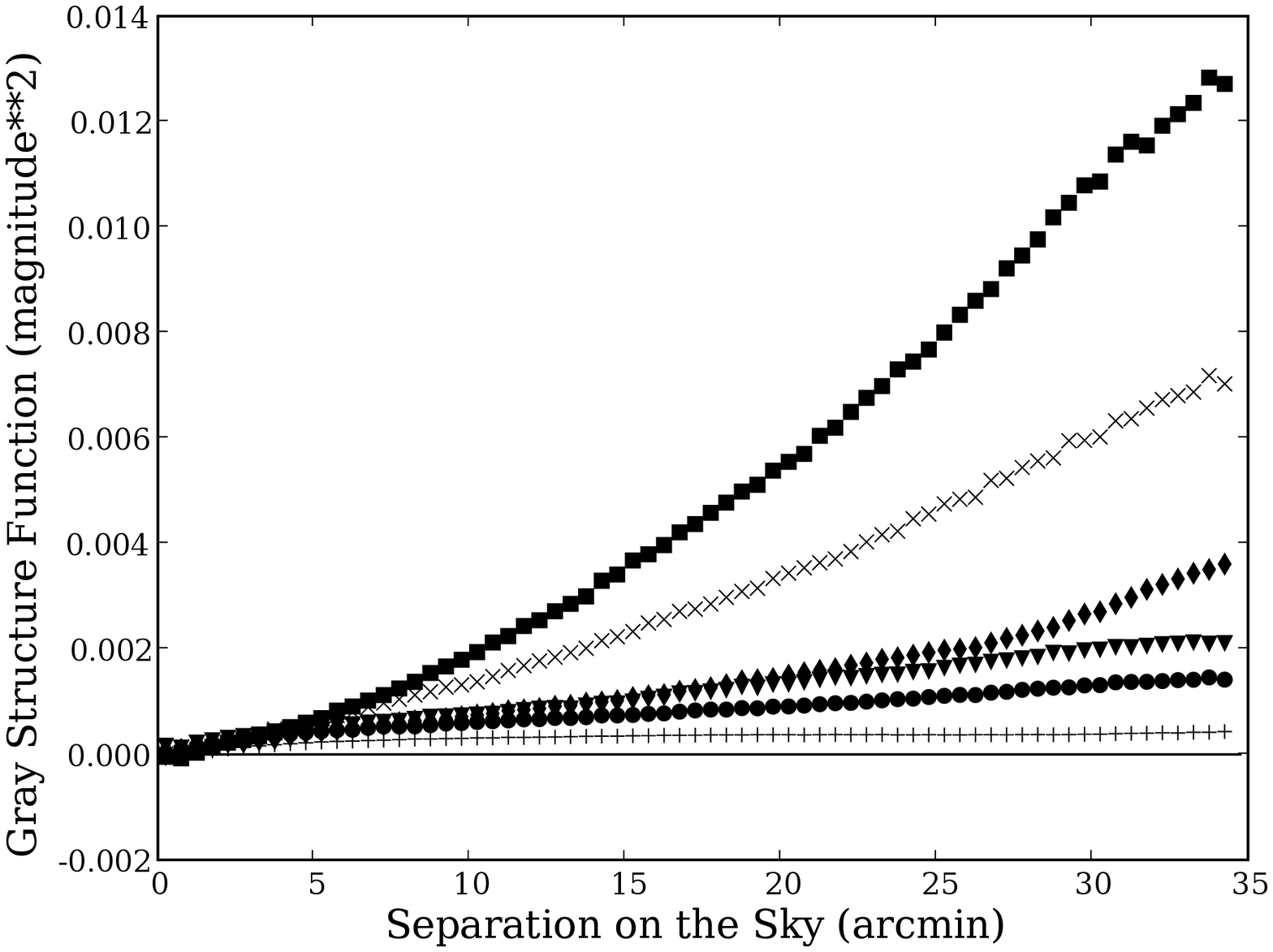}
\plotone{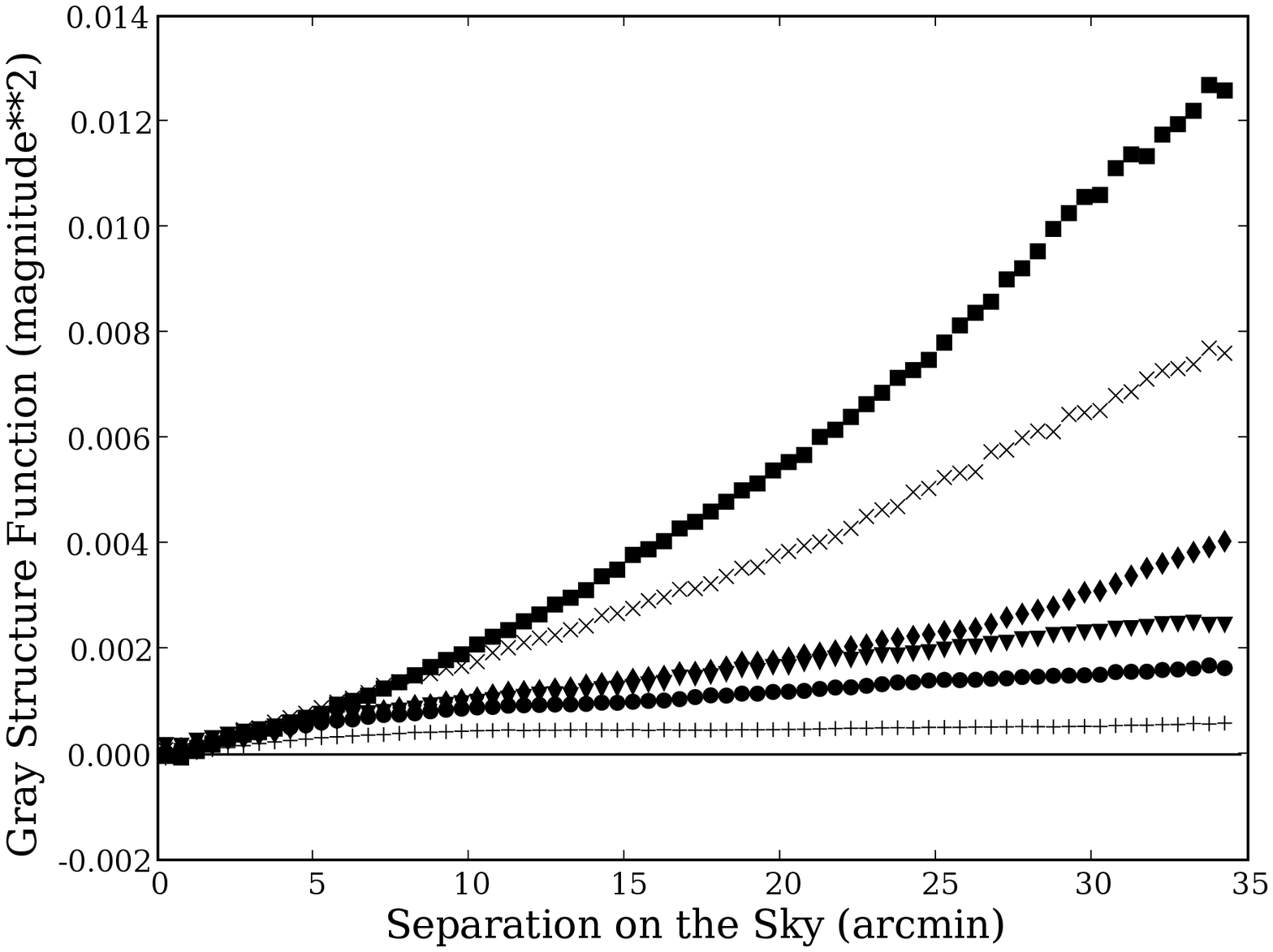}
\caption{Gray extinction structure functions measured after subtraction of bias from random photometric errors,
    and computed with calibration object magnitudes from (top) the second-order model fit, and (bottom) the third-order model fit.
    Shown separately in each plot are the zero-point structure function (+) $ZPSF$, and the gray structure function $GSF$ averaged over
    images with mean gray values
    ($\bullet$) 0.02 - 0.10, ($\blacklozenge$) 0.10 - 0.45, ($\blacktriangledown$) 0.45 - 0.80, ($\blacksquare$) 0.80 - 1.15, and (x) 1.15-1.50.    
          \label{fig:GSFplot_NoZPminus}}
\epsscale{1.0}
\end{figure}

The calculation of the $ZPSF^{meas-ran}$ is carried out for images in each photometric sample,
and the structure function $GSF^{meas-ran}$ separately computed for non-photometric images grouped into five ranges of $\overline{E^{gray}} - ZPT$.
The results are shown in Figure \ref{fig:GSFplot_NoZPminus}.
The structure functions extracted from the second and third-order gray models are nearly identical, 
and all approach zero as $d^{sky} \rightarrow 0$, so contributions from random statistical errors are reasonably well subtracted.
The curves are similar at small separations where they include significant zero-point structure.
The zero-point correction becomes rather flat at separations $\ge 5\arcmin$, while the remaining structure functions grow monotonically.

\begin{figure}
\epsscale{1.0}
\plotone{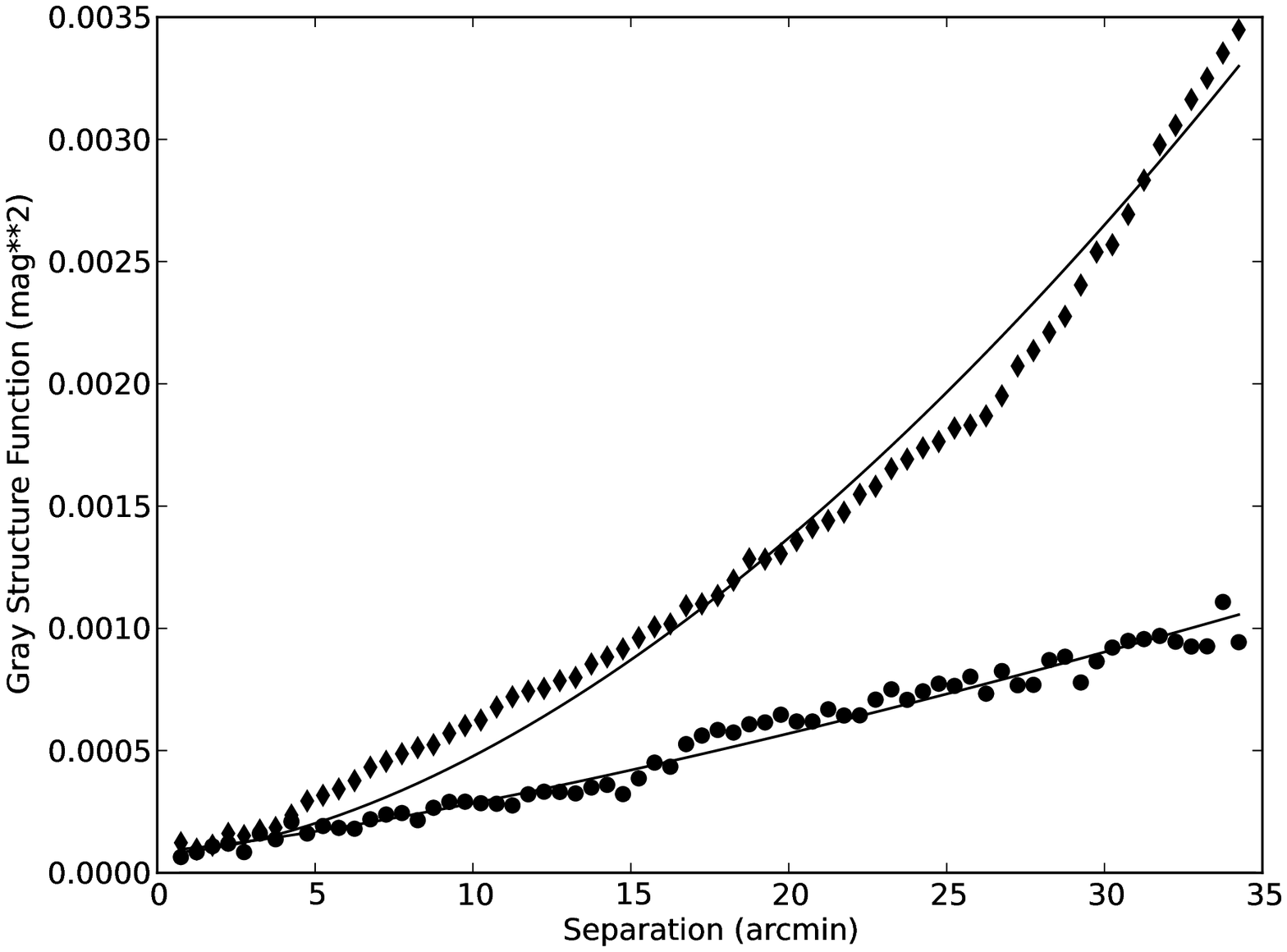}
\plotone{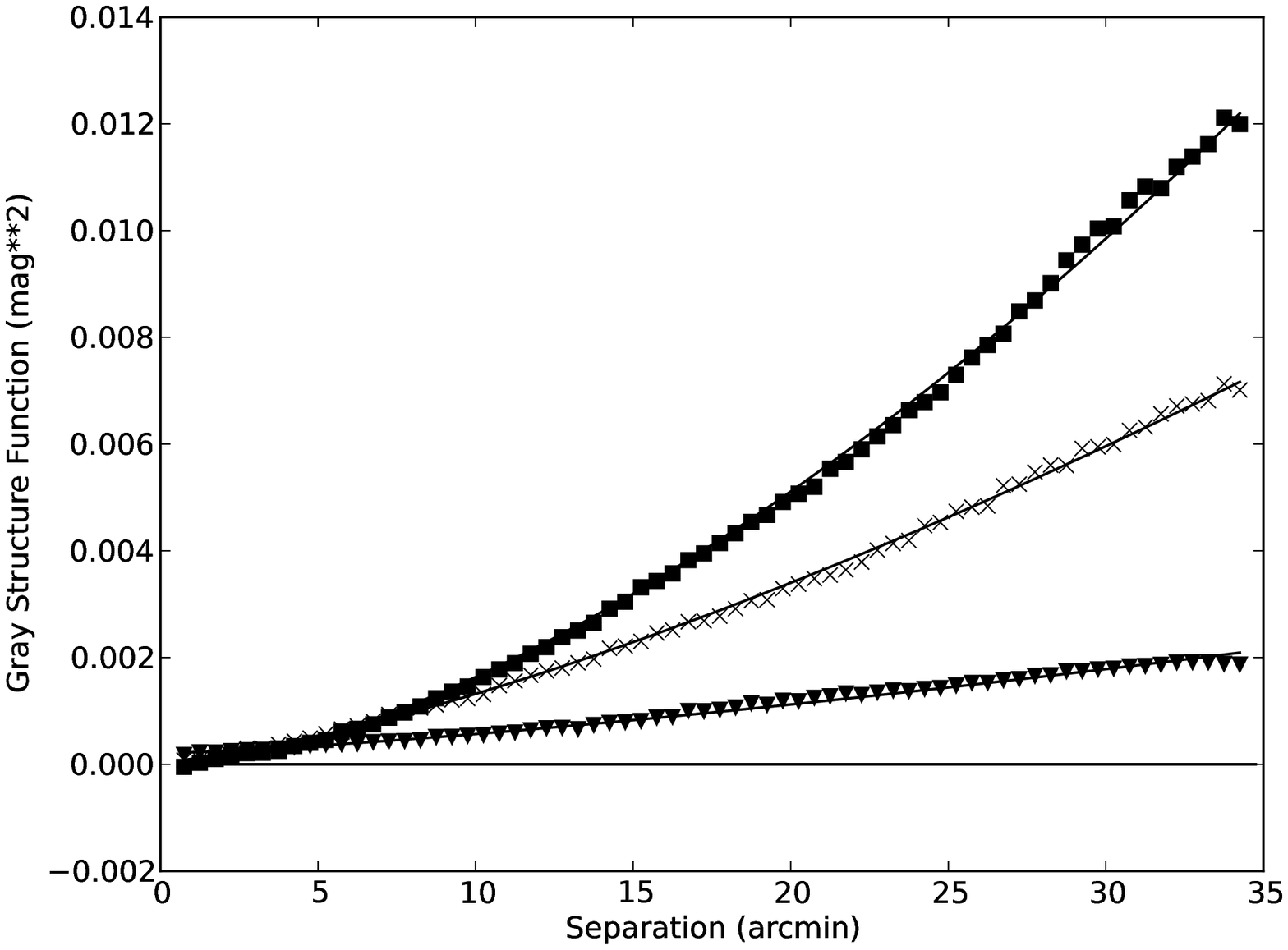}
\caption{Gray extinction structure functions measured with the third-order model after subtraction of biases from random photometric errors
     and zero-point structure.  Also shown are fits of power-law models summarized in Table \ref{table:GSFsum}.
     Structure functions are averaged over images with mean gray values (top) ($\bullet$) 0.02 - 0.10, ($\blacklozenge$) 0.10 - 0.45,
     and (bottom) ($\blacktriangledown$) 0.45 - 0.80, ($\blacksquare$) 0.80 - 1.15, and (x) 1.15-1.50. 
     Uncertainties estimated from variances of the data are shown in Figure \ref{fig:GSFerr};  
     point-to-point correlations are reflected in estimated errors in the fitted model parameters (see text).
          \label{fig:GSFplot}}
\epsscale{1.0}
\end{figure}

\begin{figure}
\epsscale{1.0}
\plotone{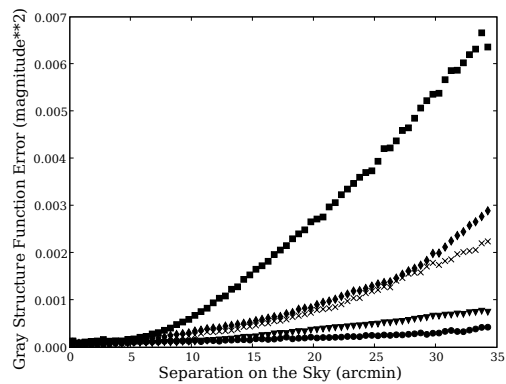}
\caption{One standard deviation errors in the averaged structure functions shown in Figure \ref{fig:GSFplot}  
    estimated from the dispersions of the $GSF$ computed from individual images in each mean gray range: 
    ($\bullet$) 0.02 - 0.10, ($\blacklozenge$) 0.10 - 0.45, ($\blacktriangledown$) 0.45 - 0.80, ($\blacksquare$) 0.80 - 1.15, and (x) 1.15-1.50.
    These uncertainties in the averaged gray structure functions are dominated by sample variance.     
          \label{fig:GSFerr}}
\epsscale{1.0}
\end{figure}

We compute the fully-corrected measured structure functions for non-photometric data using the third-order gray model,    
\begin{equation}
\label{eqn:cloudGSF}
GSF(d^{sky}) \!\! \equiv \!\! GSF^{meas-ran}(d^{sky}) - ZPSF^{meas-ran}(d^{sky}),
\end{equation}
and plot the results in Figure \ref{fig:GSFplot} overlaid with fits to the data from a model described below.
Points on a given curve are highly correlated, while the statistical photometric errors are rather small.
The interpretation of the measurements is limited by the size of the image sample.
We estimate the sampling errors for the averaged structure functions by computing the $GSF(d^{sky})$ individually for each calibration image.
The dispersions of the individual images within a fixed bin of mean gray are then 
used to compute estimates of the sampling errors on the averaged functions.
These are plotted in Figure \ref{fig:GSFerr} where they can be seen to be 30-50\% of the central values in Figure \ref{fig:GSFplot}.
It is interesting that the normalizations of the measured structure functions are not strictly ordered by the mean gray of the data bin;
but we note that the estimated sampling errors are consistent with considerable variation of cloud conditions during the observing run.

\subsection{Cloud Structure and Calibration Precision}
\label{sec:GSFs}

We look next at what can be said quantitatively about the effects of cloud structure on our ``all weather'' photometric calibration.
To do so, we use a model to connect physical atmospheric structures of clouds with our measured gray extinction.

Driven by the importance of understanding and forecasting conditions in the Earth's climate, extensive studies 
of the characteristics of cloud formation and evolution have been carried out from ground, airborne, and satellite platforms\footnote{See 
http://www.grss-ieee.org, http://www.noaa.gov, and http://modis.gsfc.nasa.gov.}. 
Most of these have been done with techniques that lack the spatial resolution needed to address the structure
of cloud extinction within the field of view of the MOSAIC II camera on the Blanco.
A few studies, however, provide relevant guidance. 
High power 94 GHz radar has been used to reconstruct full 3d structures of both cirrus and storm clouds with 15m spatial resolution \citep{fliflet06}, 
and sampling of liquid water content by aircraft has been used to measure structure at sub-meter scales \citep{davis1999}.
These studies and others find that spatial structure of clouds in the upper atmosphere (e.g. cirrus) is consistent with
formation under the influence of fluid turbulence described by well-known theory \citep{kolmogorov41}.

Our premise is that the structure functions of gray extinction reported here are determined by 2d projections of the 3d cloud structure
along the line of sight;
such a projection of Kolmogorov 3d structure function will scale with the $5/3$ power of the 2d spatial separation \citep{roddier1981}.
This has been found to be a good approximation for high cirrus, and more generally, a power $d^{\beta}$ with $1 < \beta < 2$ is found to
describe a range of formations that might be consistent with telescope operations (Davis {\it op. cit.}; Fliflet \& Manheimer {\it op. cit.}).
We note that techniques using measurements of reflected radar or direct sampling of liquid water will have different sensitivities
to variations in the sizes and shapes of water droplets and ice crystals than does transmission of optical light;
but we only use these results as a general guide to fitting our data.

Motivated by the existing data on cloud structures we define a model $GSF^{mdl}$ for our measured gray structure functions,
\begin{equation}
\label{eqn:mdlGSF}
GSF^{mdl}(d^{sky}) \equiv  GSF_0 + GSF_{20} \times (d^{sky}/20)^{\beta} ,
\end{equation}
where $d^{sky}$ is measured in arcminutes, and $GSF_0$, $GSF_{20}$, and $\beta$ are parameters fit to the data.
In this form, the $GSF_0$ parameter accounts for residual errors in the subtraction of random and zero-point correlation errors
(a nuisance parameter), and $GSF_{20}$ is the value of our model for cloud structure at $d^{sky} = 20$ arcminutes.  
We note that our data are sensitive to cloud thickness averaged over an extent and direction determined by the tracking of the telescope
and movement of clouds during the exposure of the image.
It is not our goal to study cloud formation per se, but rather to construct a model that we can use to estimate the effect of clouds on the 
calibration of our data.
While with a larger data set it might be possible to search for directional dependence of the calibration, 
we use $GSF^{mdl}$ as a fitting function for the spatially and temporally averaged gray structure that we measure.

\begin{table*} 
\caption{Summary of Cloud Structure Function Model Parameters.
         \label{table:GSFsum}}
\begin{center} 
\begin{tabular}{ccccccc} 
\tableline 
\noalign{\smallskip}
  Gray Bin  &  Gray Range     &   $  \beta  $    &     $GSF_{20}$     &    $R_{20}$    &  \multicolumn{2}{c}{$\Delta_{GSF}^{mdl} (d^{sky})$ (rms mag)}   \\
            &   (mag)         &                  &     (mag$^2$)     &               &  ($1\arcmin$)     &  ($2\arcmin$)   \\
\noalign{\smallskip}
\tableline
\noalign{\smallskip}
 1          &   0.02 - 0.10   &     1.30         &    0.00048        &   0.365       &       0.003       &   0.005                  \\
 2          &   0.10 - 0.45   &     1.75         &    0.00124        &   0.128       &       0.003       &   0.005                  \\
 3          &   0.45 - 0.80   &     1.35         &    0.00091        &   0.048       &       0.004       &   0.006                  \\
 4          &   0.80 - 1.15   &     1.60         &    0.00524        &   0.074       &       0.007       &   0.011                  \\
 5          &   1.15 - 1.50   &     1.70         &    0.00336        &   0.044       &       0.007       &   0.012                  \\
\noalign{\smallskip}
\tableline
\end{tabular} 
\end{center}
\end{table*}

We extract model parameters for each mean gray data sample by manually stepping $\beta$ in increments of 0.05 units,
and then determine $GSF_0$ and $GSF_{20}$ by minimizing a $\chi^2$ weighted with only the estimated statistical photometric errors.
The results of this procedure are included in Figure \ref{fig:GSFplot}
and the fitted values of $\beta$ and $GSF_{20}$ are given in Table \ref{table:GSFsum}.
As discussed in the previous section, the measurement is limited by the number of images available in the sample.
So, to estimate uncertainties of the fitted parameters,
we use a jackknife procedure and refit subsets of the images formed by eliminating one image at a time.
This yields an estimate of $0.20 - 0.30$ for the standard deviation of the values of $\beta$ given in the table.

The fitted $GSF^{mdl}(d^{sky})$ are seen in Figure \ref{fig:GSFplot} to be rather good descriptions of the average data,
so we use them to examine the impact of cloud structure on calibration of the photometry in our data sample.
We compute the model prediction for growth with distance on the sky of the root-mean-square of the gray extinction
\begin{equation}
\Delta_{GSF}^{mdl}(d^{sky}) \equiv \sqrt { GSF^{mdl}(d^{sky}) - GSF_0  }.
\label{eqn:cloudmag}
\end{equation}
Values of $\Delta_{GSF}^{mdl}$ computed at $d^{sky}$ = 1$\arcmin$ and 2$\arcmin$ are included in Table \ref{table:GSFsum} for each mean gray range.
Also included in the table are values for a measure of the roughness of the gray extinction relative to the mean value
\begin{equation}
R_{20} \equiv  \sqrt {GSF_{20} } / E_{mid}^{gray},
\label{eqn:R20}
\end{equation}
where $E_{mid}^{gray}$ is the central value of the mean gray range.
The total contribution to the overall calibration error grows with increasing cloud thickness (last column in Table \ref{table:GSFsum}),
while structure relative to the mean absorption represented by $R_{20}$ is seen to be greatest for the thinnest layers.

A previous study of cloud extinction in SDSS has been reported in the literature \citep{ivezic07}.
The structure function in that data was found to have a stronger dependence on separation ($\beta \sim 2$),
but with amplitudes at $d^{sky} = 30\arcmin$ separation (the column separation in the SDSS imager) $\sim 2-5$ times smaller.
This may reflect differences of the observatory site, exposure times, or may simply be variance of
atmospheric conditions during the acquisition of the particular data samples.

\section{Discussion and Conclusions}
\label{sec:discussion}

The principal results of this paper are those in Tables \ref{table:models}, \ref{table:fitresults}, \ref{table:calgray}, and \ref{table:GSFsum},
and in Figures \ref{fig:GSFplot} and \ref{fig:GSFerr}.
Our observations span a reasonable range of atmospheric conditions, but certainly do not include all possible types or thicknesses of cloud cover.
Nevertheless, the data do allow a number of important general issues to be addressed, and we can draw some conclusions from our analyses:

\begin{itemize}
\item We have demonstrated a method to calibrate imaging data to relative precisions of a per cent or better through cloud layers
that produce as much as a magnitude of wavelength-independent extinction.  The technique relies on obtaining a sufficiently large number
of repeated observations of each field, at least some of which are made in photometric conditions.
But it does not rely on identification of the images that were taken in photometric conditions; it is an ``all weather'' calibration.
\item The standard deviation $\sigma^{cal}$ of calibrated observations of test stars made in conditions with
minimal cloud cover (deemed ``photometric'' after the fact)
was found to be $\approx 0.005$ magnitude in our fields at lower Galactic latitudes (Table \ref{table:calgray}).
\item The degradation of the reproducability of the calibration with increasing cloud thickness (Table \ref{table:calgray})
is consistent with the measured structure of clouds when the calibration is interpolated over the effective
spacings of calibration stars on the sky $d^{sky} \sim 1\arcmin-2\arcmin$ at lower Galactic latitudes (Table \ref{table:GSFsum}).
\item For the conditions encountered during the observing reported here, the structure functions of thin cloud layers were greater
relative to their mean thicknesses than those of thicker (and apparently relatively smoother) cloud layers.
So thin broken clouds can produce substantial instability in calibration results,     
though the precision of the calibration will degrade steadily with increasing cloud thickness (Tables \ref{table:calgray} and \ref{table:GSFsum}).
This seems to support the observer's adage that atmospheric conditions are either photometric or not, and implies that it will be
important for surveys to identify ``photometric'' observing opportunities as rapidly and accurately as possible.  
\item To fully utilize observing time, it may be advantagous for surveys to implement auxiliary instrumentation
specifically dedicated to characterization of thin cloud structure on relatively fine spatial scales (e.g. sensitive thermal IR cameras or
lidar systems) \cite{benzvi06}.
\item To take full advantage of the techniques studied here these surveys will want to account for the amount and character of cloud cover.
For example, it will be advantagous to concentrate high Galactic latitude observations during photometric conditions, and reserve
observing during cloudy conditions for low Galactic fields in order to maintain a sufficient density of calibration stars in view.
\item There are several possible improvements to the calibration procedure presented here that could be explored with larger data sets.
Our choice of polynomial models for cloud structure may not be optimum;
for example, fields with sufficiently high densities of stars might be calibrated with pixelated corrections.
It might also be possible to pre-select a ``photometric'' sample of images for each field that could be fit with a simpler
geometric or temporally smoothed model to determine reference magnitudes for the calibration stars;
these values could then be used in re-calibration of the overall survey (including ``non-photometric'' data). 
\item We have not addressed the spatial uniformity or absolute accuracy of the photometric zero-points or measured colors.
Calibration of absolute broad-band scales will require different techniques.
The wavelength dependence of the instrumental throughput can be measured with good precision using controlled light sources and specially
designed dispersal systems \citep{sandt06}.
And proposals have been made to use observations of well characterized white dwarf stars \citep{holberg06},
or analysis of the color distributions of main sequence stars \citep{macdonald04, ivezic04, sale09, high09} to determine
the accuracy of measured colors.
Calibration of the absolute accuracy of the flux scale remains yet another problem, though one upon which relatively few science analyses depend.
\end{itemize}

\acknowledgments
The authors thank St\'ephane Blondin for reductions of the spectroscopic data taken with the SMARTS 1.5m Cassegrain.
This work has been done as part of the design and development activity of the Large Synoptic Survey Telescope (LSST).
LSST project activities are supported in part by the National Science Foundation through Governing Cooperative Agreement 0809409
managed by the Association of Universities for Research in Astronomy (AURA),
and the Department of Energy under contract DE-AC02-76-SFO0515 with the SLAC National Accelerator Laboratory.
Additional LSST funding comes from private donations, grants to universities, and in-kind support from LSSTC Institutional Members.

Facilities: \facility{SMARTS 1.5m Cassegrain Spectrograph and Blanco/MOSAIC II Imager}


\end{document}